\documentclass[preprint,prc,aps]{revtex4}
\usepackage{subfigure}
\usepackage{graphicx}
\usepackage{pdfpages}
\usepackage{graphicx,epstopdf}

\begin{document}

\title{Nonlinear waves in strongly interacting relativistic fluids}

\author{D.A. Foga\c{c}a\dag\,  F.S. Navarra\dag\ and L.G. Ferreira Filho\ddag\ }
\address{\dag\ Instituto de F\'{\i}sica, Universidade de S\~{a}o Paulo\\
 C.P. 66318, 05315-970 S\~{a}o Paulo, SP, Brazil}
\address{\ddag\ Faculdade de Tecnologia, Universidade do Estado do Rio de Janeiro \\
Via Dutra km 298, 27537-000 Resende, RJ, Brazil}

\begin{abstract}

During the past decades the study of strongly interacting fluids experienced a tremendous progress. In the relativistic heavy ion accelerators,
specially the RHIC and LHC colliders, it became possible to study not only fluids made of hadronic matter but also fluids of quarks and gluons.
Part of the physics program of these machines is the observation of waves in this strongly interacting medium. From the theoretical point of view,
these waves are often treated with li-nearized hydrodynamics. In this text we review the attempts to go beyond
linearization. We show how to use the Reductive Perturbation Method  to expand the  equations of (ideal and viscous) relativistic hydrodynamics to
obtain nonlinear wave equations.  These nonlinear wave equations govern the evolution of energy density perturbations (in hot quark gluon plasma)
or baryon density perturbations (in cold quark gluon plasma and nuclear matter).  Different nonlinear wave equations, such as the breaking wave,
Korteweg-de Vries
and Burgers equations, are obtained from different equations of state (EOS).  In nuclear matter, the  Walecka EOS may lead to a KdV equation.
We explore equations of state such as those extracted from the MIT Bag Model and from QCD in the mean field theory approach.
Some of  these equations are integrable and have analytical solitonic solutions.  We derive these equations also in spherical and cylindrical
coordinates. We extend the analysis to two
and three dimensions to obtain the Kadomtsev-Petviashvili (KP) equation, which is the generalization of the KdV.  The KP is also integrable and
presents analytical solitonic solutions. In  viscous relativistic hydrodynamics we have second order patial derivatives which physically represent
dissipation terms. We present numerical solutions and their corresponding algorithms for the cases where the equations are not integrable.

\end{abstract}

\maketitle

\section{Introduction}

The elementary particles and their interactions are well described  by the Standard Model (SM), which is an extremely successful theory
\cite{text}.  In this theory matter is composed by quarks and leptons and their interactions are due to the exchange of gauge bosons.
The sector of the SM which describes the strong interactions at the fundamental level is called Quantum Chromodynamics, or QCD \cite{muta}.
According to QCD there are six types, or flavors, of quarks: up, down, strange, charm, bottom and top, and they interact exchanging
 gluons. Quarks and gluons have a special charge called color, responsible for the strong interaction. They do not exist as individual
particles but, due to the
property of color confinement, quarks and gluons form clusters called hadrons, which can be grouped in baryons and mesons. The former are
made of three quarks,  as the proton, and the latter are made of a quark and an antiquark, as the pion, for example. The quarks carry a
fraction of the elementary electric charge and a fraction of the baryon number. The electric and color charges and the baryon number are
conserved quantities in QCD.

The hadrons also interact strongly and their interactions have been traditionally described by a quantum field theory called Quantum Hadrodynamics,
or QHD, in its different versions \cite{wal,bender}. According to QHD, in nuclear matter neutrons and protons interact exchanging scalar and vector mesons.
This potentially complicated theory becomes quite simple in the mean field approximation, in which the meson fields are treated as classical fields.
The different versions of QHD in the mean field approximation are called relativistic mean field models (RMF)\cite{wal,bender}.
More recently, nuclear matter has been
studied  with Effective Field Theories, which incorporate the fundamental symmetries of QCD in hadron physics.

In the phase diagram of QCD, we can  observe that under extreme conditions of very large temperatures and/or very large densities, the
normal hadronic matter undergoes a phase transition to a deconfined phase, a new state of matter called the quark gluon plasma (QGP)
\cite{qgp-theo}.
Together with  deconfinement, a second phase transition takes place: the chiral phase transition, during which chiral symmetry is restored and the
light quarks (up and down) become massless.  The hot QGP is produced in relativistic heavy ion collisions
in the Relativistic Heavy Ion Collider (RHIC) at the Brookhaven National Laboratory (BNL) \cite{qgp-exp1,qgp-exp2}  and even more in the
Large Hadron Collider (LHC) at CERN.  The cold QGP  may  exist in the core of  compact stars \cite{qgp-stars}.

According to our present understanding of the RHIC measurements,  the QGP  behaves as an almost perfect fluid and its
space-time evolution can be very well described by relativistic hydrodynamics \cite{wein,land,hidro1}.
The discovery of this new fluid motivated inumerous theoretical works addressing viscosity in relativistic hydrodynamics
\cite{roma,visc0,visc}. At the same time, more sophisticated
measurements made possible to study the propagation of perturbations in the QGP. We may, for example, study the effect of a fast quark traversing the
hot QGP medium. As it moves supersonically throughout the fluid, it may  generate waves of energy density or baryon density \cite{ondas}.
It may be even possible that these  waves may pile up and form Mach cones, which would affect the angular  distribution of the
produced  particles, fluid fragments which are experimentally observed.

The study of waves in the quark-gluon fluid has been mostly performed with the assumption that the amplitude of the perturbations is small enough
to justify the linearization of the Euler and continuity equations \cite{hidro1}. The analysis of perturbations with
the linearized relativistic  hydrodynamics  leads to the
standard second order wave equations and their traveling wave solutions, such as acoustic waves in the QGP. While  linearization is justified in
many cases, in others it should be  replaced by another technique to treat perturbations keeping the nonlinearities of the theory.
Since long ago there is a technique which preserves nonlinearities in the derivation of the differential equations which govern the evolution of
perturbations. This is the reductive perturbation method (RPM) \cite{rpm,davidson,leblond,loke}.

\begin{figure}[ht!]
\centering
\includegraphics[scale=0.35]{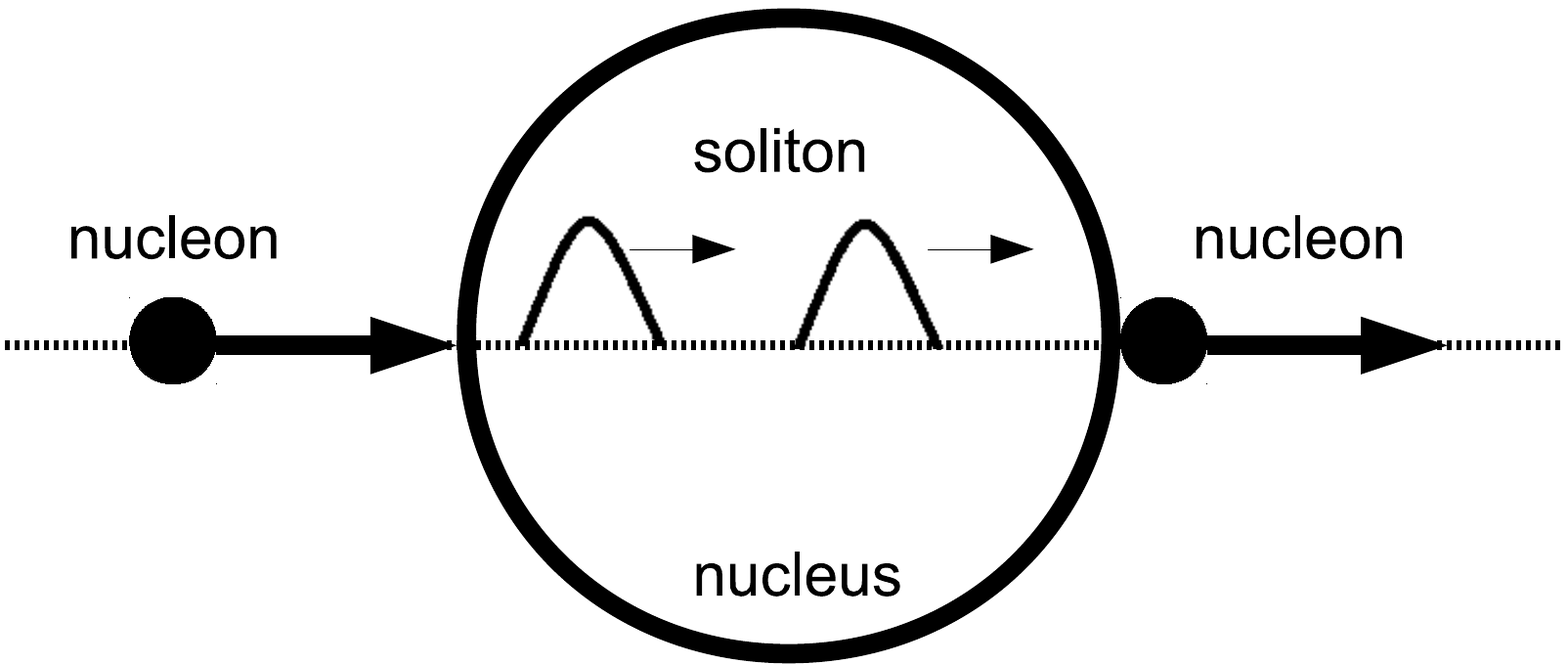}
\caption[]{KdV soliton in the baryon density simulating ``nuclear transparency''.}
\label{figsoliton}
\end{figure}

Nonlinearities may lead, as they do in other domains of  physics, to new and interesting phenomena. In a pioneering work \cite{frsw}, with the use of
nonrelativistic ideal hydrodynamics combined with the RPM and with an appropriate equation of state of cold nuclear matter, it was shown that it is
possible to derive a Korteweg- de Vries (KdV) equation for the baryon density, which has analytic solitonic solutions.
This suggests that a pulse in baryon density
(the KdV soliton) can propagate without dissipation through the nuclear medium. If this occurs this pulse  might be responsible for an interesting
phenomena: the apparent nuclear transparency in  proton nucleus collisions at low energies. The incoming proton would be absorbed by the fluid
(target nucleus) and turned into a density pulse. In the fluid the pulse satisfies the KdV equation and it is able to traverse
the target nucleus emerging on the other side. This  can be called  ``nuclear transparency'' and it is illustrated in the Fig \ref{figsoliton}.

Perturbations in fluids with  different equations of state (EOS) generate different nonlinear wave equations: the breaking wave equation,
KdV, Burgers... etc. Among these equations we find
the Kadomtsev-Petviashvili (KP) equation  \cite{KPorigin}, which  is a nonlinear wave equation in three spatial and one temporal coordinate.
It is the generalization of the KdV equation to higher dimensions. This equation has been found with the application of the reductive perturbation
method \cite{rpm} to several different problems such as the propagation of solitons in multicomponent plasmas, dust acoustic waves in hot dust
plasmas and dense electro-positron-ion plasma
\cite{dassen,wen,jukui,kp2004,maiwen,yue1,yunliang,Wangetal,mushtaq,yue2,guang,kp2010,moslempp17}.
Previous  studies on  nonlinear waves in cold and warm nuclear matter can be found in \cite{fn1,fn2,fn3,fn4,nos2010,frsw,abu}.
Works on nonlinear waves in cold QGP in the mean field approach were published in \cite{nos2011} and their extension
to three dimensions was published in \cite{nos2013}.

In this text we review the  applications of the RPM \cite{rpm,dassen,wen,jukui,kp2004,maiwen,yue1,yunliang,Wangetal,mushtaq,yue2,guang,kp2010}
to relativistic fluid dynamics \cite{wein,land}. In the next section we review the basic equations of relativistic hydrodynamics.
In section III we present the RPM and give the coordinate transformations to be used in the subsequent sections. In section IV we introduce the
equations of state of hadronic matter and of the quark-gluon plasma, giving special attention to the ingredients which may generate solitons.
In section V we list the wave equations which follow from the application of the RPM to hydrodynamics. In section VI we present the analytical
solutions of some of these differential equations and in section VII we show the time evolution of several nonlinear waves obtained by numerical
integration. In section VIII we make some final remarks.

\section{Relativistic  Hydrodynamics}

\subsection{Definitions and basic equations}

As we have mentioned before, the hot and dense  medium created in heavy ion collisions at RHIC behaves approximately as a perfect fluid
and ideal hydrodynamics can applied to describe its space-time evolution. Moreover the  study of  perturbations in the fluid, such as the
waves created by fast partons,  can be studied in the context of hydrodynamics as well.
In this section we present the formalism of relativistic hydrodynamics. Pedagogical texts on this subject can be found in \cite{wein,land}.
Here we give special attention to the variables which are relevant for strongly interacting fluids, such as the baryon density. We write the final
equations in a more extended form, which allows the reader to make a direct comparison with the corresponding  non-relativistic versions of these
equations.  In what follows we use $c=1$, $\hbar=1$ and the Boltzmann constant is taken to be one, i.e., $k_{B}=1$. All the relativistic equations are
written in terms of 4-vectors and the metric tensor is given  by $g_{\mu \nu}$, with $g_{00}=-g_{11}=-g_{22}=-g_{33}=1$ and
$g_{\mu \nu}=0$ if $\mu \neq \nu$. From \cite{wein,land,roma} the fundamental equations of a relativistic ideal fluid are:
\begin{equation}
D\varepsilon + (\varepsilon+p)\partial_{\mu}u^{\mu}=0
\label{f1}
\end{equation}
and
\begin{equation}
(\varepsilon+p)Du^{\alpha} - \nabla^{\alpha}p=0
\label{f1}
\end{equation}
where:
\begin{equation}
D \equiv u^{\mu}\partial_{\mu} \hspace{2cm} \textrm{and} \hspace{2cm}  \nabla^{\alpha} \equiv \Delta^{\mu\alpha} \partial_{\mu}
\label{opers}
\end{equation}
and the projection operator on the orthogonal direction to the fluid velocity $u^{\mu}$ is given by the tensor:
\begin{equation}
\Delta^{\mu\nu} = g^{\mu\nu}-u^{\mu}u^{\nu}
\label{dten}
\end{equation}
which has the properties $\Delta^{\mu\nu}u_{\mu} = \Delta^{\mu\nu}u_{\nu}=0$ and
$ \Delta^{\mu\nu} \Delta^{\alpha}_{\nu}= \Delta^{\mu\alpha}$.

The velocity 4-vector of the fluid element is given by \cite{wein,land,roma}: $u^{\mu}=(\gamma,\gamma\vec{v})$ where $\gamma$ is the
Lorentz factor $\gamma=(1-v^{2})^{-1/2}$ and thus $u^{\mu}u_{\mu}=1$ and $u_{\mu}\partial_{\nu}u^{\mu}=(1/2)\partial_{\nu}(u^{\mu}u_{\mu})=
(1/2)\partial_{\nu}(1)=0$.

In an  ideal fluid  all dissipative (viscous) effects are neglected and in order to introduce the  effects of viscosity, it is necessary
to add the viscous stress tensor $\Pi^{\mu\nu}$ to the energy-momentum tensor.  In a simple way we may write for a viscous fluid:
\begin{equation}
T^{\mu\nu} = T_{(0)}^{\mu\nu}+\Pi^{\mu\nu}
\label{emtensor}
\end{equation}
where
\begin{equation}
T_{(0)}^{\mu\nu}=\varepsilon u^{\mu}u^{\nu}-p \Delta^{\mu\nu}
\label{emtensor_2}
\end{equation}
is the ideal relativistic
fluid energy-momentum tensor \cite{wein,land}.  We also consider for simplicity \cite{roma} a system without conserved charges (or at zero chemical potential) and so the total momentum density is due to the
flow of energy density $u_{\mu}T^{\mu\nu} =  \varepsilon u^{\nu}$ and hence we must assume that:
\begin{equation}
u_{\mu}\Pi^{\mu\nu} =0
\label{eflow}
\end{equation}
We take the appropriate projections
of the conservation equations of the energy momentum tensor: the parallel $(u_{\nu}\partial_{\mu}T^{\mu\nu})$ and perpendicular $(\Delta^{\alpha}_{\nu}\partial_{\mu}T^{\mu\nu})$ to the fluid velocity.  The results are:
\begin{equation}
u_{\nu}\partial_{\mu}T^{\mu\nu}=D\varepsilon + (\varepsilon+p)\partial_{\mu}u^{\mu}
+u_{\nu}\partial_{\mu}\Pi^{\mu\nu}=0
\label{afv1}
\end{equation}
and
\begin{equation}
\Delta^{\alpha}_{\nu}\partial_{\mu}T^{\mu\nu}=(\varepsilon+p)Du^{\alpha} - \nabla^{\alpha}p
+\Delta^{\alpha}_{\nu}\partial_{\mu}\Pi^{\mu\nu}=0
\label{afv2}
\end{equation}
Using the symmetrization notation:
\begin{equation}
A_{(\mu}B_{\nu)}={\frac{1}{2}}(A_{\mu}B_{\nu}+A_{\nu}B_{\mu})
\label{symm}
\end{equation}
we are able to rewrite the $u_{\nu}\partial_{\mu}\Pi^{\mu\nu}$ term in (\ref{afv1}) as
$u_{\nu}\partial_{\mu}\Pi^{\mu\nu}=\partial_{\mu}(u_{\nu}\Pi^{\mu\nu})-\Pi^{\mu\nu}
\partial_{(\mu}u_{\nu)}$.  We also use the identity
\begin{equation}
\partial_{\mu}=u_{\mu}D+\nabla_{\mu}
\label{delmu}
\end{equation}
and the choice of frame $u_{\mu}\Pi^{\mu\nu}=0$.  The fundamental equations of relativistic viscous fluid dynamics are finally given by:
\begin{equation}
D\varepsilon + (\varepsilon+p)\partial_{\mu}u^{\mu}
-\Pi^{\mu\nu}\nabla_{(\mu}u_{\nu)}=0
\label{fv1}
\end{equation}
and:
\begin{equation}
(\varepsilon+p)Du^{\alpha} - \nabla^{\alpha}p
+\Delta^{\alpha}_{\nu}\partial_{\mu}\Pi^{\mu\nu}=0
\label{fv2}
\end{equation}

\subsection{The viscous stress tensor}

The viscous stress tensor $\Pi^{\mu\nu}$ has not been specified yet.  We shall derive it with the help of the second law of thermodynamics,
which states  that entropy density must always increase locally.  Moreover we shall consider a system in thermodynamic equilibrium $(dp=0)$
with zero chemical potential. Based on these statements, the thermodynamic relations are given by:
\begin{equation}
\varepsilon+p=Ts \hspace{2cm} \textrm{and} \hspace{2cm}  Tds=d\varepsilon
\label{termos}
\end{equation}
The second law of thermodynamics is considered in the covariant form:
\begin{equation}
\partial_{\mu} s^{\mu} \geq 0
\label{seclaw}
\end{equation}
where the 4-current $s^{\mu}$ is given by:
\begin{equation}
s^{\mu} =s u^{\mu}
\label{4s}
\end{equation}
Inserting relations (\ref{termos}) into the second law (\ref{seclaw}) and assuming that $Dp=0$ we find:
\begin{equation}
\partial_{\mu} s^{\mu}=Ds+s\partial_{\mu} u^{\mu}={\frac{D\varepsilon}{T}}+{\frac{(\varepsilon+p)}{T}}\partial_{\mu} u^{\mu}
\label{aseclawr}
\end{equation}
Using now  equation (\ref{fv1}) we obtain:
\begin{equation}
\partial_{\mu} s^{\mu}={\frac{1}{T}}\Pi^{\mu\nu}\nabla_{(\mu}u_{\nu)} \geq 0
\label{seclawr}
\end{equation}
The viscous tensor  may be decomposed into a traceless part $\pi^{\mu\nu}$ $(\pi^{\mu}_{\mu}=0)$ and a non-vanishing trace part, so that:
\begin{equation}
\Pi^{\mu\nu}=\pi^{\mu\nu}+\Delta^{\mu\nu}\Pi
\label{vistsplit}
\end{equation}
In terms of the notation  introduced in \cite{roma}:
\begin{equation}
\nabla_{\langle\mu}u_{\nu\rangle} \equiv 2\nabla_{(\mu}u_{\nu)}-{\frac{2}{3}}\Delta_{\mu\nu}\nabla_{\alpha}u^{\alpha}
\label{tracelessnot}
\end{equation}
it can be shown that:
\begin{equation}
u_{\mu}\pi^{\mu\nu}=0
\label{re1}
\end{equation}
\begin{equation}
u^{\mu}u^{\nu}\nabla_{\langle\mu}u_{\nu\rangle}=0
\label{re2}
\end{equation}
\begin{equation}
g^{\mu\nu}\nabla_{\langle\mu}u_{\nu\rangle}=0
\label{re3}
\end{equation}
and
\begin{equation}
\Delta^{\mu\nu}\Delta_{\mu\nu}=3
\label{re4}
\end{equation}
Inserting (\ref{vistsplit}) to (\ref{re4}) into (\ref{seclawr}) we find:
\begin{equation}
\partial_{\mu} s^{\mu}={\frac{1}{2T}}\pi^{\mu\nu}\nabla_{\langle\mu}u_{\nu\rangle}+{\frac{1}{T}}\Pi \nabla_{\alpha}u^{\alpha}  \geq 0
\label{slvi}
\end{equation}
and this inequality (a positive sum of squares) is satisfied if:
\begin{equation}
\pi^{\mu\nu}=\eta \nabla^{\langle\mu}u^{\nu\rangle} \hspace{0.1cm},\hspace{0.5cm}
\Pi=\zeta \nabla_{\alpha}u^{\alpha} \hspace{0.1cm},\hspace{0.5cm} \eta \geq 0  \hspace{0.4cm} \textrm{and} \hspace{0.4cm} \zeta \geq 0
\label{viscousitens}
\end{equation}
where $\eta$ is the {\it shear viscosity coefficient} and $\zeta$ is the {\it bulk viscosity coefficient}.
The final expression for the viscous tensor is given by the substitution of (\ref{viscousitens}) in (\ref{vistsplit}):
\begin{equation}
\Pi^{\mu\nu}=\eta \nabla^{\langle\mu}u^{\nu\rangle} +\zeta \Delta^{\mu\nu} \nabla_{\alpha}u^{\alpha}
\label{vistsplittensor}
\end{equation}

\subsection{The relativistic Navier-Stokes equation}

The system of equations (\ref{fv1}), (\ref{fv2}) and (\ref{vistsplittensor}) is called relativistic Navier-Stokes equation.
The temporal component $(\alpha=0)$ of (\ref{fv2}) multiplied by $v^{i}$ is given by:
\begin{equation}
(\varepsilon+p)(D\gamma) v^{i} -v^{i}\nabla^{0}p+v^{i}\Delta^{0}_{\nu}\partial_{\mu}\Pi^{\mu\nu}=0
\label{fv2tc}
\end{equation}
and the spatial component $(\alpha=i)$  of (\ref{fv2}):
\begin{equation}
(\varepsilon+p)(D\gamma) v^{i}=-(\varepsilon+p)\gamma (D v^{i})+\nabla^{i}p-\Delta^{i}_{\nu}\partial_{\mu}\Pi^{\mu\nu}
\label{fv2sc}
\end{equation}
Inserting (\ref{fv2sc}) into (\ref{fv2tc}) we find:
\begin{equation}
(\varepsilon+p)\gamma (D v^{i})+(v^{i}\nabla^{0}-\nabla^{i})p-(v^{i}\Delta^{0}_{\nu}-\Delta^{i}_{\nu})\partial_{\mu}\Pi^{\mu\nu} = 0
\label{ans}
\end{equation}
which can be rewritten with the help of (\ref{opers}) and (\ref{dten}) as:
\begin{equation}
(\varepsilon+p)\gamma^{2} \bigg({\frac{\partial}{\partial t}}+\vec{v} \cdot \vec{\nabla} \bigg) v^{i}+
v^{i}{\frac{\partial p}{\partial t}} - \partial^{i}p
-(v^{i}\Delta^{0}_{\nu}-\Delta^{i}_{\nu})\partial_{\mu}\Pi^{\mu\nu} = 0
\label{aans}
\end{equation}
The ideal relativistic fluid limit $\Pi^{\mu\nu} = 0$ $(\eta=\zeta=0)$ in the last equation provides the relativistic version of the Euler
equation \cite{wein,land,roma}. For future use we  rewrite (\ref{aans}) in detail. To do so, we recall (\ref{opers}), (\ref{dten}) and
(\ref{vistsplittensor}) to obtain the expression for $\partial_{\mu}\Pi^{\mu\nu}$ in (\ref{aans}):
$$
\partial_{\mu}\Pi^{\mu\nu} =\eta \bigg\{ \partial_{\mu}\partial^{\mu}u^{\nu}+\partial_{\mu}\partial^{\nu}u^{\mu}-\partial_{\mu}
\bigg[\gamma\bigg({\frac{\partial}{\partial t}}+\vec{v} \cdot \vec{\nabla}\bigg)(u^{\mu}u^{\nu}) \bigg]\bigg\}
$$
\begin{equation}
+\bigg(\zeta-{\frac{2}{3}}\eta \bigg)\partial^{\nu}\bigg[{\frac{\partial \gamma}{\partial t}}+\vec{\nabla}\cdot (\gamma \vec{v}) \bigg]
-\bigg(\zeta-{\frac{2}{3}}\eta \bigg)\partial_{\mu}\bigg\{u^{\mu}u^{\nu}\bigg[{\frac{\partial \gamma}{\partial t}}+\vec{\nabla}\cdot (\gamma \vec{v}) \bigg] \bigg\}
\label{delmip}
\end{equation}
and then insert this last result into (\ref{aans}). We find after some algebra \cite{nos2012}:
$$
(\varepsilon+p)\gamma^{2} \bigg({\frac{\partial}{\partial t}}+\vec{v} \cdot \vec{\nabla} \bigg) \vec{v}+
\vec{v}{\frac{\partial p}{\partial t}} + \vec{\nabla}p
$$
$$
-\eta \vec{v} \bigg\{\partial_{\mu}\partial^{\mu}\gamma+\partial_{\mu}{\frac{\partial u^{\mu}}{\partial t}}-\partial_{\mu}\bigg[\gamma\bigg({\frac{\partial}{\partial t}}+\vec{v} \cdot \vec{\nabla}\bigg)(\gamma u^{\mu}) \bigg]\bigg\}
-\vec{v}\bigg(\zeta-{\frac{2}{3}}\eta \bigg){\frac{\partial}{\partial t}}\bigg[{\frac{\partial \gamma}{\partial t}}+\vec{\nabla}\cdot (\gamma \vec{v}) \bigg]
$$
$$
+\vec{v}\bigg(\zeta-{\frac{2}{3}}\eta \bigg)\partial_{\mu} \bigg\{ \gamma u^{\mu} \bigg[{\frac{\partial \gamma}{\partial t}}+\vec{\nabla}\cdot (\gamma \vec{v}) \bigg]\bigg\}
$$
$$
+\eta \bigg\{\partial_{\mu}\partial^{\mu}(\gamma \vec{v})-\partial_{\mu}\vec{\nabla}u^{\mu}
-\partial_{\mu}\bigg[\gamma\bigg({\frac{\partial}{\partial t}}+\vec{v} \cdot \vec{\nabla}\bigg)(\gamma \vec{v} u^{\mu}) \bigg]\bigg\}
-\bigg(\zeta-{\frac{2}{3}}\eta \bigg)\vec{\nabla} \bigg[{\frac{\partial \gamma}{\partial t}}+\vec{\nabla}\cdot (\gamma \vec{v}) \bigg]
$$
\begin{equation}
-\bigg(\zeta-{\frac{2}{3}}\eta \bigg)\partial_{\mu} \bigg\{ \gamma \vec{v} u^{\mu} \bigg[{\frac{\partial \gamma}{\partial t}}+\vec{\nabla}\cdot (\gamma \vec{v}) \bigg] \bigg\}=0
\label{rnsagain}
\end{equation}
which is the relativistic version of the Navier-Stokes equation.
The perfect fluid, described by $\eta=\zeta=0$ in (\ref{rnsagain}) gives the relativistic version of Euler equation
\cite{wein,land}:
\begin{equation}
{\frac{\partial {\vec{v}}}{\partial t}}+(\vec{v} \cdot \vec{\nabla})\vec{v}=
-{\frac{1}{(\varepsilon + p)\gamma^{2}}}
\bigg({\vec{\nabla} p +\vec{v} {\frac{\partial p}{\partial t}}}\bigg)
\label{relateul}
\end{equation}

\subsection{Causality and  the relativistic Navier-Stokes equation}

The relativistic Navier-Stokes equation (\ref{rnsagain}) does not constitute a causal theory \cite{roma}.  This fact can be understood
when small perturbations are considered in a system in equilibrium  with  energy density $\varepsilon_{0}$, pressure $p_{0}$ and
at rest $(\vec{v}=\vec{0})$.

Such perturbations can be described by:
\begin{equation}
\varepsilon=\varepsilon_{0}+\delta \varepsilon({t,\vec{x}}) \,\, , \hspace{1.0cm}  p=p_{0}+\delta p({t,\vec{x}})
\hspace{0.8cm} and  \hspace{0.8cm} u^{\mu}=(1,\vec{0})+\delta u^{\mu}({t,\vec{x}})
\label{pert}
\end{equation}
where `` $\delta$ '' denotes small deviation from equilibrium.
For simplicity we consider  perturbations which   depend only on the $x$ space coordinate. For the particular direction $\alpha=y$ we insert
(\ref{pert}) into  (\ref{fv2}) to find:
$$
(\varepsilon_{0}+\delta \varepsilon+p_{0}+\delta p)D(\delta u^{y}) - \nabla^{y}(p_{0}+\delta p)
+\Delta^{y}_{\nu}\partial_{\mu}\Pi^{\mu\nu}=0
$$
Considering only the dependence on the $x$ coordinate  it becomes:
\begin{equation}
(\varepsilon_{0}+p_{0}){\frac{\partial}{\partial t}} \delta u^{y}
+{\frac{\partial}{\partial x}} \Pi^{xy} + \mathcal{O}({\delta^{2}})=0
\label{fv2pert}
\end{equation}
Analogously we insert (\ref{pert}) into the viscous tensor (\ref{vistsplit}) for the particular direction $\nu=y$ and considering only
the $x$ coordinate dependence:
\begin{equation}
\Pi^{x y}=-\eta {\frac{\partial }{\partial x}}\delta u^{y} + \mathcal{O}({\delta^{2}})
\label{viscyxp}
\end{equation}
Substituting (\ref{viscyxp}) in (\ref{fv2pert}) we find:
\begin{equation}
(\varepsilon_{0}+p_{0}){\frac{\partial}{\partial t}} \, \delta u^{y}
-\eta {\frac{\partial^{2}}{\partial x^{2}}} \, \delta u^{y}= \mathcal{O}({\delta^{2}})
\label{almostde}
\end{equation}
which provides, after performing the linearization approximation (i.e., neglecting $\mathcal{O}({\delta^{2}})$) the
diffusion-type evolution equation:
\begin{equation}
{\frac{\partial}{\partial t}} \, \delta u^{y}
-{\frac{\eta}{(\varepsilon_{0}+p_{0})}} {\frac{\partial^{2}}{\partial x^{2}}} \, \delta u^{y}= 0
\label{diffusion}
\end{equation}
for the perturbation $\delta u^{y}=\delta u^{y}(t,x)$.

The acausality property of the Navier-Stokes equation can be studied from the Laplace-Fourier wave ansatz for $\delta u^{y}$:
\begin{equation}
\delta u^{y}(t,x)=f_{(w,k)} \, e^{ikx-wt}
\label{ansatz}
\end{equation}
where $f_{(w,k)}$ are the coefficient for a given frequency $w$ and a given wave number $k$.
When inserting (\ref{ansatz}) in (\ref{diffusion}) we obtain the dispersion-relation of the diffusion equation:
\begin{equation}
w={\frac{\eta}{(\varepsilon_{0}+p_{0})}} k^{2}
\label{omega}
\end{equation}
which gives the speed of diffusion of a mode with wavenumber $k$ :
\begin{equation}
V_{diff}(k)={\frac{dw}{dk}}=2{\frac{\eta}{(\varepsilon_{0}+p_{0})}} k
\label{speeddiff}
\end{equation}
The causality violation occurs when
\begin{equation}
\lim_{k\to\infty} V_{diff}(k) \rightarrow \infty
\label{acausality}
\end{equation}
which exceeds the speed of light, i.e , the speed of diffusion grows without bound for sufficiently large wavenumber due its linear dependence
on the wavenumber.

A possible way to regulate the viscous fluid theory is given by  the ``Maxwell-Cattaneo law''. In this approach  a new transport coefficient
 called relaxation time $(\tau_{\pi})$ is added to  equation (\ref{viscyxp}) as follows:
\begin{equation}
\tau_{\pi}{\frac{\partial}{\partial t}} \, \Pi^{x y}+\Pi^{x y}=-\eta {\frac{\partial }{\partial x}}\delta u^{y} + \mathcal{O}({\delta^{2}})
\label{viscyxprelatau}
\end{equation}
The derivative of the above equation  with respect to $x$ provides:
\begin{equation}
\tau_{\pi}{\frac{\partial}{\partial t}} \, {\frac{\partial}{\partial x}} \, \Pi^{x y}+
{\frac{\partial}{\partial x}} \ \Pi^{x y}=-\eta {\frac{\partial^{2} }{\partial x^{2}}}\delta u^{y} + \mathcal{O}({\delta^{2}})
\label{viscyxprelatauxagain}
\end{equation}
which becomes, after substituting ${\frac{\partial}{\partial x}} \Pi^{xy}$ from (\ref{fv2pert}) and performing the linearization approximation,
the following evolution equation:
\begin{equation}
\tau_{\pi}{\frac{\partial^{2}}{\partial t^{2}}} \, \delta u^{y}
+{\frac{\partial}{\partial t}} \, \delta u^{y}
-{\frac{\eta}{(\varepsilon_{0}+p_{0})}} {\frac{\partial^{2}}{\partial x^{2}}} \, \delta u^{y}= 0
\label{diffusioncausal}
\end{equation}
where the term with second derivative with respect to time provides the following dispersion relation when (\ref{ansatz}) is  used:
$$
w={\frac{1}{2\tau_{\pi}}} \pm \sqrt{{\frac{1}{4{\tau_{\pi}}^{2}}}-{\frac{\eta \, k^{2}}{(\varepsilon_{0}+p_{0})\tau_{\pi}}}}
={\frac{1}{2\tau_{\pi}}} \pm i k \sqrt{{\frac{\eta}{(\varepsilon_{0}+p_{0})\tau_{\pi}}}-{\frac{1}{4{{\tau_{\pi}}^{2}}k^{2}}}}
$$
\begin{equation}
|w|^{2}=w\bar{w}=k^{2}{\frac{\eta}{(\varepsilon_{0}+p_{0})\tau_{\pi}}}
\label{omegacausal}
\end{equation}
The  speed of diffusion of a mode with wavenumber $k$ is now:
\begin{equation}
V_{diff_{causal}}(k)={\frac{d|w|}{dk}}=\sqrt{{\frac{\eta}{(\varepsilon_{0}+p_{0})\tau_{\pi}}}}
\label{speeddiffcausal}
\end{equation}
which does not violate causality:
\begin{equation}
\lim_{k\to\infty} V_{diff}(k) = \sqrt{{\frac{\eta}{(\varepsilon_{0}+p_{0})\tau_{\pi}}}}
\label{causality}
\end{equation}
where $\tau_{\pi}\neq 0$.  Beyond the Maxwell-Cattaneo law, the more complete formulation of viscous hydrodynamics is the
M\"uller-Israel-Stewart theory \cite{roma}, which contains the Maxwell-Cattaneo law as a limit.  A more precise study is found in \cite{roma}.
Here we just  wish to point out that there are alternative approaches, in which  causality is preserved.

We present a different approximation scheme, which goes beyond the linear approximation,  preserves nonlinear terms and does not violate
causality: the reductive perturbation method, a technique that will be presented in the next section.  When this method is applied to the
relativistic  Navier-Stokes equation (\ref{rnsagain}) we arrive at the conclusion that, for the purpose of studying perturbations, the equation:
\begin{equation}
{\frac{\partial \vec{v}}{\partial t}} +(\vec{v} \cdot \vec{\nabla}) \vec{v}=
-{\frac{1}{(\varepsilon+p)}}\bigg[\vec{\nabla} p + \vec{v}{\frac{\partial p}{\partial t}} \bigg] +
{\frac{1}{(\varepsilon+p)}}
\bigg[\eta \, {\vec{\nabla}}^{2}\vec{v}+
\bigg(\zeta+{\frac{1}{3}}\eta \bigg)
\vec{\nabla}(\vec{\nabla}\cdot \vec{v}) \bigg]
\label{nsorcendalter}
\end{equation}
is equivalent to (\ref{rnsagain}).  In other words, considering  (\ref{nsorcendalter}) or (\ref{rnsagain}) leads to the same wave equation.

\subsection{Continuity equations}

\subsubsection{Entropy}

The continuity equation for the entropy density  is given by (\ref{slvi}):
\begin{equation}
\partial_{\mu} s^{\mu}={\frac{1}{2T}}\eta \, \nabla^{\langle\mu}u^{\nu\rangle}\nabla_{\langle\mu}u_{\nu\rangle}+{\frac{1}{T}}\zeta \,
(\nabla_{\alpha}u^{\alpha})^{2}
\label{conts}
\end{equation}
Using (\ref{opers}), (\ref{dten}), $u_{\nu}\partial_{\mu}u^{\nu}=0$
and $u^{\nu}u_{\nu}=1$ we can rewrite  the above  equation  in the form:
\begin{equation}
\partial_{\mu} s^{\mu}=-{\frac{\eta}{T}}(\partial^{\mu}u^{\nu})\partial_{\nu}u_{\mu}
+{\frac{1}{T}}\bigg({\frac{2}{3}}\eta+\zeta \bigg) \, (\partial_{\mu}u^{\mu})^{2}
\label{contss}
\end{equation}
Using (\ref{4s}) and $u^{\mu}=(\gamma,\gamma\vec{v})$ the last equation can also be rewritten as:
$$
\gamma {\frac{\partial s}{\partial t}}+ \gamma \vec{\nabla}s \cdot \vec{v} + s{\frac{\partial \gamma}{\partial t}}+s\vec{\nabla}\gamma \cdot
\vec{v}+\gamma s \vec{\nabla} \cdot \vec{v}=
-{\frac{\eta}{T}}\bigg({\frac{\partial \gamma}{\partial t}}\bigg)^{2}
-2{\frac{\eta}{T}}\bigg[\vec{\nabla} \gamma \cdot {\frac{\partial}{\partial t}}(\gamma\vec{v})
\bigg]
$$
\begin{equation}
-{\frac{\eta}{T}}(\partial^{i}u^{j})\partial_{j}u_{i}
+{\frac{1}{T}}\bigg({\frac{2}{3}}\eta+\zeta \bigg) \, \bigg[ {\frac{\partial \gamma}{\partial t}}
+\gamma  \vec{\nabla} \cdot \vec{v} +\vec{\nabla}\gamma \cdot \vec{v} \bigg]^{2}
\label{relcontss}
\end{equation}
which is the relativistic version of the continuity equation for the entropy density $s$. In the case of an ideal fluid ($\eta=\zeta=0$) we
recover the entropy density conservation:
$$
\gamma {\frac{\partial s}{\partial t}}+ \gamma \vec{\nabla}s \cdot \vec{v} + s{\frac{\partial \gamma}{\partial t}}+s\vec{\nabla}\gamma
\cdot \vec{v}+\gamma s \vec{\nabla} \cdot \vec{v}=0
$$
which, with the use of:
\begin{equation}
{\frac{\partial \gamma}{\partial t}}=\gamma^{3}v{\frac{\partial v}{\partial t}}
\hspace{0.9cm} \textrm{and} \hspace{0.9cm} \vec{\nabla} \gamma=\gamma^{3}v\vec{\nabla} v
\label{derivgama}
\end{equation}
becomes \cite{wein,land}:
\begin{equation}
{\frac{\partial s}{\partial t}}+\gamma^{2}v s\Bigg({\frac{\partial v}
{\partial t}}+ \vec{v}\cdot \vec{\nabla} v\Bigg)+\vec{\nabla} \cdot (s\vec{v})=0
\label{idealrelcontss}
\end{equation}

\subsubsection{Baryon density}

The relativistic version of the continuity equation for the baryon density $\rho_{B}$
in ideal relativistic hydrodynamics is:
\begin{equation}
\partial_{\nu}{j_{B}}^{\nu}=0
\label{conucleon}
\end{equation}
Since ${j_{B}}^{\nu}=u^{\nu} \rho_{B}$ and using (\ref{derivgama}) the last equation can be rewritten as \cite{wein}:
\begin{equation}
{\frac{\partial \rho_{B}}{\partial t}}+\gamma^{2}v \rho_{B}\Bigg({\frac{\partial v}
{\partial t}}+ \vec{v}\cdot \vec{\nabla} v\Bigg)+\vec{\nabla} \cdot (\rho_{B}\vec{v})=0
\label{rhobcons}
\end{equation}

\subsection{Non-relativistic limit}

In what follows  we recover the non-relativistic limit of the continuity equation for the entropy density, continuity equation for the baryon
density and also for the Navier-Stokes equation.  The non-relativistic limit is essentially given by
$v^{2}<<1$ and  $\gamma \cong 1$.

\subsubsection{Continuity equation}

Using $v^{2}<<1$ and $\gamma \cong 1$ in the equation for entropy density (\ref{relcontss}) we find:
\begin{equation}
{\frac{\partial s}{\partial t}}+ \vec{\nabla} \cdot ( s \vec{v} )=
-{\frac{\eta}{T}}(\partial^{i}u^{j})\partial_{j}u_{i}
+{\frac{1}{T}}\bigg({\frac{2}{3}}\eta+\zeta \bigg) \, (\vec{\nabla} \cdot \vec{v})^{2}
\label{nonrelcontss}
\end{equation}
For a perfect fluid $\eta=\zeta=0$ we obtain the usual continuity equation \cite{wein}:
$$
{\frac{\partial s}{\partial t}}+ \vec{\nabla} \cdot ( s \vec{v} )=0
$$
For the baryon density continuity equation  we have, after applying $v^{2}<<1$ and $\gamma \cong 1$ in (\ref{rhobcons}):
\begin{equation}
{\frac{\partial \rho_{B}}{\partial t}}+\vec{\nabla} \cdot (\rho_{B}\vec{v})=0
\label{nonrelrhobcons}
\end{equation}

\subsubsection{Navier-Stokes equation}

The energy density of the fluid element of mass $M$ and momentum $P$ is:
$$
\varepsilon={\frac{E}{Vol}}={\frac{\sqrt{M^{2}+P^{2}}}{Vol}}
$$
In the non-relativistic limit: $M>>P$ we have:
$$
\varepsilon \cong {\frac{\sqrt{M^{2}}}{Vol}}={\frac{M}{Vol}}
$$
The volumetric density of fluid matter is given by ${\frac{M}{Vol}}=\rho$ and hence:
\begin{equation}
\varepsilon \cong \rho
\label{massmom}
\end{equation}
The pressure is defined as the ratio of the force per area of the fluid element $A$.
The force is given by the time derivative of the  momentum $P$.  The pressure is then:
\begin{equation}
p={\frac{\textrm{force}}{A}}={\frac{1}{A}}{\frac{dP}{dt}}
\label{predef}
\end{equation}
and so
$$
{\vec{\nabla} p +\vec{v} {\frac{\partial p}{\partial t}}}= {\frac{1}{A}}\vec{\nabla}{\frac{dP}{dt}} +
{\frac{1}{A}}\vec{v}{\frac{d^{2}P}{dt^{2}}}
$$
For $v^{2}<<1 $ we find:
$$
|\vec{v}|{\frac{d^{2}P}{d t^{2}}}<< \bigg|\vec{\nabla}{\frac{dP}{dt}}\bigg|
$$
and consequently:
$$
{\vec{\nabla} p +\vec{v} {\frac{\partial p}{\partial t}}}\cong{\frac{1}{A}}\vec{\nabla}{\frac{dP}{dt}}
$$
With the use of (\ref{predef})   we finally obtain:
\begin{equation}
{\vec{\nabla} p +\vec{v} {\frac{\partial p}{\partial t}}} \cong \vec{\nabla}p
\label{relaps}
\end{equation}
According to these last approximations and using (\ref{massmom}) we conclude that $\varepsilon +p$ takes the form:
$$
\varepsilon + p \cong \rho + {\frac{1}{A}}{\frac{dP}{d t}}
$$
and with $\rho >> {\frac{1}{A} \frac{dP}{dt}}$ (because $M>>P$) we find:
\begin{equation}
\varepsilon + p \cong \rho
\label{soma}
\end{equation}
Inserting $\gamma \cong 1$, (\ref{relaps}) and (\ref{soma}) in (\ref{rnsagain}) we find:
$$
\rho\bigg[{\frac{\partial \vec{v}}{\partial t}} +(\vec{v} \cdot \vec{\nabla}) \vec{v}\bigg] + \vec{\nabla} p
$$
$$
\eta \bigg\{-{\vec{\nabla}}^{2}\vec{v}-\vec{\nabla}(\vec{\nabla}\cdot \vec{v})-\bigg(
{\frac{\partial \vec{v}}{\partial t}} \cdot \vec{\nabla} \bigg) \vec{v}
-(\vec{v}\cdot \vec{\nabla}){\frac{\partial \vec{v}}{\partial t}}
-\bigg(\vec{\nabla} \cdot {\frac{\partial \vec{v}}{\partial t}}\bigg)\vec{v}
$$
$$
-\vec{\nabla}\cdot \vec{v} \,\, {\frac{\partial \vec{v}}{\partial t}}
-\vec{v} \cdot \vec{\nabla}
\Big[(\vec{v}\cdot \vec{\nabla})\vec{v}\Big] \bigg\}
$$
\begin{equation}
+ \bigg(\zeta-{\frac{2}{3}}\eta \bigg)
\bigg\{-\vec{\nabla}(\vec{\nabla}\cdot \vec{v}) -{\frac{\partial \vec{v}}{\partial t}} \, \vec{\nabla}\cdot \vec{v} -\vec{v} \,\, \vec{\nabla} \cdot {\frac{\partial \vec{v}}{\partial t}}
-\Big[(\vec{v}\cdot \vec{\nabla})\vec{v}\Big]\vec{\nabla}\cdot \vec{v} \bigg\}=0
\label{nonrnsagain}
\end{equation}
Neglecting the terms of order $v^{3}$, which are  $\vec{v} \cdot \vec{\nabla} \Big[(\vec{v}\cdot \vec{\nabla})\vec{v}\Big]$ and
$ \Big[(\vec{v}\cdot \vec{\nabla})\vec{v}\Big]\vec{\nabla}\cdot \vec{v}$ we have:
$$
\rho\bigg[{\frac{\partial \vec{v}}{\partial t}} +(\vec{v} \cdot \vec{\nabla}) \vec{v}\bigg] + \vec{\nabla} p -\eta{\vec{\nabla}}^{2}\vec{v}-\bigg(\zeta+{\frac{1}{3}}\eta \bigg)
\vec{\nabla}(\vec{\nabla}\cdot \vec{v})
$$
\begin{equation}
-\bigg(\zeta+{\frac{1}{3}}\eta \bigg) \, {\frac{\partial}{\partial t}}\Big[\vec{v} \,(\vec{\nabla}\cdot \vec{v})  \Big]
-\eta \, {\frac{\partial}{\partial t}}\Big[ (\vec{v} \cdot \vec{\nabla})\vec{v} \Big]=0
\label{nonrnsagainn}
\end{equation}
We still need to estimate the relative size of the terms in last equation.  This is made by the comparison
of terms without viscosity to others with viscosity.
From (\ref{nonrnsagainn}):
$$
\rho(\vec{v} \cdot \vec{\nabla}) \vec{v}-\eta \, {\frac{\partial}{\partial t}}\Big[ (\vec{v} \cdot \vec{\nabla})\vec{v} \Big]={\frac{M}{Vol}}(\vec{v} \cdot \vec{\nabla}) \vec{v}
-{\frac{\eta}{M^{2}}}{\frac{\partial}{\partial t}}\Big[ (\vec{P} \cdot \vec{\nabla})\vec{P} \Big]
$$
where we have used the volumetric density of the fluid matter ${\frac{M}{Vol}}=\rho$
and the non-relativistic momentum $\vec{P}=M\vec{v}$ for the element with mass $M$ for the fluid matter.  In the non-relativistic limit we have, as usual: $M>>P$ and the last expression becomes:
\begin{equation}
\rho(\vec{v} \cdot \vec{\nabla}) \vec{v}-\eta \, {\frac{\partial}{\partial t}}\Big[ (\vec{v} \cdot \vec{\nabla})\vec{v} \Big]
\cong \rho(\vec{v} \cdot \vec{\nabla}) \vec{v}
\label{otherap}
\end{equation}
since
$$
{\frac{M}{Vol}}(\vec{v} \cdot \vec{\nabla}) \vec{v} >>
-{\frac{\eta}{M^{2}}}{\frac{\partial}{\partial t}}\Big[ (\vec{P} \cdot \vec{\nabla})\vec{P} \Big]
$$
Analogously we also have from (\ref{nonrnsagainn}):
\begin{equation}
\rho {\frac{\partial \vec{v}}{\partial t}}-\bigg(\zeta+{\frac{1}{3}}\eta \bigg) \, {\frac{\partial}{\partial t}}\Big[\vec{v} \,(\vec{\nabla}\cdot \vec{v})  \Big]
\cong \rho {\frac{\partial \vec{v}}{\partial t}}
\label{otherapp}
\end{equation}
since
$$
{\frac{M}{Vol}} {\frac{\partial \vec{v}}{\partial t}} >> -\bigg(\zeta+{\frac{1}{3}}\eta \bigg) \, {\frac{1}{M^{2}}}{\frac{\partial}{\partial t}}\Big[\vec{P} \,(\vec{\nabla}\cdot \vec{P})  \Big]
$$
The non-relativistic Navier-Stokes equation is then given by using (\ref{otherap}) and (\ref{otherapp}) in (\ref{nonrnsagainn}):
\begin{equation}
{\frac{\partial \vec{v}}{\partial t}} +(\vec{v} \cdot \vec{\nabla}) \vec{v}=
-{\frac{1}{\rho}}\vec{\nabla} p +{\frac{\eta}{\rho}}{\vec{\nabla}}^{2}\vec{v}+
{\frac{1}{\rho}}\bigg(\zeta+{\frac{1}{3}}\eta \bigg)
\vec{\nabla}(\vec{\nabla}\cdot \vec{v})
\label{nonrnsagainnfinal}
\end{equation}
Again, the perfect fluid is described by $\eta=\zeta=0$ in (\ref{nonrnsagainnfinal}) and gives the non-relativistic version of Euler equation
\cite{wein,land}:
\begin{equation}
{\frac{\partial \vec{v}}{\partial t}} +(\vec{v} \cdot \vec{\nabla}) \vec{v}=
-{\frac{1}{\rho}}\vec{\nabla} p
\label{nonrelateul}
\end{equation}

\section{The Reductive Perturbation Method }

\subsection{Linearization}

We start from the equations of ideal relativistic hydrodynamics and, using the linearization approximation, we derive a wave equation for
perturbations in the pressure. This equation has traveling wave solutions which represent acoustic waves. In the derivation presented here we follow
closely the reference \cite{hidro1}. In the presence of perturbations the energy density and pressure for the relativistic fluid are written
as (\ref{pert}):
\begin{equation}
\varepsilon(\vec{r},t)=\varepsilon_{0}+\delta \varepsilon(\vec{r},t)
\label{15}
\end{equation}
and
\begin{equation}
p(\vec{r},t)=p_{0}+\delta p(\vec{r},t)
\label{16}
\end{equation}
respectively.  The uniform relativistic fluid is defined by $\varepsilon_{0}$ and $p_{0}$,  while
$\delta \varepsilon$ and $\delta p$ correspond to perturbations in this fluid.  Energy-momentum conservation implies that for an ideal fluid $\Pi^{\mu\nu}=0$ and from
(\ref{emtensor}) we have:
\begin{equation}
\partial_{\mu}T^{\mu\nu}=0
\label{7}
\end{equation}
where $T^{\mu\nu}$ is the energy-momentum tensor given by:
\begin{equation}
T^{\mu\nu}=(\varepsilon+p)u^{\mu}u^{\nu}-pg^{\mu\nu}
\label{enermomtensor}
\end{equation}
Linearization consists in keeping only first order terms such as $\delta \varepsilon$, $\delta P$ and $\vec{v}$ and
neglect terms proportional to:
\begin{equation}
{{v}}^{2}, \hspace{0.2cm} v\delta \varepsilon, \hspace{0.2cm} v \delta P, \hspace{0.2cm}
\vec{v} \cdot \vec{\nabla}v, \hspace{0.2cm} (\vec{v} \cdot \vec{\nabla}) \vec{v}
\label{aplin}
\end{equation}
and also neglect higher powers of these products or other combinations  of  them.
Naturally we have $\gamma \sim 1$ .
From (\ref{7}) we have:
\begin{equation}
u^{\mu}\partial_{\nu}[(\varepsilon + p)u^{\nu}]+(\varepsilon + p)u^{\nu}\partial_{\nu}u^{\mu}-\partial_{\nu}(pg^{\nu\mu})=0
\label{8ag}
\end{equation}
The temporal component ($\mu=0$) of the above equation is given by:
\begin{equation}
\gamma\partial_{0}[(\varepsilon + p)\gamma]+\gamma\partial_{i}[(\varepsilon + p)u^{i}]+
(\varepsilon + p)u^{0}\partial_{0}\gamma+(\varepsilon + p)u^{i}\partial_{i}\gamma-\partial_{0}p=0
\label{17}
\end{equation}
which, after using (\ref{aplin}) and $\gamma \sim 1$,   becomes:
$$
\partial_{0}(\varepsilon + p)+\partial_{i}[(\varepsilon + p)v^{i}]-\partial_{0}p=0
$$
or
\begin{equation}
{\frac{\partial \varepsilon}{\partial t}}+ \vec{\nabla} \cdot [(\varepsilon + p)\vec{v}]=0
\label{energcons}
\end{equation}
For the $j$-th spatial component ($\mu=j$)  in (\ref{8ag}) we have:
$$
u^{j}\partial_{0}[(\varepsilon + p)u^{0}]
+u^{j}\partial_{i}[(\varepsilon + p)u^{i}]+(\varepsilon + p)u^{0}\partial_{0}u^{j}+
(\varepsilon + p)u^{i}\partial_{i}u^{j}-\partial^{j}p=0
$$
which, with the use of (\ref{aplin}) and $\gamma \sim 1$,  becomes:
\begin{equation}
{\frac{\partial}{\partial t}}[(\varepsilon + p)\vec{v}]+ \vec{\nabla}p=0
\label{newtonl}
\end{equation}
Substituting the expansions (\ref{15}) and (\ref{16}) in (\ref{energcons}) and (\ref{newtonl}) we find:
\begin{equation}
{\frac{\partial }{\partial t}}[\varepsilon_{0}+\delta \varepsilon]+
\vec{\nabla} \cdot [(\varepsilon_{0}+\delta \varepsilon + p_{0}+\delta p) \vec{v}]=0
\label{energconsexp}
\end{equation}
and
\begin{equation}
{\frac{\partial}{\partial t}}[(\varepsilon_{0}+\delta \varepsilon + p_{0}+\delta p)\vec{v}]+
\vec{\nabla}[p_{0}+\delta p]=0
\label{newtonlexp}
\end{equation}
Neglecting the terms listed in  (\ref{aplin}) in (\ref{energconsexp}) and (\ref{newtonlexp}) these equations become:
\begin{equation}
{\frac{\partial (\delta \varepsilon)}{\partial t}}+
(\varepsilon_{0}+p_{0})\vec{\nabla} \cdot  \vec{v}=0
\label{enerlin}
\end{equation}
and
\begin{equation}
(\varepsilon_{0}+p_{0}){\frac{\partial \vec{v}}{\partial t}}+
\vec{\nabla}(\delta p)=0
\label{newtolin}
\end{equation}
Equation (\ref{enerlin}) expresses  energy conservation and equation (\ref{newtolin}) is  Newton's second law.
Integrating (\ref{newtolin}) with respect to the time and setting the integration constant to zero we find:
\begin{equation}
\vec{v}=-{\frac{1}{(\varepsilon_{0}+p_{0})}}\int{\vec{\nabla}(\delta p)} dt
\label{velolin}
\end{equation}
which inserted in (\ref{enerlin}) yields:
\begin{equation}
{\frac{\partial (\delta \varepsilon)}{\partial t}}-
\int \vec{\nabla}^{2}(\delta p)dt=0
\label{enerlina}
\end{equation}
Performing the time derivative we obtain:
\begin{equation}
{\frac{\partial^{2} (\delta \varepsilon)}{\partial t^{2}}}-
\vec{\nabla}^{2}(\delta p)=0
\label{enerlinaa}
\end{equation}
Assuming that
\begin{equation}
\delta \varepsilon={\frac{\partial \varepsilon}{\partial p}} \delta p
\label{presom}
\end{equation}
with $\partial \varepsilon / \partial p$ being a constant, we have (\ref{enerlinaa}) rewritten  as:
\begin{equation}
{\frac{\partial \varepsilon}{\partial p}}{\frac{\partial^{2} (\delta p)}{\partial t^{2}}}
-\vec{\nabla}^{2}(\delta p)=0
\label{enerlinaaa}
\end{equation}
The above expression  is a wave equation from where we can identify the velocity of propagation as:
\begin{equation}
c_{s}=\bigg({\frac{\partial p}{\partial \varepsilon}}\bigg)^{1/2}
\label{som}
\end{equation}
where $c_{s}$ is the speed of sound. Equation  (\ref{enerlinaaa}) can then be finally  written as:
\begin{equation}
\vec{\nabla}^{2}(\delta p)
-{\frac{1}{{c_{s}}^{2}}}{\frac{\partial^{2} (\delta p)}{\partial t^{2}}}=0
\label{enerlinfinal}
\end{equation}
which describes the  propagation of a pressure wave in the fluid.
This study ensures that the analysis of perturbations with
the linearized relativistic  hydrodynamics  leads to the
standard second order linear wave equations and their traveling wave solutions, such as acoustic waves in the hadronic medium.

\subsection{Beyond linearization}

While  linearization is justified in many cases, in others, where perturbations are not so small,  it should be  replaced by another
technique to treat perturbations keeping the nonlinearities of the theory.  This is where
a physical theory, in our case  relativistic hydrodynamics, may benefit from developments in applied mathematics. Indeed, since long ago
there is a technique which
preserves nonlinearities in the derivation of the differential equations which govern the evolution of perturbations. This is the
reductive perturbation method (RPM) \cite{rpm,leblond,loke,davidson}.

We start the RPM description by considering the simple linear wave equation:
\begin{equation}
\frac{\partial F}{\partial t}+ \alpha \frac{\partial F}{\partial x}=0
\label{wave}
\end{equation}
which describes one-dimensional waves in an ideal fluid. In the above equation $\alpha$ is a constant. Let us expand of $F(x,t)$ around the
constant value $F_{0}$ in terms of  the small expansion parameter
$\sigma$  ($0 < \sigma < 1$) :
\begin{equation}
F(x,t)=F_{0}+\sigma F_{1}(x,t)+\sigma^{2}F_{2}(x,t)+\sigma^{3}F_{3}(x,t)+ \dots
\label{exp}
\end{equation}
Inserting (\ref{exp}) into (\ref{wave}) we obtain a series:
\begin{equation}
\sum_{j} \, \sigma^{j}\Bigg\{\frac{\partial F_{j}}{\partial t}+ \alpha \frac{\partial F_{j}}{\partial x} \Bigg\}=0 \hspace{1cm}  \,\,\,
\textrm{for \,\,\,\,\,  $j=1,2,3,\dots$}
\label{linear}
\end{equation}
The coefficients of each power of $\sigma$ must vanish independently, i.e., each bracket must vanish. This condition yields a set of differential
equations for the $F_j$:
\begin{equation}
\frac{\partial F_{j}}{\partial t}+ \alpha \frac{\partial F_{j}}{\partial x} =0
\label{lineara}
\end{equation}
Now we consider the simplest nonlinear wave equation, the so called breaking wave equation:
\begin{equation}
\frac{\partial F}{\partial t}+\alpha F\frac{\partial F}{\partial x}=0
\label{bw}
\end{equation}
where $\alpha$ is a real coefficient.  Performing the expansion (\ref{exp}) the above equation becomes:
\begin{equation}
\sigma\frac{\partial F_{1}}{\partial t}+\sigma^{2}\frac{\partial F_{2}}{\partial t}+
\sigma \alpha F_{0}\frac{\partial F_{1}}{\partial x}
+\sigma^{2} \alpha F_{0} \frac{\partial F_{2}}{\partial x}
+\sigma^{2} \alpha F_{1}\frac{\partial F_{1}}{\partial x} +  \dots  =0
\label{bwe}
\end{equation}
Let us look for simple nonlinear differential equations for the perturbations $F_1$, $F_2$, ... .
We expect to find  algebraic structures similar to  (\ref{bw}) in  equation (\ref{bwe}) in order $\sigma$ and $\sigma^{2}$.
To do so, we introduce the new coordinates $\tau$ and $\xi$, which are connected to $t$ and $x$. This coordinate transformation is
such that:
\begin{equation}
\frac{\partial }{\partial t}=\sigma^{n}\frac{\partial }{\partial \tau}
\hspace{1cm}  \textrm{and} \hspace{1cm}
\frac{\partial }{\partial x}=\sigma^{m}\frac{\partial }{\partial \xi}
\label{stre}
\end{equation}
where $n$ and $m$ will be determined. We note that a highly nontrivial aspect of this transformation is that it contains the
same small  parameter $\sigma$ used in the expansion (\ref{bwe}).
Inserting (\ref{stre}) into (\ref{bwe}) we find the following equation:
\begin{equation}
\sigma^{n+1} \frac{\partial F_{1}}{\partial \tau}
+\sigma^{n+2}\frac{\partial F_{2}}{\partial \tau}
+\sigma^{m+2} \alpha F_{1}\frac{\partial F_{1}}{\partial \xi}
+\sigma \alpha F_{0} \Bigg[
\sigma^{m}\frac{\partial F_{1}}{\partial \xi}
+\sigma^{m+1} \frac{\partial F_{2}}{\partial \xi}\Bigg]=0
\label{bwes}
\end{equation}
where, for the sake of simplicity,  we have chosen  $F_0 = 0$ and we have truncated the sum keeping only the lowest order terms in $\sigma$.
We observe that if we choose:
\begin{equation}
n+1=m+2 \hspace{1cm}  \textrm{and}  \hspace{1cm} n+2 \geq 3
\label{rel}
\end{equation}
the first and third terms can be grouped together and, for the particular choice $n=3/2$ and $m=1/2$, equation (\ref{bwes}) becomes:
\begin{equation}
\sigma^{2} \Bigg[ \frac{\partial F_{1}}{\partial \tau}
+\alpha F_{1}\frac{\partial F_{1}}{\partial \xi} \Bigg]
+\sigma^{3} \Bigg[ \frac{\partial F_{2}}{\partial \tau}\Bigg]=0
\label{bwefas}
\end{equation}
where, just before arriving at the above expression,  we have divided the whole equation by $ \sigma^{1/2}$.  From the term proportional to
$\sigma^{2}$ we have:
\begin{equation}
 \frac{\partial F_{1}}{\partial \tau}
+\alpha F_{1}\frac{\partial F_{1}}{\partial \xi}=0
\label{bwefasfinal}
\end{equation}
as expected. Including all the higher order terms, $F_2$, $F_3$, ... etc, we would obtain a set of differential equations. Solving them we
would be able to write the complete series (\ref{exp}). In practice, since $\sigma$ is small, this series is dominated by the first terms and
it is often sufficient to compute only $F_1$.

From  this simple exercise we conclude that for a nonlinear wave equation such as (\ref{bw}) it is  necessary to introduce the
``stretching'' operators (\ref{stre}):
\begin{equation}
\frac{\partial }{\partial t}=\sigma^{3/2}\frac{\partial }{\partial \tau}
\hspace{1cm}  \textrm{and also} \hspace{1cm}
\frac{\partial }{\partial x}=\sigma^{1/2}\frac{\partial }{\partial \xi}
\label{stref}
\end{equation}
which are related to the ``stretched'' coordinates:
\begin{equation}
\xi=\sigma^{1/2}x
\hspace{0.2cm} \hspace{1.5cm}   \textrm{and}            \hspace{1.5cm}
\tau=\sigma^{3/2}t
\label{prototypestret}
\end{equation}
If there was a dissipative term in (\ref{bw}), the wave equation would have the following form:
\begin{equation}
\frac{\partial F}{\partial t}+\alpha F\frac{\partial F}{\partial x}=\nu
\frac{\partial^{2} F}{\partial x^{2}}
\label{burg}
\end{equation}
which is called Burgers equation with a dissipative (viscous) coefficient $\nu$.   It becomes, after performing the  expansion (\ref{exp}) and
keeping only terms up to order $\sigma^2$:
\begin{equation}
\sigma\frac{\partial F_{1}}{\partial t}+\sigma^{2}\frac{\partial F_{2}}{\partial t}+
\sigma \alpha F_{0}\frac{\partial F_{1}}{\partial x}
+\sigma^{2} \alpha F_{0} \frac{\partial F_{2}}{\partial x}
+\sigma^{2} \alpha F_{1}\frac{\partial F_{1}}{\partial x}
=\nu \sigma \frac{\partial^{2} F_{1}}{\partial x^{2}}+
\nu \sigma^{2} \frac{\partial^{2} F_{2}}{\partial x^{2}}
\label{burge}
\end{equation}
As before, we try to find in equation (\ref{burge}) a structure similar to (\ref{burg}). Keeping the previous choice $n=3/2$\,\, , $m=1/2$ and
applying  (\ref{stref}) to (\ref{burge}) we find:
\begin{equation}
\sigma^{5/2} \Bigg[ \frac{\partial F_{1}}{\partial \tau}
+\alpha F_{1}\frac{\partial F_{1}}{\partial \xi} \Bigg]
+\sigma^{7/2} \Bigg[ \frac{\partial F_{2}}{\partial \tau}\Bigg]=
\nu \sigma^{2} \frac{\partial^{2} F_{1}}{\partial \xi^{2}}
+\nu \sigma^{3} \frac{\partial^{2} F_{2}}{\partial \xi^{2}}
\label{burgefa}
\end{equation}
When dissipative effects are included we perform the following transformation in the dissipation coefficient \cite{stvis1,stvis2}:
\begin{equation}
\nu=\sigma^{1/2} \, \tilde{\nu}
\label{viscstreta}
\end{equation}
in (\ref{burgefa}), resulting in:
\begin{equation}
\sigma^{5/2} \Bigg[ \frac{\partial F_{1}}{\partial \tau}
+\alpha F_{1}\frac{\partial F_{1}}{\partial \xi} \Bigg]
+\sigma^{7/2} \Bigg[ \frac{\partial F_{2}}{\partial \tau}\Bigg]=
\tilde{\nu} \sigma^{5/2} \frac{\partial^{2} F_{1}}{\partial \xi^{2}}
+\tilde{\nu} \sigma^{7/2} \frac{\partial^{2} F_{2}}{\partial \xi^{2}}
\label{burgefaa}
\end{equation}
As before, we divide the last equation by $\sigma^{1/2}$, obtaining:
\begin{equation}
\sigma^{2} \Bigg[ \frac{\partial F_{1}}{\partial \tau}
+\alpha F_{1}\frac{\partial F_{1}}{\partial \xi} \Bigg]
+\sigma^{3} \Bigg[ \frac{\partial F_{2}}{\partial \tau}\Bigg]=
\tilde{\nu} \sigma^{2} \frac{\partial^{2} F_{1}}{\partial \xi^{2}}
+\tilde{\nu} \sigma^{3} \frac{\partial^{2} F_{2}}{\partial \xi^{2}}
\label{burgesaag12}
\end{equation}
The terms proportional to $\sigma^{3}$ in (\ref{burgesaag12}) are neglected and from the $\sigma^{2}$ terms we have:
\begin{equation}
 \frac{\partial F_{1}}{\partial \tau}
+\alpha F_{1}\frac{\partial F_{1}}{\partial \xi} =
\tilde{\nu}\frac{\partial^{2} F_{1}}{\partial \xi^{2}}
\label{burgefaag12final}
\end{equation}
which is a Burgers equation like (\ref{burg}).
In addition, we conclude that when a dissipative term is present in a nonlinear wave equation, apart from (\ref{stref}) and (\ref{prototypestret}),
we also need the transformation (\ref{viscstreta}) for the dissipation coefficient.

If a more complete wave equation is considered, i.e., when  nonlinear, dissipative and dispersive terms are present, we have the
Korteweg-de Vries Burgers (KdV-B) equation:
\begin{equation}
\frac{\partial F}{\partial t}+\alpha F\frac{\partial F}{\partial x}+
\beta\frac{\partial^{3} F}{\partial x^{3}}=\nu\frac{\partial^{2} F}{\partial x^{2}}
\label{kdvb}
\end{equation}
Now, $\beta$ is the dispersive coefficient.  Performing the same calculation we naturally find:
\begin{equation}
\sigma^{2} \Bigg[ \frac{\partial F_{1}}{\partial \tau}
+\alpha F_{1}\frac{\partial F_{1}}{\partial \xi}+
\beta\frac{\partial^{3} F_{1}}{\partial \xi^{3}}\Bigg] =
\sigma^{2} \Bigg[\tilde{\nu}\frac{\partial^{2} F_{1}}{\partial \xi^{2}}\Bigg]
\label{kdvb12}
\end{equation}
which gives the KdV-B:
\begin{equation}
\frac{\partial F_{1}}{\partial \tau}
+\alpha F_{1}\frac{\partial F_{1}}{\partial \xi}+
\beta\frac{\partial^{3} F_{1}}{\partial \xi^{3}} =\tilde{\nu}\frac{\partial^{2} F_{1}}{\partial \xi^{2}}
\label{kdvb12final}
\end{equation}
When more dimensions and different coordinate systems are considered, the procedure can be systematically improved.

\subsection{Some special cases}

In the framework of the RPM we transport the  equations of  hydrodynamics from the space of cartesian, spherical
or cylindrical coordinates to the space of the  ``stretched  coordinates''.  Some well known equations and the coordinate transformations and
expansions required to obtain them  are given below.

\subsubsection{ One dimensional KdV equation}

We  write this equation in cartesian coordinates ($x$), in radial cylindrical coordinate and in radial spherical coordinate ($r$).
The corresponding ``stretched coordinates'' are given by $\xi$ for space and $\tau$ for time as:
\cite{fn1,fn2,fn3,fn4,frsw,abu,nos2010,nos2011a}:
\begin{equation}
\xi={\frac{\sigma^{1/2}}{L}}(\mathcal{X}-{c_{s}}t)
\label{xi}
\end{equation}
and
\begin{equation}
\tau={\frac{\sigma^{3/2}}{L}}{c_{s}}t
\label{tau}
\end{equation}
where $\mathcal{X}=x$ or $\mathcal{X}=r$.  The above transformation of  coordinates must be made simultaneously with the following expansions of
the baryon density, energy density and fluid velocity around their equilibrium values:
\begin{equation}
\hat\rho={\frac{\rho_{B}}{\rho_{0}}}=1+\sigma \rho_{1}+ \sigma^{2} \rho_{2}+ \dots
\label{oneroexp}
\end{equation}
\begin{equation}
\hat\varepsilon={\frac{\varepsilon}{\varepsilon_{0}}}=1+\sigma \varepsilon_{1}+ \sigma^{2} \varepsilon_{2}+ \dots
\label{oneenergexpansdedd}
\end{equation}
\begin{equation}
\hat v={\frac{v}{c_{s}}}=\sigma v_{1}+ \sigma^{2} v_{2}+ \dots
\label{onedvexp}
\end{equation}
For two and three dimensional nonlinear wave equations we will use the  ``stretched coordinates'' presented in \cite{nos2013,nos2012}
and references therein.

\subsubsection{Two dimensional cylindrical breaking wave equation}

In this case the coordinate transformation is given by:
\begin{equation}
R={\frac{\sigma^{1/2}}{L}}(r-{c_{s}}t)
\hspace{0.2cm}, \hspace{0.5cm}
Z={\frac{\sigma}{L}}z
\hspace{0.2cm}, \hspace{0.5cm}
T={\frac{{\sigma^{3/2}}}{L}}{c_{s}}t
\label{streta23cdim}
\end{equation}
The energy density and fluid velocity components are expanded around their equilibrium values:
\begin{equation}
\hat\varepsilon={\frac{\varepsilon}{\varepsilon_{0}}}=1+\sigma \varepsilon_{1}+ \sigma^{2} \varepsilon_{2}
+\sigma^{3} \varepsilon_{3}+ \dots
\label{roexpa23c}
\end{equation}
\begin{equation}
\hat {v_{r}}={\frac{v_{r}}{c_{s}}}=\sigma v_{{r}_1}+ \sigma^{2} v_{{r}_2}+ \sigma^{3} v_{{r}_3}+ \dots
\label{vexpa23c}
\end{equation}
\begin{equation}
\hat {v_{z}}={\frac{v_{z}}{c_{s}}}=\sigma^{3/2} v_{{z}_1}+ \sigma^{5/2} v_{{z}_2}+ \sigma^{7/2} v_{{z}_3}+ \dots
\label{vexpar23c}
\end{equation}

\subsubsection{Three dimensional cylindrical KP equation}

In this case  we have:
\begin{equation}
R={\frac{\sigma^{1/2}}{L}} (r-{c_{s}}t)
\hspace{0.2cm}, \hspace{0.5cm}
\Phi={\sigma^{-1/2}}\varphi
\hspace{0.2cm}, \hspace{0.5cm}
Z={\frac{\sigma}{L}}z
\hspace{0.2cm}, \hspace{0.5cm}
T={\frac{{\sigma^{3/2}}}{L}}c_{s}t
\label{stretaaagain}
\end{equation}
and for the baryonic density and fluid velocity components expansions are:
\begin{equation}
\hat\rho={\frac{\rho_{B}}{\rho_{0}}}=1+\sigma \rho_{1}+ \sigma^{2} \rho_{2} + \sigma^{3} \rho_{3}+ \dots
\label{roexpargain}
\end{equation}
\begin{equation}
\hat {v_{r}}={\frac{v_{r}}{c_{s}}}=\sigma v_{{r}_1}+ \sigma^{2} v_{{r}_2} + \sigma^{3} v_{{r}_3}+ \dots
\label{vexpargain}
\end{equation}
\begin{equation}
\hat {v_{\varphi}}={\frac{v_{\varphi}}{c_{s}}}=\sigma^{3/2} v_{{\varphi}_1}+ \sigma^{5/2} v_{{\varphi}_2}+ \sigma^{7/2} v_{{\varphi}_3}+ \dots
\label{phivexpaargain}
\end{equation}
\begin{equation}
\hat {v_{z}}={\frac{v_{z}}{c_{s}}}=\sigma^{3/2} v_{{z}_1}+ \sigma^{5/2} v_{{z}_2}+ \sigma^{7/2} v_{{z}_3}+ \dots
\label{vexpaargain}
\end{equation}
\begin{equation}
{\hat\rho}\,^{4/3}=\big[1+(\sigma \rho_{1}+ \sigma^{2} \rho_{2}+ \dots)\big]^{4/3}
\cong 1+{\frac{4}{3}}\sigma\rho_{1}+{\frac{4}{3}} \sigma^{2} \rho_{2}+ \dots
\label{roexpaqtrgain}
\end{equation}
\begin{equation}
{\hat\rho}\,^{1/3}=\big[1+(\sigma \rho_{1}+ \sigma^{2} \rho_{2}+ \dots)\big]^{1/3}
\cong 1+{\frac{1}{3}}\sigma\rho_{1}+{\frac{1}{3}} \sigma^{2} \rho_{2}+ \dots
\label{roexpautrgain}
\end{equation}

\subsubsection{Three dimensional cartesian KP equation}

Here the stretched coordinates are:
\begin{equation}
X={\frac{\sigma^{1/2}}{L}} (x-{c_{s}}t)
\hspace{0.2cm}, \hspace{0.5cm}
Y={\frac{\sigma}{L}}y
\hspace{0.2cm}, \hspace{0.5cm}
Z={\frac{\sigma}{L}}z
\hspace{0.2cm}, \hspace{0.5cm}
T={\frac{{\sigma^{3/2}}}{L}}c_{s}t
\label{streta}
\end{equation}
and the expansions are:
\begin{equation}
\hat\rho={\frac{\rho_{B}}{\rho_{0}}}=1+\sigma \rho_{1}+ \sigma^{2} \rho_{2} + \sigma^{3} \rho_{3}+ \dots
\label{roexpa}
\end{equation}
\begin{equation}
\hat {v_{x}}={\frac{v_{x}}{c_{s}}}=\sigma v_{{x}_1}+ \sigma^{2} v_{{x}_2} + \sigma^{3} v_{{x}_3}+ \dots
\label{vexpa}
\end{equation}
\begin{equation}
\hat {v_{y}}={\frac{v_{y}}{c_{s}}}=\sigma^{3/2} v_{{y}_1}+ \sigma^{2} v_{{y}_2}+ \sigma^{5/2} v_{{y}_3}+ \dots
\label{vexpaa}
\end{equation}
\begin{equation}
\hat {v_{z}}={\frac{v_{z}}{c_{s}}}=\sigma^{3/2} v_{{z}_1}+ \sigma^{2} v_{{z}_2}+ \sigma^{5/2} v_{{z}_3}+ \dots
\label{vexpaa}
\end{equation}
\begin{equation}
{\hat\rho}\,^{4/3}
\cong 1+{\frac{4}{3}}\sigma\rho_{1}+{\frac{4}{3}} \sigma^{2} \rho_{2}+ \dots
\label{roexpaqt}
\end{equation}
\begin{equation}
{\hat\rho}\,^{1/3}
\cong 1+{\frac{1}{3}}\sigma\rho_{1}+{\frac{1}{3}} \sigma^{2} \rho_{2}+ \dots
\label{roexpaut}
\end{equation}
In all cases $L$ is a characteristic length scale of the problem,
$\varepsilon_{0}$ is the equilibrium (or reference) energy density, $\rho_{0}$ is the equilibrium (or reference) baryon density and
$c_{s}$ is the speed of sound.  When viscosity is included, we perform the transformation (\ref{viscstreta}) as in \cite{stvis1,stvis2}:
\begin{equation}
\zeta=\sigma^{1/2} \, \tilde{\zeta} \hspace{2cm} \textrm{and also}  \hspace{2cm}   \eta=\sigma^{1/2} \, \tilde{\eta}
\label{stv}
\end{equation}
Once the equations are in the ``stretched spaces'' and expanded, we neglect terms proportional to $\sigma^{n}$ for $n > 2$ and
organize the equations as series in powers of $\sigma$, $\sigma^{3/2}$ and $\sigma^{2}$.  These equations form a system of differential
equations which are combined to yield the final nonlinear equation for the relevant perturbation.   Finally we transform the nonlinear
equation back to the cartesian (cylindrical or spherical) space and solve it.

\section{The equation of state}

The equations of hydrodynamics discussed in the previous sections must be supplemented with an equation of state (EOS), i.e., a relation between
pressure ($p$) and energy density ($\varepsilon$) or matter density ($\rho$). In relativistic hydrodynamics, we can write the EOS as:
\begin{equation}
p = {{c_s}^{2}} \varepsilon
\label{eosrel}
\end{equation}
where the speed of sound $c_s$ must be smaller than one. The equation of state is derived from microscopic theories of the strongly interacting system.
As mentioned above, the fundamental theory of strong interactions (QCD) predicts that cold and/or dilute systems are in the hadronic phase, where the
degrees of freedom are baryons (proton, neutron, $\Delta$, ...) and  mesons ($\pi$, $\rho$, ...). At higher densities and/or temperatures, there is a
phase transition and the formation of  the quark gluon plasma (QGP),  a phase where these particles are free and all hadrons are dissolved.  In all phases
we may have thermal equilibrium and hydrodynamical behavior. In both hadron and quark-gluon fluids we may have excitations and the nonlinear propagation
of perturbations. In what follows we shall study the formation of these nonlinear waves and the role played by the equation of state.

In the pioneering study of Refs. \cite{frsw,abu} the authors studied the equations of non-relativistic hydrodynamics and, with the help of the RPM,
they were able to derive a  KdV (Korteweg de-Vries) equation for a perturbation in the nuclear density.  A key ingredient in that work was the equation
of state, which established the following relation between pressure ($p$) and density ($\rho$):
\begin{equation}
\frac{\vec{\nabla} p}{\rho} = \vec{\nabla} \phi
\label{gradp}
\end{equation}
where $\phi$ is a potential, playing the role of heat function, given by:
\begin{equation}
\phi  = \frac{1}{\rho_0} \left[ c_1 \rho' + c_2 \vec{\nabla}^2 \rho' + ... \right]
\label{potfi}
\end{equation}
where $c_1$ and $c_2$ are constants and $\rho' $ is the deviation of the density from its equilibrium value $\rho' = \rho - \rho_0$. The Laplacian in
(\ref{potfi}), combined with the gradient in (\ref{gradp}) gives origin to the cubic derivative of the KdV equation. As a result the authors arrived at the conclusion that a KdV soliton may exist in cold nuclear matter, when, for example, a light nucleus is impinged on a heavy  nucleus.  This conclusion relied
strongly on (\ref{potfi}) and (\ref{gradp}), which come from an oversimplified description of the nuclear interactions.  In \cite{fn1,fn2,fn3,fn4} the
nuclear interactions were described in terms of a relativistic  mean field model, which is a variant of the Walecka model. Within this framework we may try
to answer the question: what is the microscopic origin of higher order derivative terms appearing in the equation of state? This will be discussed in the
following sections.

\subsection{Hadronic  Matter}

In this subsection we present an equation of state derived from a successful relativistic mean field model: Quantum Hadrodynamics (QHD) or Walecka Model. For a modern approach, using chiral power counting in an effective field theory for nuclear matter with nucleons and
pions as degrees of freedom, see Ref. \cite{lac}.

This model is well established and is the subject of textbooks as  Refs.  \cite{wal}.  The Lagrangian density of nonlinear QHD is
given by:
$$
\mathcal{L}=\bar{\psi}[\gamma_{\mu}(i \partial^{\mu} - g_{V}V^{\mu})-(M-g_{S} \phi)]\psi +
{\frac{1}{2}}\Big(\partial_{\mu} \phi \partial^{\mu} \phi - {m_{S}}^{2} \phi^{2}\Big)
-{\frac{1}{4}}F_{\mu \nu}F^{\mu \nu}+
$$
\begin{equation}
+{\frac{1}{2}}{m_{V}}^{2}V_{\mu}V^{\mu}-{\frac{b}{3}}\phi^{3}-{\frac{c}{4}}\phi^{4}
+ \mathcal{L}_d
\label{lagra}
\end{equation}
where $F_{\mu \nu} =  \partial_{\mu} V_{\nu} - \partial_{\nu} V_{\mu}$ and:
\begin{equation}
\mathcal{L}_d =  d{\frac{g_{V}}{{m_{V}}^{2}}}\bar{\psi}(\partial_{\nu} \partial^{\nu} V_{\mu})\gamma^{\mu} \psi
\label{lagra-d}
\end{equation}
Except for the last term, this is the standard nonlinear QHD Lagrangian which is able to reproduce all the main features of nuclear matter and
finite nuclei. This last term was added only in \cite{fn1,fn2,fn3,fn4} and illustrates how to include higher order derivative terms in the equation
of state.

The Lagrangian (\ref{lagra-d}) is the modern version of  (\ref{potfi}) and (\ref{gradp}).
In (\ref{lagra}) the degrees of freedom are the baryon field $\psi$, the neutral scalar meson field $\phi$
and the neutral vector meson field $V_{\mu}$, with the respective couplings and masses.
The last and new term  (\ref{lagra-d}) is  designed to be small in comparison with the main
baryon-vector meson interaction term $g_{v} \bar{\psi} \gamma_{\mu} V^{\mu}  \psi$. Because of the derivatives, it is of the order of:
\begin{equation}
\frac{p^2}{m_V^2} \sim \frac{k_F^2}{m_V^2} \sim 0.12
\label{estimate}
\end{equation}
where the Fermi momentum is $k_F\simeq 0.28$ GeV and $m_V \simeq 0.8$ GeV.
The form chosen for the new interaction term is not dictated by any symmetry argument, has no
other deep justification and is just one possible interaction term among many others.  It
is used here as a prototype to study the effects of  higher derivative terms, which, as it will be seen,
may generate more complex wave equations, such as KdV. The parameter  $d$ is free and plays the role of  a ``marker''. Setting $d$ equal to zero
switches off the new term and we recover the usual QHD. On the other hand $d=1$ means that the coupling $g_V$ is the standard
one. Other values imply a correction in this coupling.
As mentioned in the beginning of this section, interesting phenomena, such as KdV solitons, may appear as a consequence of the use of equations of
state with higher order derivative terms.

Baryon number propagation in nuclear matter satisfies the diffusion equation:
\begin{equation}
{\frac{\partial \rho_{B}}{\partial t}} = D \,  \nabla^{2}\rho_{B}
\label{diffusion}
\end{equation}
where the diffusion constant $D$ has been  numerically calculated and studied as a function of density and
temperature. For example, in \cite{shin} it was found that $D \simeq 0.35 \, fm$ at
densities comparable to the equilibrium nuclear density and temperatures of the order of
$80$ $MeV$.  This number is small compared to any nuclear size scale and can be interpreted
as indicating that
\begin{equation}
{\frac{\partial \rho_{B}}{\partial t}} <<  \nabla^{2}\rho_{B}
\label{compara}
\end{equation}
and therefore the density gradients do not disappear very rapidly in nuclear matter. Because of the above  inequality we can neglect the
time derivatives in (\ref{lagra-d}).

The usual mean field theory (MFT) approximation
is based on two assumptions: a) the baryonic sources are intense and their coupling to the  meson fields is strong and
b) infinite nuclear matter is static, homogeneous in space  and isotropic.

The first assumption above is  implemented  performing the substitutions:
\begin{equation}
V_{\mu} \rightarrow <V_{\mu}> \equiv \delta_{\mu 0} V_{0}
\label{apum}
\end{equation}
and
\begin{equation}
\phi \rightarrow <\phi> \equiv \phi_{0}
\label{apdois}
\end{equation}
in ({\ref{lagra}}) and obtaining
$\mathcal{L^{*}}=\mathcal{L} \Big(V_{\mu} \rightarrow <V_{\mu}> \equiv \delta_{\mu 0} V_{0}\hspace{0.1cm}
\hspace{0.2cm}\phi \rightarrow <\phi> \equiv \phi_{0}\Big) $:
$$
\mathcal{L^{*}}=\bar{\psi}[(i\gamma_{\mu} \partial^{\mu} - g_{V}\gamma_{0}V_{0})
-(M-g_{S} \phi_{0})]\psi +{\frac{1}{2}} \Big(\partial_{\mu}\phi_{0}\partial^{\mu}\phi_{0}-{m_{S}}^{2}
{\phi_{0}}^{2} \Big)+
$$
\begin{equation}
+{\frac{1}{2}}(\vec{\nabla} V_{0})^{2}+{\frac{1}{2}}{m_{V}}^{2}{V_{0}}^{2}-{\frac{b}{3}}{\phi_{0}}^{3}
-{\frac{c}{4}}{\phi_{0}}^{4}
\label{lagratf}
\end{equation}
The equations of motion are given by \cite{fn4,vietnan,giamb,kons}:
\begin{equation}
{\frac{\partial \mathcal{L^{*}}}{\partial \eta_{i}}}
-\partial_{\mu}{\frac{\partial \mathcal{L^{*}}}{\partial(\partial_{\mu} \eta_{i})}}
+\partial_{\nu}\partial_{\mu}\bigg[{\frac{\partial \mathcal{L^{*}}}
{\partial(\partial_{\mu}\partial_{\nu}\eta_{i})}}\bigg]=0
\label{eulerlagra}
\end{equation}
where $\eta_{i}= \Psi, \, \phi_{0}, \, V_{0}$ and read:
\begin{equation}
-\vec{\nabla}^{2}V_{0}+
{m_{V}}^{2}V_{0}=g_{V}\bar{\psi}\gamma^{0}\psi
\label{veap}
\end{equation}
\begin{equation}
(\partial_{\mu}\partial^{\mu}+{m_{S}}^{2} )\phi_{0}=
g_{S}\bar{\psi}{\psi}-b{\phi_{0}}^{2}-c{\phi_{0}}^{3}
\label{fiap}
\end{equation}
\begin{equation}
\bigg[ i\gamma_{\mu}\partial^{\mu}-g_{V}\gamma_{0}V_{0}
-(M-g_{S} \phi_{0})\bigg]\Psi=0
\label{psiap}
\end{equation}
The effective mass of the nucleon is given by $ M^{*}=M-g_{S} \, \phi_{0}$.
In the case of pure QHD (\ref{lagratf}), because of the interaction between the nucleon and the vector meson,  we can
anticipate that the second order derivative term in the field $V^{\mu}$, which comes from its equation of motion, may be transferred to the term
$\bar{\psi} \gamma_{\mu} \psi$ and hence to the baryon density $\rho_B$. In the mean field approximation, only the time component of the field, $V^0$,
contributes, yielding the term $\vec{\nabla}^{2}V_{0}$.  It is possible to estimate $ \,\, \vec{\nabla}^{2}V_{0}$
from the equation of motion (\ref{veap}). To do so, we first rewrite the equation (\ref{veap}) as function of the baryon density:
\begin{equation}
-\vec{\nabla}^{2}V_{0}+
{m_{V}}^{2}V_{0}=g_{V}\rho_{B}
\label{veapdn}
\end{equation}
Next we assume that $\vec{\nabla}^{2}V_{0}/m^2_V << V_0$,
neglecting the Laplacian in (\ref{veapdn}) to find the first order estimate of $V_0$:
\begin{equation}
V_{0}={\frac{g_{V}}{{m_{V}}^{2}}}\rho_{B}
\label{vz}
\end{equation}
An improvement of this estimate of $V_0$ is obtained taking the Laplacian of (\ref{vz}) and
substituting it  in the Laplacian present in (\ref{veapdn}). After this simple algebraic procedure, we solve the resulting equation
for $V_0$ and obtain:
\begin{equation}
V_{0}={\frac{g_{V}}{{m_{V}}^{2}}}\rho_{B} +
{\frac{g_{V}}{{m_{V}}^{4}}}\vec{\nabla}^{2}\rho_{B}
\label{vehm}
\end{equation}
The energy-momentum tensor is given by \cite{fn4,vietnan,giamb,kons}:
\begin{equation}
T^{\mu \nu}={\frac{\partial \mathcal{L^{*}}}{\partial(\partial_{\mu}\eta_{i})}}(\partial^{\nu}\eta_{i})
-g^{\mu \nu}\mathcal{L^{*}}-\bigg[\partial_{\beta}
{\frac{\partial \mathcal{L^{*}}}{\partial (\partial_{\mu}\partial_{\beta} \eta_{i})}}\bigg]
(\partial^{\nu}\eta_{i})+
{\frac{\partial \mathcal{L^{*}}}{\partial (\partial_{\mu}\partial_{\beta} \eta_{i})}}
(\partial_{\beta}\partial^{\nu} \eta_{i})
\label{tensorem}
\end{equation}
The energy density is:
\begin{equation}
\varepsilon=<T_{00}>
\label{eps}
\end{equation}
which becomes \cite{fn4} :
\begin{equation}
\varepsilon={\frac{1}{2}}(\partial_{0}\phi_{0})^{2}+{\frac{1}{2}}(\vec{\nabla}\phi_{0})^{2}
+{\frac{1}{2}}g_{V}\rho_{B}V_{0}
+{\frac{1}{2}}{m_{S}}^{2}{\phi_{0}}^{2}+{\frac{b}{3}}{\phi_{0}}^{3}+{\frac{c}{4}}
{\phi_{0}}^{4}
+{\frac{\gamma_{s}}{(2\pi)^{3}}}\int_{0}^{k_{F}} d^3{k} ({\vec{k}}^{2}+{M^{*}}^{2})^{1/2}
\label{energyqfinal}
\end{equation}
Inserting (\ref{vehm}) in the expression of the energy density (\ref{energyqfinal}) and also writing $\phi_0$ in terms
of $M^*$ we find the following expression:
$$
\varepsilon=
{\frac{{g_{V}}^{2}}{2{m_{V}}^{2}}}{\rho_{B}}^{2}+
{\frac{{g_{V}}^{2}}{2{m_{V}}^{4}}}\rho_{B}{\vec{\nabla}}^{2}\rho_{B}
+{\frac{{m_{S}}^{2}}{2{g_{S}}^{2}}}(M-M^{*})^{2}
$$
\begin{equation}
+b{\frac{(M-M^{*})^{3}}{3{g_{S}}^{3}}}+
c{\frac{(M-M^{*})^{4}}{4{g_{S}}^{4}}}
+{\frac{\gamma_{s}}{(2\pi)^{3}}}\int_{0}^{k_{F}} d^3{k} ({\vec{k}}^{2}+{M^{*}}^{2})^{1/2}
\label{energyfinal}
\end{equation}
In a first approximation, the variables  $\rho_B$,   ${\vec{\nabla}}^{2}\rho_{B}$ and  $M^*$  are
independent from each other and therefore,   taking the derivative
of the above expression with respect to $\rho_B$ we have:
\begin{equation}
{\frac{\partial \varepsilon}{\partial \rho_{B}}}={\frac{{g_{V}}^{2}}{{m_{V}}^{2}}}{\rho_{B}}
+{\frac{{g_{V}}^{2}}{2{m_{V}}^{4}}}{\vec{\nabla}}^{2}\rho_{B}
\label{derenerd}
\end{equation}
We will add the two sources of inhomogeneities in $\rho_B$ (which are responsible for a non-vanishing $\vec{\nabla}^{2} \rho_{B}$).
For  cold nuclear matter we have then \cite{fn4}:
$$
\varepsilon={\frac{1}{2}}\bigg\lbrace{\frac{\partial}{\partial t}}\bigg[{\frac{(M-M^{*})}{g_{S}}}\bigg]\bigg \rbrace^{2}
+{\frac{1}{2}}\bigg\lbrace\vec{\nabla}\bigg[{\frac{(M-M^{*})}{g_{S}}}\bigg]\bigg\rbrace^{2}+
{\frac{{m_{S}}^{2}}{2{g_{S}}^{2}}}(M-M^{*})^{2}+
$$
$$
b{\frac{(M-M^{*})^{3}}{3{g_{S}}^{3}}}+
c{\frac{(M-M^{*})^{4}}{4{g_{S}}^{4}}}+{\frac{{g_{V}}^{2}}{2{m_{V}}^{2}}}{\rho_{B}}^{2}+
\bigg(d+{\frac{1}{2}}\bigg){\frac{{g_{V}}^{2}}{{m_{V}}^{4}}}\rho_{B}{\vec{\nabla}}^{2}\rho_{B}+
$$
\begin{equation}
+{\frac{\gamma_{s}}{(2\pi)^{3}}}\int_{0}^{k_{F}} d^3{k} ({\vec{k}}^{2}+{M^{*}}^{2})^{1/2}
\label{energyfinalapp}
\end{equation}
where $\gamma_s=4$ is the nucleon degeneracy factor. For  hot nuclear matter, the energy density as described in \cite{fn4} is given by:
$$
\varepsilon={\frac{1}{2}}\bigg\lbrace{\frac{\partial}{\partial t}}\bigg[{\frac{(M-M^{*})}{g_{S}}}\bigg]\bigg \rbrace^{2}
+{\frac{1}{2}}\bigg\lbrace\vec{\nabla}\bigg[{\frac{(M-M^{*})}{g_{S}}}\bigg]\bigg\rbrace^{2}+
{\frac{{m_{S}}^{2}}{2{g_{S}}^{2}}}(M-M^{*})^{2}+
$$
$$
b{\frac{(M-M^{*})^{3}}{3{g_{S}}^{3}}}+
c{\frac{(M-M^{*})^{4}}{4{g_{S}}^{4}}}+{\frac{{g_{V}}^{2}}{2{m_{V}}^{2}}}{\rho_{B}}^{2}+
\bigg(d+{\frac{1}{2}}\bigg){\frac{{g_{V}}^{2}}{{m_{V}}^{4}}}\rho_{B}{\vec{\nabla}}^{2}\rho_{B}+
$$
\begin{equation}
+{\frac{\gamma_s}{(2\pi)^{3}}}\int d^3{k}\hspace{0.2cm}h_{+}\hspace{0.2cm}[n_{\vec{k}}(T,\nu)
+\bar{n}_{\vec{k}}(T,\nu)]
\label{energyfinalappT}
\end{equation}
The baryon density is:
\begin{equation}
\rho_{B}={\frac{\gamma_s}{(2\pi)^{3}}}\int d^3{k}\hspace{0.2cm}
[n_{\vec{k}}(T,\nu)-\bar{n}_{\vec{k}}(T,\nu)]
\label{032}
\end{equation}
with
\begin{equation}
n_{\vec{k}}(T,\nu)\equiv{\frac{1}{1+e^{(h_{+}-\nu)/T}}}
\label{031}
\end{equation}
\begin{equation}
\bar{n}_{\vec{k}}(T,\nu)\equiv{\frac{1}{1+e^{(h_{+}+\nu)/T}}}
\label{031aaa}
\end{equation}
\begin{equation}
\nu \equiv \mu_{B}-g_{V}V_{0}+d{\frac{g_{V}}{{m_{V}}^{2}}}(\partial^{\mu}\partial_{\nu}V_{0})
\label{026}
\end{equation}
and
\begin{equation}
h_{+}\equiv({\vec{k}}^{2}+{M^{*}}^{2})^{1/2}
\label{019}
\end{equation}
It is important to note that the term  $\mathcal{L}_d$  provides the ``$(d+1/2)$-term'' in the energy densities
(\ref{energyfinalappT}) and (\ref{energyfinalapp}) that generates the KdV equation.
The nucleon effective mass $(M^{*})$ at zero temperature
is obtained through the minimization of $\varepsilon$ with respect to $M^{*}$:
$$
{\frac{\partial \varepsilon}{\partial M^{*}}}=0
$$
which, with the help of (\ref{energyfinalapp})  yields:
$$
M^{*}=M-{\frac{{g_{S}}^{2}}{{m_{S}}^{2}}}
{\frac{\gamma_{s}}{(2\pi)^{3}}}\int_{0}^{k_{F}} d^3{k}{\frac{M^{*}}{({\vec{k}}^{2}+{M^{*}}^{2})^{1/2}}}
$$
\begin{equation}
+{\frac{{g_{S}}^{2}}{{m_{S}}^{2}}}\bigg[{\frac{b}{{g_{S}}^{3}}}(M-M^{*})^{2}+{\frac{c}{{g_{S}}^{4}}}(M-M^{*})^{3}\bigg]
\label{efmasstz}
\end{equation}
Analogously, at finite temperature:
$$
M^{*}=M-{\frac{{g_{S}}^{2}}{{m_{S}}^{2}}}{\frac{\gamma_{s}}{(2\pi)^{3}}}\int d^3{k}{\frac{M^{*}}{h_{+}}}
\bigg[n_{\vec{k}}(T,\nu)+\bar{n}_{\vec{k}}(T,\nu)\bigg]
$$
\begin{equation}
+{\frac{{g_{S}}^{2}}{{m_{S}}^{2}}}\bigg[{\frac{b}{{g_{S}}^{3}}}(M-M^{*})^{2}+{\frac{c}{{g_{S}}^{4}}}(M-M^{*})^{3}\bigg]
\label{efmasstdifz}
\end{equation}
and we conclude by these two last expressions that ``$d$-term'' does not affect the nucleon effective mass.

\subsection{Quark Gluon Plasma}

The idea that quarks and gluons may exist as free particles in a deconfined phase was advanced long ago \cite{collins} in the context of
compact stars. In these objects gravitation is strong enough to compress the nucleons and make them overlap with each other. The distance
between the quarks can be so small that, due to asymptotic freedom, they almost do not interact. In this picture, we may have large regions of space
populated with free quarks. This state is called the (cold) quark gluon plasma. This naive description of  QCD in extreme conditions of density
was later revisited with much more sophisticated models. Quarks and gluons may also form a hot quark gluon plasma. The hot QGP has been extensively
studied in lattice simulations and also in heavy ion experiments at CERN and LHC.  Today we know that a hot and deconfined phase is formed in the
existing accelerators, but it is far more complicated than previously imagined.  In particular the quarks and gluons are not really free. Instead,
they still interact strongly with each other forming a state called the strongly interacting QGP, or sQGP. Also the non-trivial vacuum structure
persists until relatively large temperatures.

A starting point to study in a unified way both the cold and hot QGP is the MIT bag model \cite{mit}.
According to this model in its simplest version,  massless
and non-interacting quarks live in a spherical cavity (the ``bag'') in  the physical QCD vacuum, which is a medium. The confining property is
represented by a constant term, ``B'',  called the bag constant. In the next subsection we shall briefly show how to calculate the QGP equation of state
with the help of this model.

\subsubsection{The MIT Bag Model}

In what follows we consider only quarks $u$ and $d$.
Each quark has three color states and they are massless.  We also have eight massless gluons and we neglect the interactions in the  QGP.
The confinement property is included in the model through the introduction of  a constant, positive energy per unit volume in the vacuum:
\begin{equation}
\mathcal{B}=\bigg({\frac{E}{V}}\bigg)_{vac}
\label{bs}
\end{equation}
that can be interpreted as the energy needed to create a bubble or bag in the vacuum, in which the noninteracting quarks and gluons
are confined. $\mathcal{B}$ is known as ``bag constant''.

The parameter $\mathcal{B}$ can be extracted from a  phenomenological analysis of hadron spectroscopy  or from lattice QCD calculations.
There is a relationship between
$\mathcal{B}$ and the critical temperature of quark-hadron transition $T_{c}$ which
is determined by considering that during the  phase transition the pressure is zero.

The quarks have baryon number $1/3$ and the chemical potential for gluons is zero.
The baryon density, energy density and pressure for the QGP are given by:
\begin{equation}
\rho_{B}={\frac{1}{3}}{\frac{\gamma_Q}{(2\pi)^{3}}}\int d^3{k}\hspace{0.2cm}
[n_{\vec{k}}-\bar{n}_{\vec{k}}]
\label{rodensTq}
\end{equation}
where
\begin{equation}
n_{\vec{k}} \equiv n_{\vec{k}}(T)={\frac{1}{1+e^{(k-{\frac{1}{3}}\mu_{B})/ T}}}
\label{qdis}
\end{equation}
and
\begin{equation}
\bar{n}_{\vec{k}} \equiv \bar{n}_{\vec{k}}(T)={\frac{1}{1+e^{(k+{\frac{1}{3}}\mu_{B})/ T}}}
\label{aqdis}
\end{equation}
where $\mu_{B}$ is the baryon chemical potential.
The energy density is given by:
\begin{equation}
\varepsilon=\mathcal{B}+{\frac{\gamma_G}{(2\pi)^{3}}}\int d^3{k}\hspace{0.2cm}k\hspace{0.2cm}(e^{k/T}-1)^{-1}
+{\frac{\gamma_Q}{(2\pi)^{3}}}\int d^3{k}\hspace{0.2cm}k\hspace{0.2cm}[n_{\vec{k}}
+\bar{n}_{\vec{k}}]
\label{esdTqg}
\end{equation}
where the first term of the expression above is the gluon contribution.
The pressure is given by:
\begin{equation}
p=-\mathcal{B}+{\frac{1}{3}}\Bigg\lbrace {\frac{\gamma_G}{(2\pi)^{3}}}\int d^3{k}\hspace{0.2cm}k\hspace{0.2cm}(e^{k/T}-1)^{-1}+
{\frac{\gamma_Q}{(2\pi)^{3}}}\int d^{3}k\hspace{0.1cm}k\bigg[{n}_{\vec{k}}+\bar{n}_{\vec{k}}\bigg] \Bigg\rbrace
\label{psdTqg}
\end{equation}
The degeneracy factors are:
\begin{equation}
\gamma_G=2\textrm{(polarizations)}\times 8\textrm{(colors)}=16 \hspace{1cm} \textrm{for gluons}
\label{gamaG}
\end{equation}
and
\begin{equation}
\gamma_Q=2\textrm{(spins)}\times 2\textrm{(flavours)}\times 3\textrm{(colors)}=12 \hspace{1cm} \textrm{for quarks}
\label{gamaQ}
\end{equation}
The integral of the gluon distribution function can be calculated analytically  and the thermodynamics
of QGP can be summarized in the following expressions:
\begin{equation}
\rho_{B}={\frac{2}{\pi^{2}}}\int_{0}^{\infty} d{k}\hspace{0.2cm}k^{2}
[n_{\vec{k}}-\bar{n}_{\vec{k}}]
\label{rodensTqfinal}
\end{equation}
and
\begin{equation}
3(p+\mathcal{B})=\varepsilon-\mathcal{B}={\frac{8\pi^{2}}{15}} \ T^{4}+{\frac{6}{\pi^{2}}}\int_{0}^{\infty} d{k}\hspace{0.2cm}k^{3}
[n_{\vec{k}}+\bar{n}_{\vec{k}}]
\label{bacana}
\end{equation}
which provide us with the EOS of QGP with all baryon densities and at all temperatures:
\begin{equation}
p={\frac{1}{3}}\varepsilon-{\frac{4}{3}}\mathcal{B}
\label{eosqg}
\end{equation}
The sound speed $c_{s}$ is given by:
\begin{equation}
{c_{s}}^{2}={\frac{\partial p}{\partial \varepsilon}}={\frac{1}{3}}
\label{soundone}
\end{equation}

\subsubsection{The cold QGP}

In  the core of a neutron star, the temperature is zero and baryon density is considerable. This is a good place to
study baryon density perturbations.
The quark distribution function becomes the step function with $\mu_{B}=3k_{F}$ and  (\ref{rodensTqfinal}) becomes:
\begin{equation}
{\rho_{B}}={\frac{\gamma_{s}}{6\pi^{2}}}{k_{F}}^{3}
\label{a12pa}
\end{equation}
Expression (\ref{bacana}) becomes:
\begin{equation}
3(p+\mathcal{B})=\varepsilon-\mathcal{B}={\frac{6}{\pi^{2}}}\bigg({\frac{k^{4}}{4}}\bigg{|}_{0}^{k_{F}} \bigg)
={\frac{3}{2\pi^{2}}}{k_{F}}^{4}
\label{bacanatz}
\end{equation}
Using (\ref{a12pa}) we find:
$$
3(p+\mathcal{B})=\varepsilon-\mathcal{B}=\bigg({\frac{3}{2}}\bigg)^{7/3}\pi^{2/3}
{\rho_{B}}^{4/3}
$$
and so
\begin{equation}
\varepsilon(\rho_{B})=\bigg({\frac{3}{2}}\bigg)^{7/3}\pi^{2/3}{\rho_{B}}^{4/3}+\mathcal{B}
\label{bacanatrh}
\end{equation}
Inserting (\ref{bacanatrh}) into (\ref{eosqg}) we find:
\begin{equation}
p(\rho_{B})={\frac{1}{3}}\bigg({\frac{3}{2}}\bigg)^{7/3}\pi^{2/3}{\rho_{B}}^{4/3}-\mathcal{B}
\label{prerho}
\end{equation}
From (\ref{bacanatrh}) and (\ref{prerho}) is possible to rewrite the sum of energy density and pressure as:
\begin{equation}
\varepsilon+p={\frac{4}{3}}\bigg({\frac{3}{2}}\bigg)^{7/3}\pi^{2/3}{\rho_{B}}^{4/3}
\label{emaisprerho}
\end{equation}
From (\ref{eosqg}) we have:
\begin{equation}
\vec{\nabla}p={\frac{1}{3}}\vec{\nabla}\varepsilon \hspace{1.8cm} \textrm{and \ \ also} \hspace{1.8cm} {\frac{\partial p}{\partial t}}=
{\frac{1}{3}}{\frac{\partial \varepsilon}{\partial t}}
\label{petvar}
\end{equation}
Using (\ref{bacanatrh}) in (\ref{petvar}) we find:
\begin{equation}
\vec{\nabla}p={\frac{4}{9}}\bigg({\frac{3}{2}}\bigg)^{7/3}\pi^{2/3}{\rho_{B}}^{1/3} \ \vec{\nabla}{\rho_{B}}
\label{pevar}
\end{equation}
and
\begin{equation}
{\frac{\partial p}{\partial t}}={\frac{4}{9}}\bigg({\frac{3}{2}}\bigg)^{7/3}\pi^{2/3}{\rho_{B}}^{1/3} \
{\frac{\partial {\rho_{B}}}{\partial t}}
\label{ptvar}
\end{equation}

\subsubsection{The hot QGP}

The hot QGP is formed in  heavy ion collisions  in the central rapidity region
where  we have $\rho_{B}=0$. For our purposes the most relevant physical quantity is the energy density $\varepsilon$.
Since $\rho_{B}=0$, the baryon chemical potential is zero $(\mu_{B}=0)$ and so the distribution functions given by (\ref{qdis}) and (\ref{aqdis})
are the same:
$$
{n}_{\vec{k}}=\bar{n}_{\vec{k}} ={\frac{1}{1+e^{k/ T}}}
$$
and then (\ref{bacana}) takes the form:
\begin{equation}
3(p+\mathcal{B})=\varepsilon-\mathcal{B}={\frac{8\pi^{2}}{15}} \
T^{4}+{\frac{12}{\pi^{2}}}\int_{0}^{\infty} d{k}\hspace{0.2cm}
{\frac{k^{3}}{[1+e^{k/ T}]}}
\label{bacanaa}
\end{equation}
We can perform the above integral analytically  to find:
\begin{equation}
3(p+\mathcal{B})=\varepsilon-\mathcal{B}={\frac{37}{30}} \pi^{2} T^{4}
\label{bacanaaend}
\end{equation}
From basic thermodynamics  we know that
\begin{equation}
s=\Bigg({\frac{\partial p}{\partial T}}\Bigg)_{V}
\label{sbuscaepsilon}
\end{equation}
which, with the use of (\ref{bacanaaend}) leads to the specific form for $s$:
\begin{equation}
s={\frac{\partial }{\partial T}}\bigg(-\mathcal{B}+{\frac{37}{90}} \pi^{2} T^{4}\bigg)=
4  \\ {\frac{37}{90}} \pi^{2} T^{3}
\label{denstemp}
\end{equation}
The parameter $\mathcal{B}$, the ``bag constant'' can be
defined at the temperature $T_{B}$.
In the bag surface, (\ref{bacanaaend}) is given by:
\begin{equation}
\mathcal{B}={\frac{37}{90}} \pi^{2} (T_{B})^{4}
\label{BT}
\end{equation}
Choosing $\mathcal{B}^{1/4}=170MeV$  corresponds to $T_B=91 \, MeV$.
Rewriting (\ref{bacanaaend}) as:
\begin{equation}
T=\Bigg[{\frac{30}{37\pi^{2}}}(\varepsilon-\mathcal{B})\Bigg]^{1/4}
\label{TfromEpsilon}
\end{equation}
and inserting it into (\ref{denstemp}) we find:
$$
s=s(\varepsilon)=4  \\ {\frac{37}{90}} \pi^{2} \Bigg[{\frac{30}{37\pi^{2}}}(\varepsilon-\mathcal{B})\Bigg]^{3/4}
$$
In a compact notation:
\begin{equation}
s(\varepsilon)=A(\varepsilon-\mathcal{B})^{3/4}
\label{densenerd}
\end{equation}
where
\begin{equation}
A \equiv 4  \\ {\frac{37}{90}} \pi^{2} \Bigg[{\frac{30}{37\pi^{2}}}\Bigg]^{3/4}
\label{Anumber}
\end{equation}

\subsection{Mean field theory for Quantum Chromodynamics}

In spite of its phenomenological success the MIT bag model gives a poor representation of the quark gluon plasma. From heavy ion collisions there is convincing
evidence  that quarks  and gluons interact strongly forming  rather a ``prefect fluid'' than an ideal gas. Therefore, the picture of free partons must be
modified. There are several ways to do that. Here we discuss the approach proposed in \cite{nos2011}, which can be called mean field QCD (MFQCD). It allows us
to start from QCD Lagrangian and derive an equation of state, which incorporates the effects of a sizeable strong coupling constant and also
residual  non -perturbative
effects from the QCD vacuum. As an interesting  by-product this equation of state supports the existence of KdV solitons in a cold QGP.
In what follows
we summarize the basic ideas of this approach and derive the expressions for the pressure and energy density, which are relevant for the hydrodynamical study.

We first  introduce the  mean field approximation for QCD, extending previous works  along the same line \cite{shakin,shakinn}. We consider a system of quarks
and gluons which are represented by the QCD Lagrangian density:
\begin{equation}
{\mathcal{L}}_{QCD}=-{\frac{1}{4}}F^{a}_{\mu\nu}F^{a\mu\nu}
+\sum_{q=1}^{N_{f}}\bar{\psi}^{q}_{i}\Big[i\gamma^{\mu}(\delta_{ij}\partial_{\mu}-
igT^{a}_{ij}G_{\mu}^{a})
- \delta_{ij} m_q \Big]\psi^{q}_{j}
\label{lqcdu}
\end{equation}
with
\begin{equation}
F^{a\mu\nu}=\partial^{\mu}G^{a\nu}-\partial^{\nu}G^{a\mu}+gf^{abc}G^{b\mu}G^{c\nu}
\label{efe}
\end{equation}
where $\psi^q_i$ and $G^a_{\mu}$ represent the quark and gluon fields respectively.
The summation on $q$ runs over all quark flavors,
$m_q$ is the mass of the quark of flavor $q$,
$i$ and $j$ are the color indices of the quarks,
$T^{a}$ are the SU(3) generators and $f^{abc}$ are the SU(3) antisymmetric
structure constants.  For simplicity we  consider only
light quarks with the same mass $m$. Moreover, we  drop the summation and consider only
one flavor. At the end of our calculation the number of flavors will be recovered.
Following \cite{shakin, shakinn}, we shall start writing the gluon field as:
\begin{equation}
G^{a\mu}={A}^{a\mu}+{\alpha}^{a\mu}
\label{amd}
\end{equation}
where ${A}^{a\mu}$  and ${\alpha}^{a\mu}$ are the low (``soft'') and high (``hard'')  momentum
components of the gluon field respectively. We will assume that ${A}^{a\mu}$
represents the soft modes which populate the vacuum and  ${\alpha}^{a\mu}$
represents the modes for which the running coupling constant is small.
In a cold quark gluon plasma the density is much larger than the ordinary nuclear matter
density. These high densities imply a very large number of sources of the gluon field.
Assu-ming that the coupling constant is not very small, the existence of  intense
sources implies that  the bosonic fields tend to have large occupation numbers at all
energy levels, and therefore they can be treated as classical fields.
This is the famous approximation for
bosonic fields used in relativistic mean field models of nuclear matter \cite{wal}.
It has been applied to QCD in the past  and amounts to assume that the
``hard'' gluon field, ${\alpha}_{\mu}^{a}$,  is simply a  function of the coordinates
\cite{wal}:
\begin{equation}
{\alpha}_{\mu}^{a}(\vec{x},t)=\delta_{\mu 0} \, {\alpha}_{0}^{a}(\vec{x},t)
\label{watype}
\end{equation}
with  $\partial_{\nu}{\alpha}^{a}_{\mu}\neq 0$. This space and time dependence goes
beyond the standard mean field approximation,  where  ${\alpha}_{\mu}^{a}$ is constant in
space and time  and consequently $\partial_{\nu}{\alpha}^{a}_{\mu}=0$.
We keep  assuming, as in \cite{nos2011},  that the soft gluon field ${A}^{a\mu}$ is
independent of position and time and thus $\partial^{\nu}{A}^{a\mu}=0$ .  Following the same steps of
\cite{nos2011} we obtain the following effective Lagrangian:
\begin{equation}
\mathcal{L}=-{\frac{1}{2}}{\alpha}^{a}_{0}\big({\vec{\nabla}}^{2}{\alpha}^{a}_{0}\big)
+{\frac{{m_{G}}^{2}}{2}}{\alpha}^{a}_{0}{\alpha}^{a}_{0}-\mathcal{B}_{QCD}
+ \bar{\psi}_{i}\Big(i\delta_{ij}\gamma^{\mu}\partial_{\mu}+g\gamma^{0}T^{a}_{ij}
{\alpha}^{a}_{0}-\delta_{ij}m\Big)\psi_{j}
\label{mfqcdf}
\end{equation}
where the constant $\mathcal{B}_{QCD}$ is the same bag constant for QCD as defined in \cite{nos2011}.
In fact, the effective Lagrangian (\ref{mfqcdf}) is quite similar to the effective Lagrangian obtained in \cite{nos2011}.
The new feature of (\ref{mfqcdf}) is the term $-{\frac{1}{2}}{\alpha}^{a}_{0}\big({\vec{\nabla}}^{2}{\alpha}^{a}_{0}\big)$.
For simplicity, we  take the quarks to be massless. From the above Lagrangian, using (\ref{tensorem}), it is straightforward to derive
the energy-momentum tensor $T_{\mu \nu}$,  which gives us the energy density $\varepsilon$ and pressure $p$:
$$
\varepsilon=\bigg({\frac{27g^{2}}{16{m_{G}}^{2}}}\bigg)  {\rho_{B}}^{2}+
\bigg({\frac{27g^{2}}{16{m_{G}}^{4}}}\bigg)
\rho_{B}  {\vec{\nabla}}^{2}\rho_{B}
+\bigg({\frac{27g^{2}}{16{m_{G}}^{6}}}\bigg)  \rho_{B}  {\vec{\nabla}}^{2}({\vec{\nabla}}^{2}\rho_{B})
$$
\begin{equation}
+\bigg({\frac{27g^{2}}{16{m_{G}}^{8}}}\bigg)  {\vec{\nabla}}^{2}\rho_{B} {\vec{\nabla}}^{2}({\vec{\nabla}}^{2}\rho_{B})
+\mathcal{B}_{QCD}
+3{\frac{\gamma_{Q}}{2{\pi}^{2}}}{\frac{{k_{F}}^{4}}{4}}
\label{epstd}
\end{equation}
and the pressure is:
$$
p=\bigg({\frac{27g^{2}}{16{m_{G}}^{2}}}\bigg) \ {\rho_{B}}^{2}+\bigg({\frac{9g^{2}}{4{m_{G}}^{4}}}\bigg) \
{\rho_{B}} \ {\vec{\nabla}}^{2}{\rho_{B}}
-\bigg({\frac{9g^{2}}{8{m_{G}}^{6}}}\bigg) \ {\rho_{B}} \ {\vec{\nabla}}^{2}({\vec{\nabla}}^{2}{\rho_{B}})
$$
$$
-\bigg({\frac{9g^{2}}{16{m_{G}}^{4}}}\bigg) \vec{\nabla}{\rho_{B}} \cdot \vec{\nabla}{\rho_{B}}
+\bigg({\frac{9g^{2}}{16{m_{G}}^{6}}}\bigg) \ {\vec{\nabla}}^{2}{\rho_{B}} \ {\vec{\nabla}}^{2}{\rho_{B}}
-\bigg({\frac{9g^{2}}{8{m_{G}}^{8}}}\bigg) \ {\vec{\nabla}}^{2}{\rho_{B}} \ {\vec{\nabla}}^{2}({\vec{\nabla}}^{2}{\rho_{B}})
$$
$$
-\bigg({\frac{9g^{2}}{16{m_{G}}^{8}}}\bigg) \vec{\nabla}({\vec{\nabla}}^{2}{\rho_{B}})
\cdot \vec{\nabla}({\vec{\nabla}}^{2}{\rho_{B}})
-\bigg({\frac{9g^{2}}{8{m_{G}}^{6}}}\bigg) \vec{\nabla}{\rho_{B}} \cdot \vec{\nabla}({\vec{\nabla}}^{2}{\rho_{B}})
$$
\begin{equation}
-\mathcal{B}_{QCD}
+{\frac{\gamma_{Q}}{2{\pi}^{2}}}{\frac{{k_{F}}^{4}}{4}}
\label{prestd}
\end{equation}
where $\gamma_{Q}$ is the quark degeneracy factor
$\gamma_{Q} = 2 (\mbox{spin}) \times 3 (\mbox{flavor}) \, =6 $ and
$k_{F}$ is the Fermi momentum defined by the baryon number density by $\rho_{B}={k_{F}}^{3}/{\pi}^{2}$ .
The other parameters $g$, $m_{G}$ and $\mathcal{B}_{QCD}$ are
the coupling of the hard gluons, the dynamical gluon mass and the bag constant in terms of the gluon condensate,
respectively.
An improved version of the EOS of (\ref{epstd}) and (\ref{prestd}) was used in the study of three dimensional solitons in cold QGP.  These solitons
are solutions of the Kadomtsev-Petviashvili (KP) equation, which is the three dimensional generalization of the KdV equation.  The complete
description of the  calculation may be found  in \cite{nos2013}.

\section{Nonlinear wave equations}

In the previous sections we  presented a review of the equations of hydrodynamics and we introduced the
equations of state, which represent the microscopic dynamics of the corresponding fluids. Furthermore,
we have presented a mathematical prescription (the RPM) to study perturbations in these fluids preserving the nonlinearities of
the differential equations. The application of  the RPM to several systems of interest leads to the nonlinear differential equations
which we discuss in this section. Most of these wave equations were developed in \cite{nos2011a,fn1,fn2,fn3,fn4,nos2010,nos2013,nos2012}.

\subsection{KdV equation in nuclear matter }

\subsubsection{Cold nuclear matter}

We insert (\ref{energyfinalapp}) in the Euler equation  given by (\ref{relateul}). We then combine this Euler equation
with (\ref{rhobcons}) and follow the RPM procedure  to obtain the following KdV equation:
\begin{equation}
{\frac{\partial {\hat{\rho}_{1}}}{\partial t}}+
{c_{s}}{\frac{\partial {\hat{\rho}_{1}}}{\partial \mathcal{X}}}+
(3-{c_{s}}^{2}){c_{s}}
{\hat{\rho}_{1}}{\frac{\partial{\hat{\rho}_{1}}}{\partial \mathcal{X}}}
+\bigg(d+{\frac{1}{2}}\bigg)\bigg({\frac{{g_{V}}^{2}\rho_{0}}{2M{{m_{V}}^{4}}{c_{s}}}}\bigg)
{\frac{\partial^{3}{\hat{\rho}_{1}}}{\partial\mathcal{X}^{3}}}+
\mathcal{G}{\frac{{\hat{\rho}_{1}}}{t}}=0
\label{esfkdvtzero}
\end{equation}
where $\mathcal{G}$ is a ``geometrical factor'':
\bigskip
\begin{equation}
\mathcal{G} \equiv \left\{ \begin{array}{ll}
0 \,\, & \textrm{for cartesian coordinates:}  \,\,\,\, \mathcal{X}=x \\
1 \,\, & \textrm{for spherical coordinates:}  \,\,\,\, \mathcal{X}=r
\end{array} \right.
\label{chavegeom}
\end{equation}
\bigskip
The non-relativistic limit is obtained by the approximation $3-{c_{s}}^{2} \cong 3$ and
so (\ref{esfkdvtzero}) becomes:
\begin{equation}
{\frac{\partial {\hat{\rho}_{1}}}{\partial t}}+
{c_{s}}{\frac{\partial {\hat{\rho}_{1}}}{\partial \mathcal{X}}}+
3{c_{s}}
{\hat{\rho}_{1}}{\frac{\partial{\hat{\rho}_{1}}}{\partial \mathcal{X}}}
+\bigg(d+{\frac{1}{2}}\bigg)\bigg({\frac{{g_{V}}^{2}\rho_{0}}{2M{{m_{V}}^{4}}{c_{s}}}}\bigg)
{\frac{\partial^{3}{\hat{\rho}_{1}}}{\partial\mathcal{X}^{3}}}+
\mathcal{G}{\frac{{\hat{\rho}_{1}}}{t}}=0
\label{esfkdvtzeronrlim}
\end{equation}
which could have been obtained with the use of the same procedure applied to the nonrelativistic Euler equation
(\ref{nonrelateul})  with $\rho=M\rho_{B}$.

\subsubsection{Hot nuclear matter}

We follow the same steps listed above, but changing
the (\ref{energyfinalapp}) by the EOS (\ref{energyfinalappT}) to find a KdV given by:
\begin{equation}
{\frac{\partial {\hat{\rho_{1}}}}{\partial t}}+{c_{s}}
{\frac{\partial {\hat{\rho_{1}}}}{\partial \mathcal{X}}}
+\bigg(2-{c_{s}}^{2}-{\frac{{\mu_{B}}{m_{V}}^{2}{c_{s}}^{2}}{2{g_{V}}^{2}{\rho_{0}}}}\bigg)
{c_{s}}\hat{\rho_{1}}{\frac{\partial{\hat{\rho_{1}}}}{\partial \mathcal{X}}}
+\bigg(d+{\frac{1}{2}}\bigg)\bigg({\frac{{c_{s}}}{2{{m_{V}}^{2}}}}\bigg)
{\frac{\partial^{3}{\hat{\rho_{1}}}}{\partial\mathcal{X}^{3}}}
+\mathcal{G}{\frac{\hat{\rho_{1}}}{t}}=0
\label{esfkdvt}
\end{equation}
In the last two wave equations ${\hat{\rho_{1}}}\equiv \sigma\rho_{1}$
as defined in (\ref{oneroexp}) and $\mathcal{G}$ is defined  by (\ref{chavegeom}).
Choosing $d = -\frac{1}{2}$ in the above equations we eliminate the cubic derivative term and obtain a breaking wave equation.

\subsection{Breaking wave equation in QGP}

\subsubsection{Cold QGP}

Inserting the cold QGP  MIT EOS, (\ref{bacanatz}) to (\ref{ptvar}),   into the ideal hydrodynamical
equations (\ref{relateul}) and (\ref{rhobcons}) and following the RPM procedure
we obtain the following breaking wave equation:
\begin{equation}
{\frac{\partial\hat\rho_{1}}{\partial t}}+
c_{s}{\frac{\partial \hat\rho_{1}}{\partial x}}+
{\frac{2}{3}}c_{s}\hat\rho_{1}{\frac{\partial \hat\rho_{1}}{\partial x}}=0
\label{bwqcdxitauXt}
\end{equation}
where again $\hat\rho_{1}\equiv \sigma\rho_{1}$.

\subsubsection{Hot QGP}

Now we insert the hot QGP MIT EOS relations, (\ref{bacanaa}) to (\ref{Anumber}),   into the ideal hydrodynamical
equations (\ref{relateul}) and (\ref{idealrelcontss}) and following the RPM procedure
we obtain the following breaking wave equation:
\begin{equation}
{\frac{\partial\hat\varepsilon_{1}}{\partial t}}+c_{s}
{\frac{\partial \hat\varepsilon_{1}}{\partial x}}+
\Bigg[1+ \Bigg({\frac{T_{B}}{T_{0}}}\Bigg)^{4}\Bigg]{\frac{c_{s}}{2}}
\hat\varepsilon_{1}{\frac{\partial \hat\varepsilon_{1}}{\partial x}}=0
\label{bwqcdTfincomrefens}
\end{equation}
where $\hat\varepsilon_{1}\equiv \sigma \varepsilon_{1}$.

\subsection{KP and  KdV  equations from MFQCD}

The energy density and pressure are given respectively by (\ref{epstd}) and (\ref{prestd}) and the wave equations are for perturbations in the
baryon density given by $\hat\rho_{1}\equiv \sigma\rho_{1}$ as defined by (\ref{roexpargain}) and (\ref{roexpa}).
We use the ideal hydrodynamical
equations (\ref{relateul}) and (\ref{rhobcons}) following the RPM procedure.  In the calculations we find the following relation:
\begin{equation}
\bigg({\frac{27g^{2}\,{\rho_{0}}^{2}}{8{m_{G}}^{2}}}\bigg){c_{s}}^{2}
+3\pi^{2/3}{\rho_{0}}^{4/3}{c_{s}}^{2}=\bigg({\frac{27g^{2}\,{\rho_{0}}^{2}}{8{m_{G}}^{2}}}\bigg)
+\pi^{2/3}{\rho_{0}}^{4/3} = A
\label{aconsaggainn}
\end{equation}
which relates the speed of sound to the  background density ${\rho_{0}}$:
\begin{equation}
{c_{s}}^{2}={\frac{\bigg({\frac{27g^{2}\,{\rho_{0}}^{2}}{8{m_{G}}^{2}}}\bigg)+\pi^{2/3}{\rho_{0}}^{4/3}}
{\bigg({\frac{27g^{2}\,{\rho_{0}}^{2}}{8{m_{G}}^{2}}}\bigg)+3\pi^{2/3}{\rho_{0}}^{4/3}}}
\label{csrelationaggainn}
\end{equation}
The wave equations are the following.

\subsubsection{Cylindrical KP equation}
$$
{\frac{\partial}{\partial r}}\Bigg\lbrace{\frac{\partial\hat\rho_{1}}{\partial t}}
+{c_{s}}{\frac{\partial\hat\rho_{1}}{\partial r}}
+\bigg[{\frac{3}{2}}(1-{c_{s}}^{2})-{\frac{\pi^{2/3}{\rho_{0}}^{4/3}}{3A}}\bigg]{c_{s}}
\hat\rho_{1}{\frac{\partial\hat\rho_{1}}{\partial r}}
+\bigg[{\frac{9g^{2}\,{\rho_{0}}^{2}c_{s}}{8{m_{G}}^{4}A}}\bigg]
{\frac{\partial^{3}\hat\rho_{1}}{\partial r^{3}}}+
{\frac{\hat\rho_{1}}{2t}}\Bigg\rbrace
$$
\begin{equation}
+{\frac{1}{2c_{s}t^{2}}}{\frac{\partial^{2}\hat\rho_{1}}{\partial \varphi^{2}}}
+{\frac{c_{s}}{2}}{\frac{\partial^{2}\hat\rho_{1}}{\partial z^{2}}}=0
\label{r3dckp}
\end{equation}

\subsubsection{KP equation}

$$
{\frac{\partial}{\partial x}}\Bigg\lbrace{\frac{\partial\hat\rho_{1}}{\partial t}}
+{c_{s}}{\frac{\partial\hat\rho_{1}}{\partial x}}
+\bigg[{\frac{3}{2}}(1-{c_{s}}^{2})-{\frac{\pi^{2/3}{\rho_{0}}^{4/3}}{3A}}\bigg]{c_{s}}
\hat\rho_{1}{\frac{\partial\hat\rho_{1}}{\partial x}}
+\bigg[{\frac{9g^{2}\,{\rho_{0}}^{2}c_{s}}{8{m_{G}}^{4}A}}\bigg]
{\frac{\partial^{3}\hat\rho_{1}}{\partial x^{3}}}\Bigg\rbrace
$$
\begin{equation}
+{\frac{c_{s}}{2}}{\frac{\partial^{2}\hat\rho_{1}}{\partial y^{2}}}
+{\frac{c_{s}}{2}}{\frac{\partial^{2}\hat\rho_{1}}{\partial z^{2}}}=0
\label{kpxyztsual}
\end{equation}

\subsubsection{KdV equation}

The one dimensional cartesian particular case
of (\ref{kpxyztsual}) is obtained by neglecting the $y$ and $z$ dependence, so that (\ref{kpxyztsual}) becomes the KdV :
$$
{\frac{\partial\hat\rho_{1}}{\partial t}}
+{c_{s}}{\frac{\partial\hat\rho_{1}}{\partial x}}+
\bigg[{\frac{(2-{c_{s}}^{2})}{2}}-\bigg({\frac{27g^{2}\,{\rho_{0}}^{2}}{8{m_{G}}^{2}}}\bigg)
{\frac{(2{c_{s}}^{2}-1)}{2A}}-{\frac{\pi^{2/3}{\rho_{0}}^{4/3}}{A}}\bigg({c_{s}}^{2}-{\frac{1}{6}}\bigg)\bigg]{c_{s}}
\hat\rho_{1}{\frac{\partial\hat\rho_{1}}{\partial x}}
$$
\begin{equation}
+\bigg[{\frac{9g^{2}\,{\rho_{0}}^{2}c_{s}}{8{m_{G}}^{4}A}}\bigg]
{\frac{\partial^{3}\hat\rho_{1}}{\partial x^{3}}}=0
\label{kdvxt}
\end{equation}
Taking the limit $m_{G} \, \rightarrow \, \infty$ we obtain from (\ref{aconsaggainn}) and (\ref{csrelationaggainn}):
$$
A=\pi^{2/3}{\rho_{0}}^{4/3}  \,\,\,\, , \hspace{2cm} {c_{s}}^{2}={\frac{1}{3}}
$$
and (\ref{kdvxt}) becomes:
\begin{equation}
{\frac{\partial\hat\rho_{1}}{\partial t}}+
c_{s}{\frac{\partial \hat\rho_{1}}{\partial x}}+
{\frac{2}{3}}c_{s}\hat\rho_{1}{\frac{\partial \hat\rho_{1}}{\partial x}}=0
\label{bwmitxitauXt}
\end{equation}
and we recover exactly the result (\ref{bwqcdxitauXt}), the breaking wave
equation for $\hat\rho_{1}$ at zero temperature in the QGP derived from the MIT equation of state.

\subsubsection{Breaking wave equation}

Neglecting the spatial derivatives in (\ref{epstd}) and (\ref{prestd}), the equation
(\ref{kdvxt}) reduces to:
\begin{equation}
{\frac{\partial\hat\rho_{1}}{\partial t}}
+{c_{s}}{\frac{\partial\hat\rho_{1}}{\partial x}}+
\bigg[{\frac{(2-{c_{s}}^{2})}{2}}-\bigg({\frac{27g^{2}\,{\rho_{0}}^{2}}{8{m_{G}}^{2}}}\bigg)
{\frac{(2{c_{s}}^{2}-1)}{2A}}-{\frac{\pi^{2/3}{\rho_{0}}^{4/3}}{A}}\bigg({c_{s}}^{2}-{\frac{1}{6}}\bigg)\bigg]{c_{s}}
\hat\rho_{1}{\frac{\partial\hat\rho_{1}}{\partial x}}=0
\label{bwqxt}
\end{equation}
which is also a breaking wave equation for $\hat\rho_{1}$ with the $\rho_{0}$, $m_{G}$ and $g$ dependence in its coefficients.

\subsection{KP-Burgers  equation in hot QGP}

The relations (\ref{bacanaa}) to (\ref{Anumber}) for the hot QGP are now inserted
in the relativistic viscous hydrodynamical
equations, i.e., the Navier-Stokes (\ref{rnsagain}) and the continuity for the entropy density (\ref{relcontss}).  The ideal case is recovered when the viscous coefficients are set to zero.  Following the RPM procedure
we obtain the following  wave equations:

\subsubsection{Cylindrical KP-Burgers}

$$
{\frac{\partial}{\partial r}}\Bigg\{{\frac{\partial\hat\varepsilon_{1}}{\partial t}}+
c_{s}{\frac{\partial\hat\varepsilon_{1}}{\partial r}}+
{\frac{c_{s}}{2}}\Bigg[1+ \Bigg({\frac{T_{B}}{T_{0}}}\Bigg)^{4}\Bigg]
\ \hat\varepsilon_{1}{\frac{\partial \hat\varepsilon_{1}}{\partial r}}
+{\frac{\hat\varepsilon_{1}}{2t}}
-{\frac{1}{2T_{0}}}\Bigg({\frac{\zeta}{s}}  + \frac{4}{3} {\frac{\eta}{s}}  \Bigg)\frac{\partial^{2} \hat\varepsilon_{1}}{\partial r^{2}}
\Bigg\}
$$
\begin{equation}
+{\frac{1}{2{c_{s}}t^{2}}}{\frac{\partial^{2} \hat\varepsilon_{1}}{\partial \varphi^{2}}}
+{\frac{c_{s}}{2}}{\frac{\partial^{2} \hat\varepsilon_{1}}{\partial z^{2}}}=0
\label{cburgers3d}
\end{equation}

\subsubsection{Cylindrical Burgers}

Neglecting the $\varphi$ and $z$ dependence, the equation (\ref{cburgers3d}) becomes:
\begin{equation}
{\frac{\partial\hat\varepsilon_{1}}{\partial t}}+
c_{s}{\frac{\partial\hat\varepsilon_{1}}{\partial r}}+
{\frac{c_{s}}{2}}\Bigg[1+ \Bigg({\frac{T_{B}}{T_{0}}}\Bigg)^{4}\Bigg]
\ \hat\varepsilon_{1}{\frac{\partial \hat\varepsilon_{1}}{\partial r}}
+{\frac{\hat\varepsilon_{1}}{2t}}
={\frac{1}{2T_{0}}}\Bigg({\frac{\zeta}{s}}  + \frac{4}{3} {\frac{\eta}{s}}  \Bigg) \frac{\partial^{2} \hat\varepsilon_{1}}{\partial r^{2}}
\label{burgers_final}
\end{equation}
Setting $\eta=\zeta=0$ (ideal fluid) the (\ref{burgers_final}) becomes:
\begin{equation}
{\frac{\partial\hat\varepsilon_{1}}{\partial t}}+
c_{s}{\frac{\partial\hat\varepsilon_{1}}{\partial r}}+
{\frac{c_{s}}{2}}\Bigg[1+ \Bigg({\frac{T_{B}}{T_{0}}}\Bigg)^{4}\Bigg]
\ \hat\varepsilon_{1}{\frac{\partial \hat\varepsilon_{1}}{\partial r}}
+{\frac{\hat\varepsilon_{1}}{2t}} =0
\label{weq}
\end{equation}

\section{Analytical solutions of nonlinear wave equations}

We present some cases where particular analytical solutions exist.  For the KdV equation we have the soliton solution.
Soliton or solitary wave is a localized pulse which propagates without change in shape.  For a detailed study of the KdV
solitons we recommend the reading of \cite{drazin}.  The soliton will be also used as an initial condition in the study of
the numerical solution of the spherical KdV and the breaking wave equations.  For the Burgers, cKP, KP and cylindrical
KP-Burgers we present the exact solutions as developed in \cite{egypt,jukui,yunliang,mushtaq,sahu11,kp2004,kpsol,gino} where
several  techniques to solve nonlinear wave equations are presented.

\subsection{KdV equation in  nuclear matter}

We first show the soliton solution of the KdV equations (\ref{esfkdvtzero}) and (\ref{esfkdvtzeronrlim}) in cold and hot nuclear matter,
respectively \cite{fn1,fn2,fn3,fn4}. We have only soliton solutions in the cartesian case $\mathcal{G}=0$, so $\mathcal{X}=x$.
We also choose $d=1/2$  as in \cite{fn4}.
The equation (\ref{esfkdvtzero}) can be integrated and solved exactly and its soliton solution is given by:
\begin{equation}
{\hat{\rho}_{1}}(x,t)={\frac{3(u-{c_{s}})}{{c_{s}}}}(3-{c_{s}}^{2})^{-1}sech^{2}\bigg[
{\frac{{m_{V}}^{2}}{{g_{V}}}}\sqrt{{\frac{(u-{c_{s}}){c_{s}}M}
{2{\rho_{0}}}}}(x-ut) \bigg]
\label{cartsolitz}
\end{equation}
and the solution of (\ref{esfkdvtzeronrlim}) is found by setting  $3-{c_{s}}^{2} \cong 3$ in (\ref{cartsolitz}):
\begin{equation}
{\hat{\rho}_{1}}(x,t)={\frac{(u-{c_{s}})}{{c_{s}}}}sech^{2}\bigg[
{\frac{{m_{V}}^{2}}{{g_{V}}}}\sqrt{{\frac{(u-{c_{s}}){c_{s}}M}
{2{\rho_{0}}}}}(x-ut) \bigg]
\label{cartsolitznrlim}
\end{equation}
Finally, the solution of  (\ref{esfkdvt})  is:
\begin{equation}
{\hat{\rho}_{1}}(x,t)={\frac{3(u-{c_{s}})}{{c_{s}}}}\bigg(2-{c_{s}}^{2}
-{\frac{{\mu_{B}}{m_{V}}^{2}{c_{s}}^{2}}{2{g_{V}}^{2}{\rho_{0}}}}\bigg)^{-1}sech^{2}\bigg[
\sqrt{{\frac{(u-{c_{s}}){m_{V}}^{2}}
{2c_{s} }}}(x-ut) \bigg]
\label{cartsolit}
\end{equation}

\subsection{KP equations in cold QGP}

The cKP equation (\ref{r3dckp}) has
the exact analytical soliton solution \cite{nos2013}:
\begin{equation}
\hat\rho_{1}(r,\varphi,z,t)={\frac{h_{1}}{h_{2}}}
sech^{2}\Bigg\{{\frac{\sqrt{h_{1}}}{2}}\Bigg[ar+bz-\Bigg(u+a{\frac{{c_{s}}\varphi^{2}}{2}} \Bigg)t\Bigg]\Bigg\}
\label{ckpsol}
\end{equation}
where $u$ is a  parameter which satisfies $u>a{c_{s}}+b^{2}{c_{s}}/2a$  and the
phase velocity given by $u+a{\frac{{c_{s}}\varphi^{2}}{2}}$.
The  constants appearing in the above expression are:
\begin{equation}
h_{1}= {\frac{u-a{c_{s}}-b^{2}{c_{s}}/2a}{{a^{3}}c_{2}}} \hspace{2.0cm} \hbox{and} \hspace{2.0cm}
h_{2}= {\frac{c_{1}}{3a^{2}c_{2}}}
\label{bees}
\end{equation}
where
\begin{equation}
c_{1} \equiv \bigg[{\frac{3}{2}}(1-{c_{s}}^{2})-{\frac{\pi^{2/3}{\rho_{0}}^{4/3}}{3A}}\bigg]c_{s}
\end{equation}
and
\begin{equation}
c_{2} \equiv \bigg[{\frac{9g^{2}\,{\rho_{0}}^{2}c_{s}}{8{m_{G}}^{4}A}}\bigg]
\end{equation}
For the KP equation (\ref{kpxyztsual}) we have the following soliton solution \cite{nos2013}:
\begin{equation}
\hat\rho_{1}(x,y,z,t)={\frac{3(U-w)}{{\mathcal{A}}c_{1}}}
sech^{2}\Bigg[{\sqrt{{\frac{(U-w)}{4{\mathcal{A}}^{3}c_{2}}}}}\Bigg({\mathcal{A}}x+{\mathcal{B}}y+{\mathcal{C}}z-Ut\Bigg)\Bigg]
\label{kpsol}
\end{equation}
where ${\mathcal{A}}$, ${\mathcal{B}}$, ${\mathcal{C}}$ are real constants and  $w$ is given by:
\begin{equation}
w={\mathcal{A}}{c_{s}}+{\frac{{\mathcal{B}}^{2}{c_{s}}}{2}}+{\frac{{\mathcal{C}}^{2}{c_{s}}}{2}}
\label{pabove}
\end{equation}
We shall consider ${\mathcal{A}}>0$ for simplicity and the parameter $U$ such that $U > w$ .
For the KdV equation (\ref{kdvxt}) we have:
\begin{equation}
\hat\rho_{1}(x,t)={\frac{3(u-c_{s})}{c_{3}}} \ sech^{2}\Bigg[{\sqrt{{\frac{(u-c_{s})}{4c_{4}}}}}(x-ut)\Bigg]
\label{solitonkdvqcd}
\end{equation}
where $u$ is an arbitrary supersonic velocity and the constants $c_{3}$ and $c_{4}$ are given by:
\begin{equation}
c_{3} \equiv \bigg[{\frac{(2-{c_{s}}^{2})}{2}}-\bigg({\frac{27g^{2}\,{\rho_{0}}^{2}}{{m_{G}}^{2}}}\bigg)
{\frac{(2{c_{s}}^{2}-1)}{2A}}-{\frac{\pi^{2/3}{\rho_{0}}^{4/3}}{A}}\bigg({c_{s}}^{2}-{\frac{1}{6}}\bigg)\bigg]c_{s}
\label{alfakdvqcd}
\end{equation}
and
\begin{equation}
c_{4} \equiv \bigg[{\frac{9g^{2}\,{\rho_{0}}^{2}{c_{s}}}{{m_{G}}^{4}A}}\bigg]
\label{betakdvqcd}
\end{equation}

\subsection{KP-Burgers equation in hot QGP}

The cKP-B (\ref{cburgers3d}) has the solutions:
$$
\hat\varepsilon_{1}(r,z,\varphi,t)={\frac{2D A}{c_{s}T_{0}}}\Bigg({\frac{\zeta}{s}}  + \frac{4}{3} {\frac{\eta}{s}}  \Bigg)
\Bigg[1+ \Bigg({\frac{T_{B}}{T_{0}}}\Bigg)^{4}\Bigg]^{-1}
$$
$$
-{\frac{2D A}{c_{s}T_{0}}}\Bigg({\frac{\zeta}{s}}  + \frac{4}{3} {\frac{\eta}{s}}  \Bigg)
\Bigg[1+ \Bigg({\frac{T_{B}}{T_{0}}}\Bigg)^{4}\Bigg]^{-1} \times
$$
\begin{equation}
\times \, tanh\Bigg\{D\Bigg[Ar+Bz-A{\frac{{c_{s}}\varphi^{2}t}{2}}-\Bigg(Ac_{s}+{\frac{B^{2}c_{s}}{2A}}+
{\frac{D A^{2}}{T_{0}}}\bigg({\frac{\zeta}{s}}  + \frac{4}{3} {\frac{\eta}{s}}  \bigg)\Bigg)t\Bigg]\Bigg\}
\label{burgimpoapair1final}
\end{equation}
\

$$
\hat\varepsilon_{1}(r,z,\varphi,t)=-{\frac{2D A}{c_{s}T_{0}}}\Bigg({\frac{\zeta}{s}}  + \frac{4}{3} {\frac{\eta}{s}}  \Bigg)
\Bigg[1+ \Bigg({\frac{T_{B}}{T_{0}}}\Bigg)^{4}\Bigg]^{-1}
$$
$$
-{\frac{2D A}{c_{s}T_{0}}}\Bigg({\frac{\zeta}{s}}  + \frac{4}{3} {\frac{\eta}{s}}  \Bigg)
\Bigg[1+ \Bigg({\frac{T_{B}}{T_{0}}}\Bigg)^{4}\Bigg]^{-1} \times
$$
\begin{equation}
\times \, tanh\Bigg\{D\Bigg[Ar+Bz-A{\frac{{c_{s}}\varphi^{2}t}{2}}-\Bigg(Ac_{s}+{\frac{B^{2}c_{s}}{2A}}-
{\frac{D A^{2}}{T_{0}}}\bigg({\frac{\zeta}{s}}  + \frac{4}{3} {\frac{\eta}{s}}  \bigg)\Bigg)t\Bigg]\Bigg\}
\label{burgimpoapair2final}
\end{equation}
\

or the following solutions:
$$
\hat\varepsilon_{1}(r,z,\varphi,t)={\frac{2D A}{c_{s}T_{0}}}\Bigg({\frac{\zeta}{s}}  + \frac{4}{3} {\frac{\eta}{s}}  \Bigg)
\Bigg[1+ \Bigg({\frac{T_{B}}{T_{0}}}\Bigg)^{4}\Bigg]^{-1}
$$
$$
-{\frac{2D A}{c_{s}T_{0}}}\Bigg({\frac{\zeta}{s}}  + \frac{4}{3} {\frac{\eta}{s}}  \Bigg)
\Bigg[1+ \Bigg({\frac{T_{B}}{T_{0}}}\Bigg)^{4}\Bigg]^{-1} \times
$$
\begin{equation}
\times \, coth\Bigg\{D\Bigg[Ar+Bz-A{\frac{{c_{s}}\varphi^{2}t}{2}}-\Bigg(Ac_{s}+{\frac{B^{2}c_{s}}{2A}}+
{\frac{D A^{2}}{T_{0}}}\bigg({\frac{\zeta}{s}}  + \frac{4}{3} {\frac{\eta}{s}}  \bigg)\Bigg)t\Bigg]\Bigg\}
\label{burgimpoapair1finalcoth}
\end{equation}
\

$$
\hat\varepsilon_{1}(r,z,\varphi,t)=-{\frac{2D A}{c_{s}T_{0}}}\Bigg({\frac{\zeta}{s}}  + \frac{4}{3} {\frac{\eta}{s}}  \Bigg)
\Bigg[1+ \Bigg({\frac{T_{B}}{T_{0}}}\Bigg)^{4}\Bigg]^{-1}
$$
$$
-{\frac{2D A}{c_{s}T_{0}}}\Bigg({\frac{\zeta}{s}}  + \frac{4}{3} {\frac{\eta}{s}}  \Bigg)
\Bigg[1+ \Bigg({\frac{T_{B}}{T_{0}}}\Bigg)^{4}\Bigg]^{-1} \times
$$
\begin{equation}
\times \, coth\Bigg\{D\Bigg[Ar+Bz-A{\frac{{c_{s}}\varphi^{2}t}{2}}-\Bigg(Ac_{s}+{\frac{B^{2}c_{s}}{2A}}-
{\frac{D A^{2}}{T_{0}}}\bigg({\frac{\zeta}{s}}  + \frac{4}{3} {\frac{\eta}{s}}  \bigg)\Bigg)t\Bigg]\Bigg\}
\label{burgimpoapair2finalcoth}
\end{equation}
where the real constants to be chosen
are $A$, $B$ and $D$.

\section{Numerical solutions }

In this section we present numerical results. The solutions of the differential equations can be grouped in those which are
smooth and those which exhibit some non-smooth behavior, such as rapid oscillations or the formation of ``walls'', specially at
later times.  This kind of behavior appears when there is a lack of balance between the different terms of the equations. Therefore,
before presenting numbers and plots, we discuss, in the next subsection, the conditions for finding stable solutions.

\subsection{Soliton stability}

As we have seen, perturbations in fluids with  different equations of state  generate different nonlinear wave equations.  Some examples are
the Kadomtsev-Petviashvili (KP) equation:
\begin{equation}
{\frac{\partial}{\partial x}}\Bigg\{\frac{\partial F}{\partial t}+\alpha_{1} F\frac{\partial F}{\partial x}+\alpha_{2}
\frac{\partial ^3F}{\partial x^3}\Bigg\}+
\alpha_{3}\frac{\partial ^2F}{\partial y^2}+\alpha_{4}\frac{\partial ^2F}{\partial z^2}=0.
\label{KPgeneral}
\end{equation}
and its particular cases, such as the KdV:
\begin{equation}
\frac{\partial F}{\partial t}+\alpha_{1} F\frac{\partial F}{\partial x}+\alpha_{2} \frac{\partial ^3F}{\partial x^3}=0
\label{KdVgeneral}
\end{equation}
and the breaking wave equation:
\begin{equation}
\frac{\partial F}{\partial t}+\alpha_{1} F\frac{\partial F}{\partial x}=0
\label{BWgeneral}
\end{equation}
Another example of a  nonlinear wave in a  dissipative system is  the Burgers equation:
\begin{equation}
\frac{\partial F}{\partial t}+\alpha_{1} F\frac{\partial F}{\partial x}
+\alpha_{2} \frac{\partial ^2F}{\partial x^2}=0
\label{Burgersgeneral}
\end{equation}
where $\alpha_{1}$ to $\alpha_{4}$ are real constants.
Having derived a particular differential equation,  we can check whether the obtained equation is
consistent with the physical picture of a small amplitude and long wave length perturbation propagating over large distances.
We shall follow the analysis performed in Ref. \cite{leblond}.  Let us assume that the nonlinear equations listed above for a generic
function $F$ have a solitary wave solution with a typical large length $L \simeq 1/\sigma \,\,$   $ \,\, (0<\sigma << 1)$. Considering the
general case, the KP equation has a dispersion $\frac{\partial^4 F}{\partial x^4}$ term that is about $ \frac{\partial^4 F}{\partial x^4}
 \simeq  \sigma^4 F$. It must arise at a propagation distance (or equivalently propagation time T) D, accounted for in the equation by the
 term $\frac{\partial^2 F} {\partial x \partial t}   \simeq \sigma  \frac{F}{T} $.
If both the dispersion and propagation terms have the same size, then $T \simeq  1/\sigma^3$. Regarding the nonlinear term, if it has the form
$\frac{\partial}{\partial x}  ( F \frac{\partial F} { \partial x }) $ its order of magnitude is
$\sigma^2 F^2$. The formation of the soliton requires that the nonlinear effect balances the dispersion. Hence it must have the same
order of magnitude and  $F^2 \sigma^2 =  F \sigma^4$. Hence $ F \simeq  \sigma^2 $. We can then conclude that
$F \, << \, L \, << \, D$ and the above equation describes the propagation of a wave with small amplitude  $(F)$ and large wave
length $(L)$ which travels  large distances $(D)$.  In the case of KP we have terms that describe the transverse evolution of  the wave. We can
estimate their sizes only if we make assumptions about the transverse length scales. In most cases the resulting flow is one-dimensional along the $x$
direction with some ``leakage'' to the transverse directions.

\subsection{Numerical analysis}

In what follows we apply the numerical tools developed in the appendix to several cases.

\subsubsection{Nuclear soliton}

\begin{figure}[ht!]
\centering
\subfigure[]{\label{fig:1a}
\includegraphics[scale=0.35]{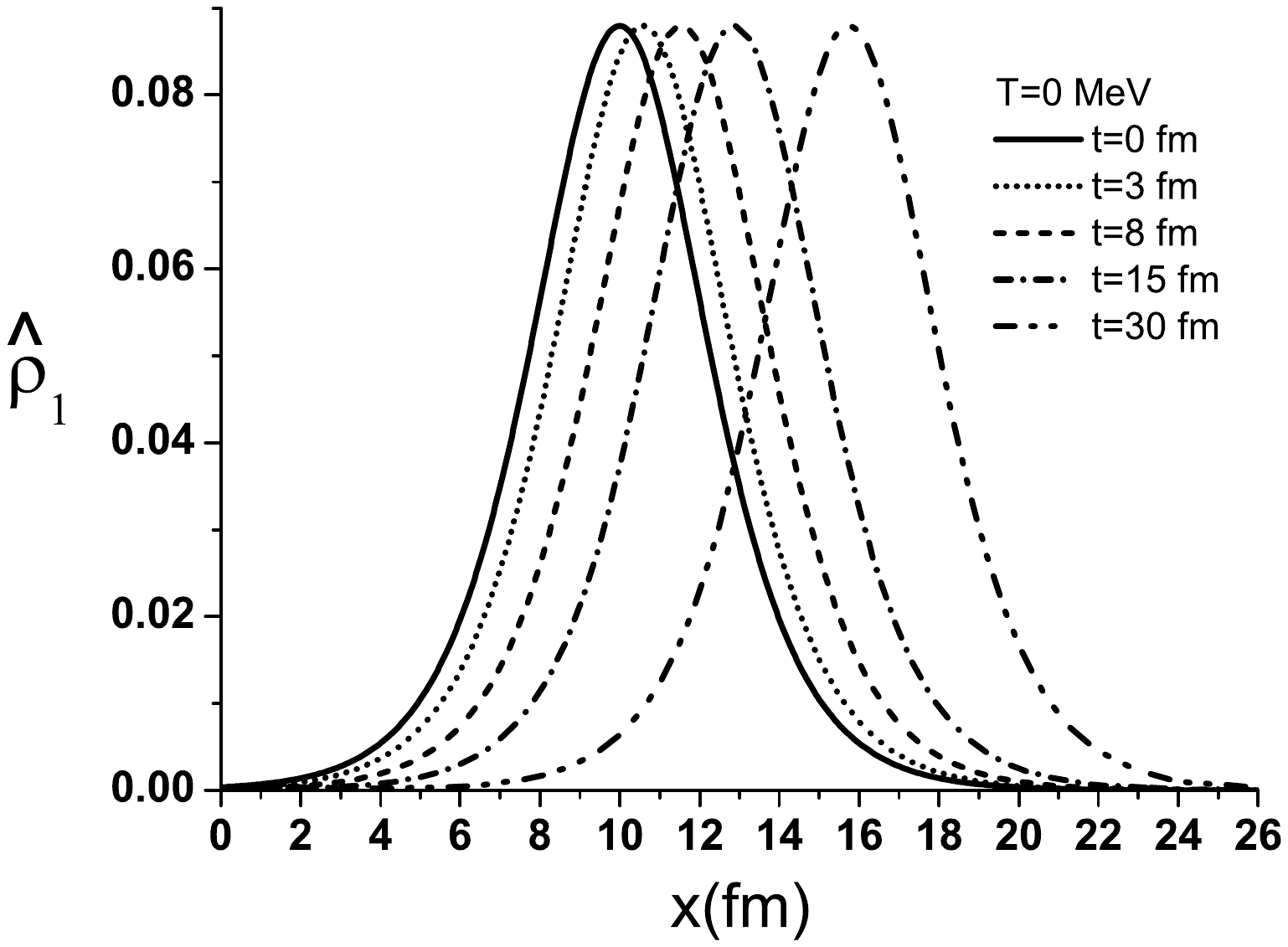}}
\subfigure[]{\label{fig:1b}
\includegraphics[scale=0.35]{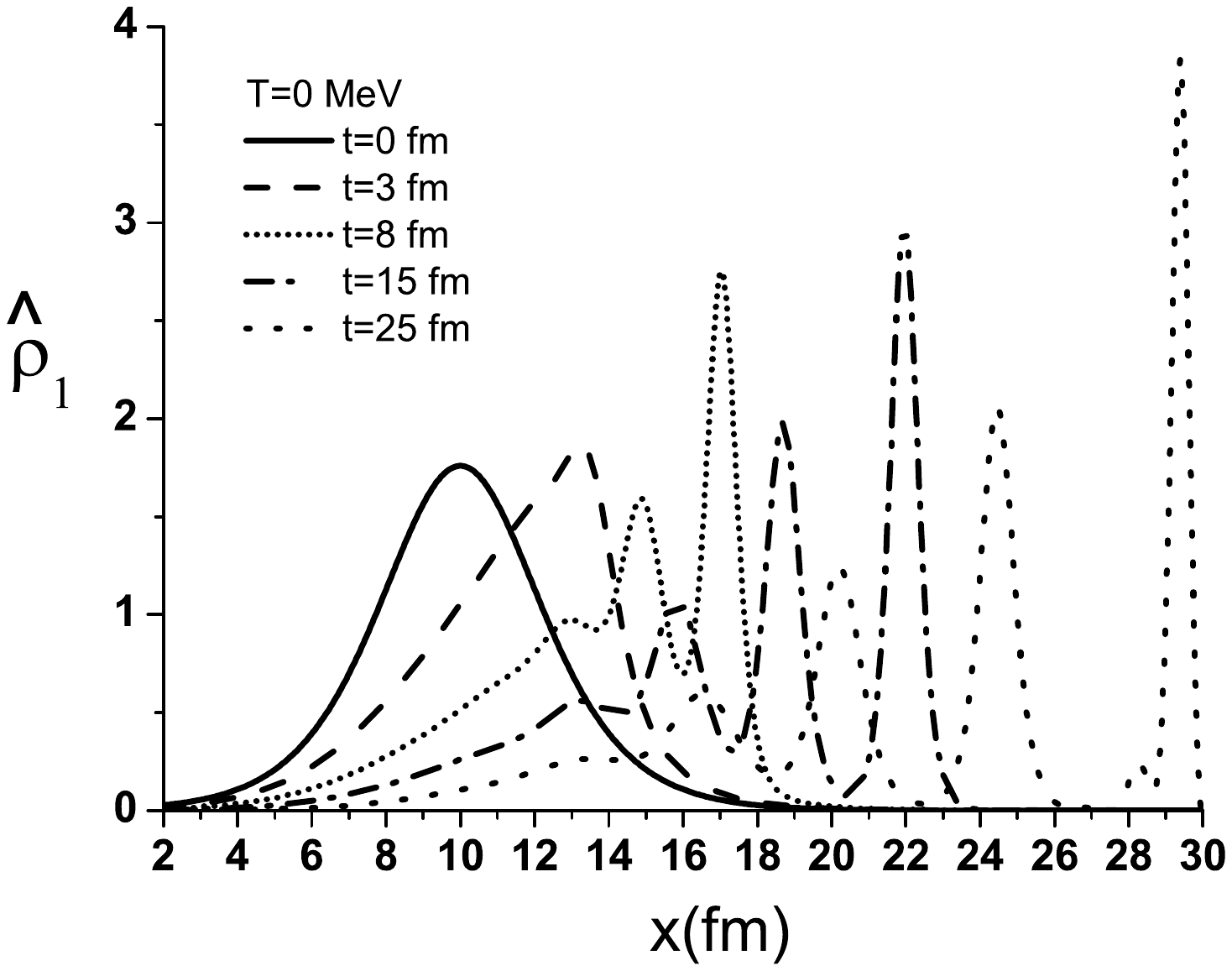}}
\caption[]{a) The evolution of the analytic solution of the
KdV equation. b) The evolution of the analytic solution
multiplied by a factor $20$.}
\label{fig1}
\end{figure}

We start our numerical analysis showing in Fig. \ref{fig1} the solution of
the linear KdV equation at $T=0$, ($\mathcal{G}=0$ and $\mathcal{X}=x$) (\ref{esfkdvtzero}) with $d={\frac{1}{2}}$. In
Fig. \ref{fig:1a}, we use the analytical solution
(\ref{cartsolitz}) as initial condition. As expected this pulse
propagates without dissipation nor dispersion: it is a soliton wave. This is the situation illustrated
in Fig. \ref{figsoliton}.  Any change in the initial
condition has noticeable consequences as it can be seen in Fig. \ref{fig:1b}, where we follow the evolution
of the numerical solution of (\ref{esfkdvtzero}) for an initial pulse given
by (\ref{cartsolitz}) multiplied by a factor $20$. As it can be seen, the
amplitude grows, the width decreases and  secondary bumps appear propagating
behind  the first.

\begin{figure}[ht!]
\centering
\subfigure[]{\label{fig:2a}
\includegraphics[scale=0.35]{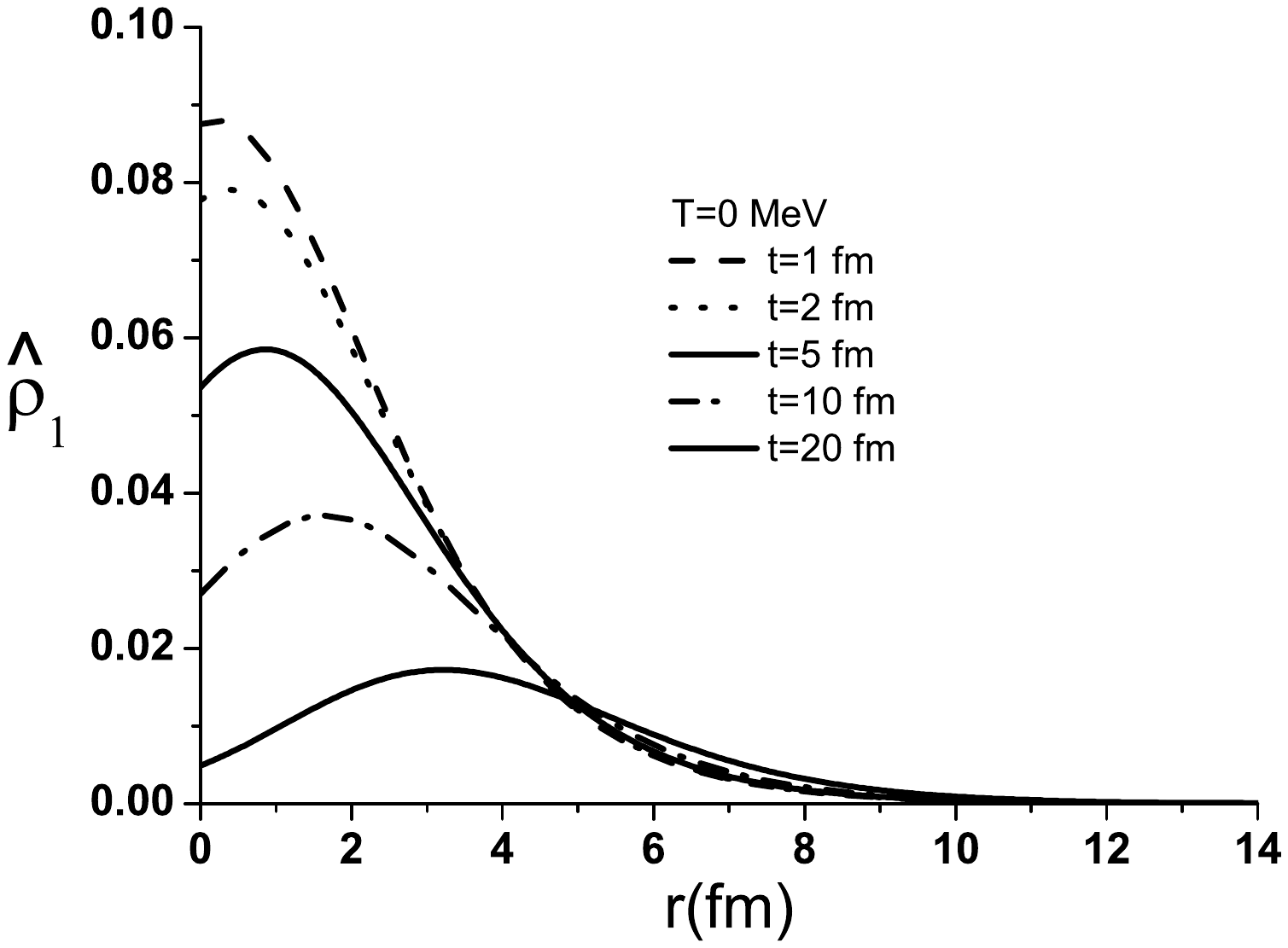}}
\subfigure[]{\label{fig:2b}
\includegraphics[scale=0.35]{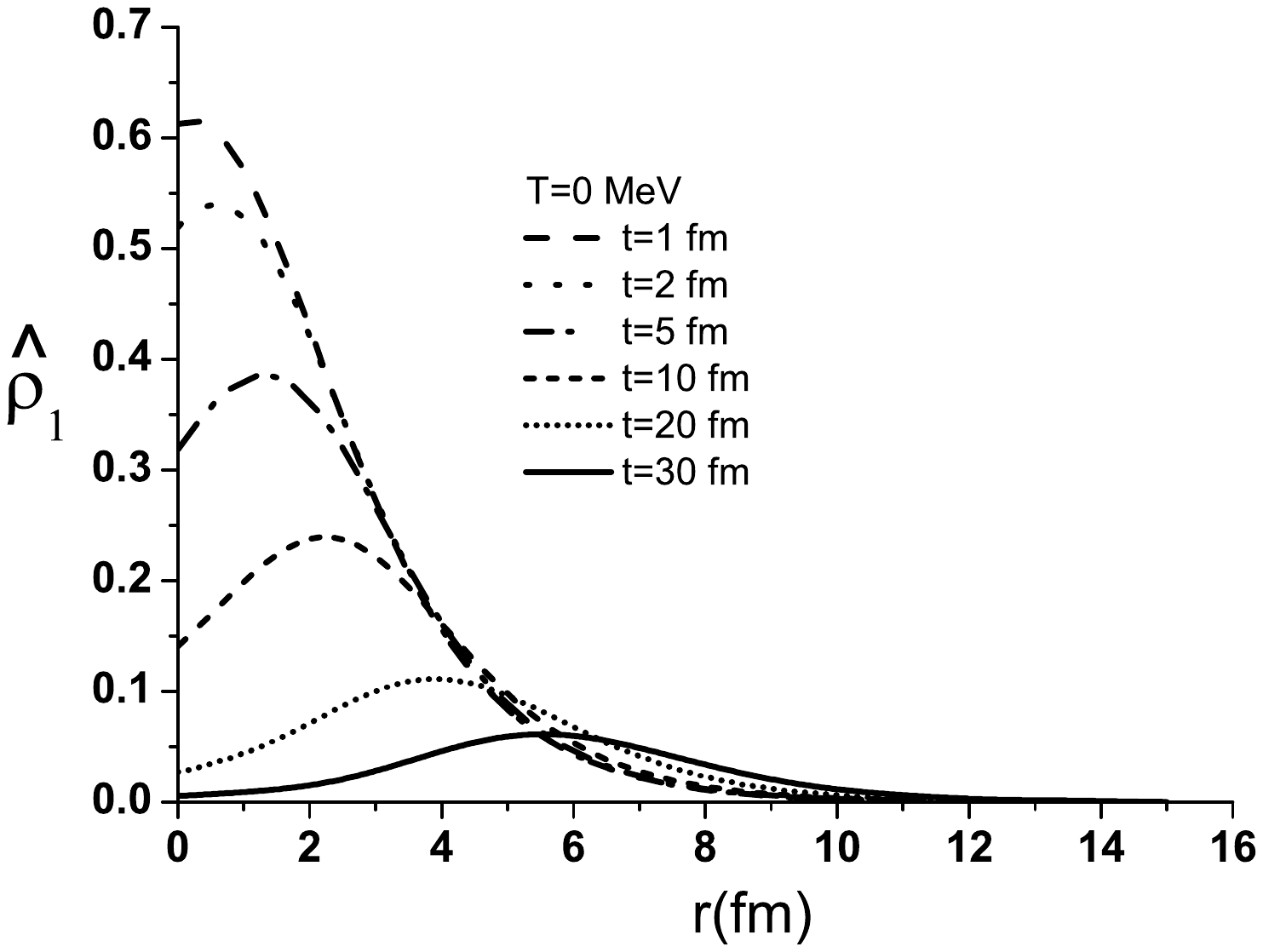}}
\caption[]{Analogous to Fig. \ref{fig1} for spherical coordinates.}
\label{fig2}
\end{figure}

In Fig. \ref{fig2}, we show the equivalent plot for the spherical
case: $\mathcal{G}=1$ and $\mathcal{X}=r$. In contrast to the linear case there is a strong damping of the pulse. The dependence
on the initial conditions is also strong.

\begin{figure}[ht!]
\centering
\subfigure[ ]{\label{fig:3a}
\includegraphics[scale=0.35]{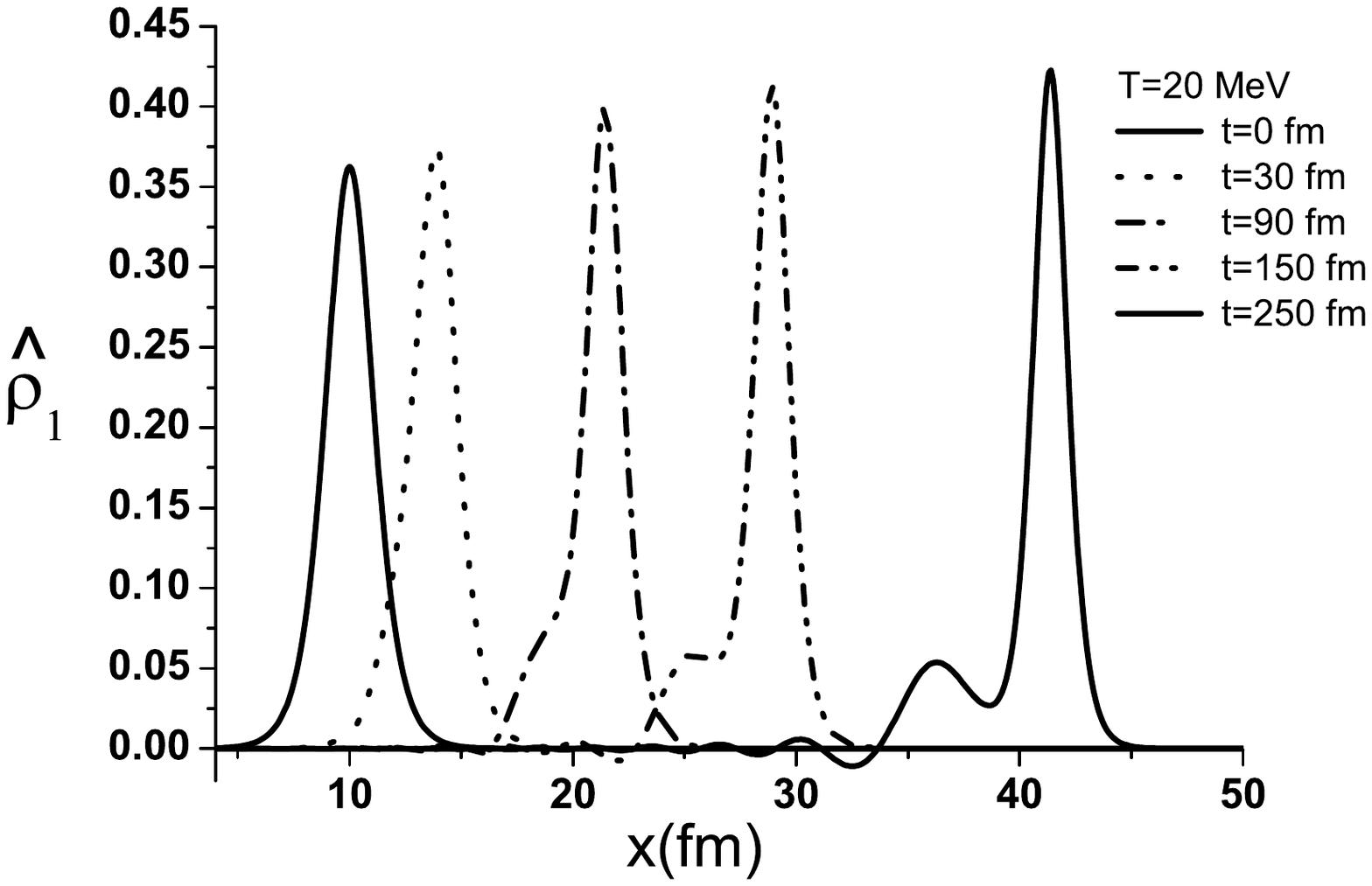}}
\subfigure[ ]{\label{fig:3b}
\includegraphics[scale=0.35]{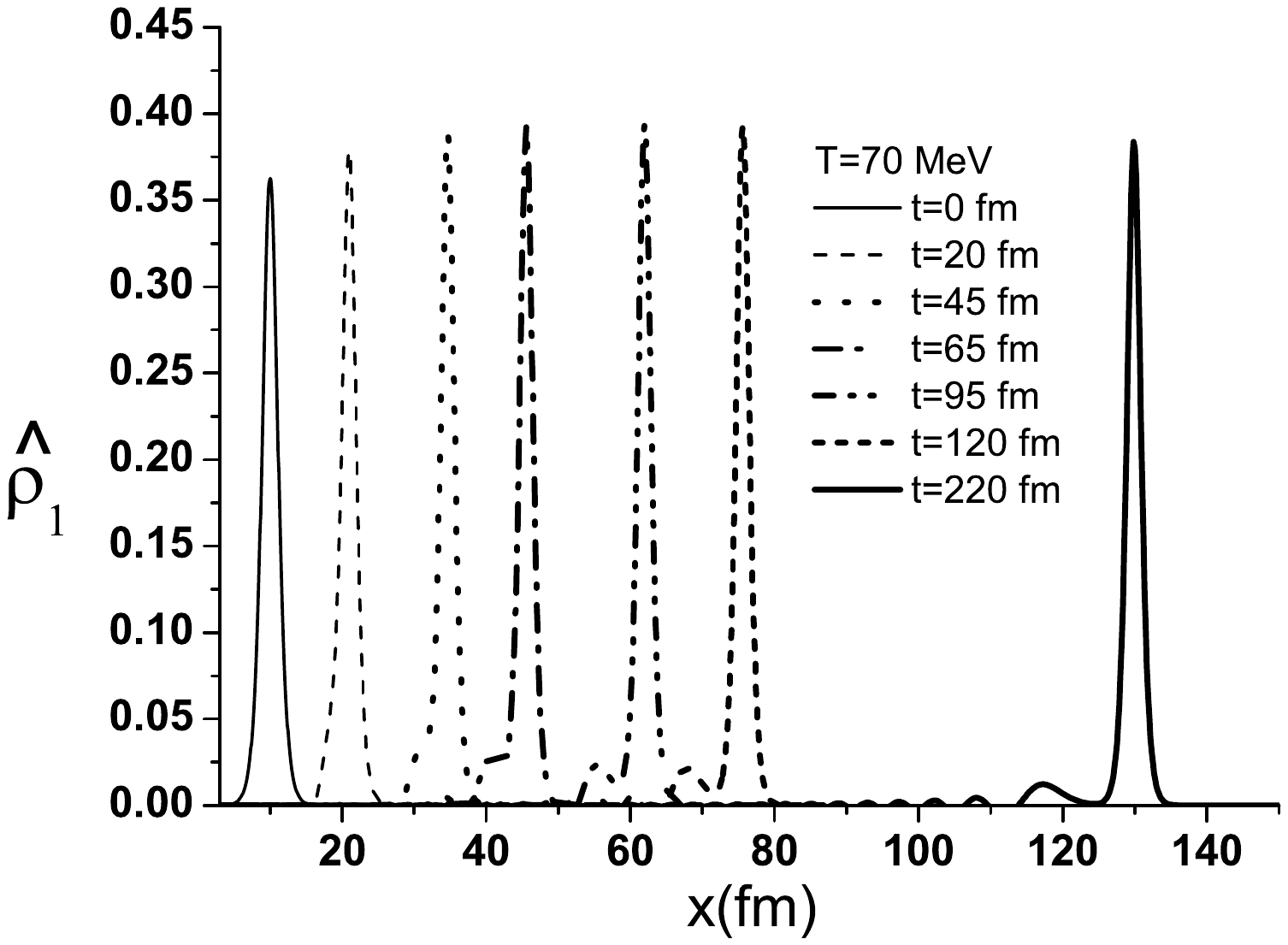}}
\caption[]{The panels show calculations with
temperatures a) $T=20$ \, and \, b) $T=70$ $MeV$ .}
\label{fig3}
\end{figure}

\begin{figure}[ht!]
\centering
\subfigure[ ]{\label{fig:4a}
\includegraphics[scale=0.35]{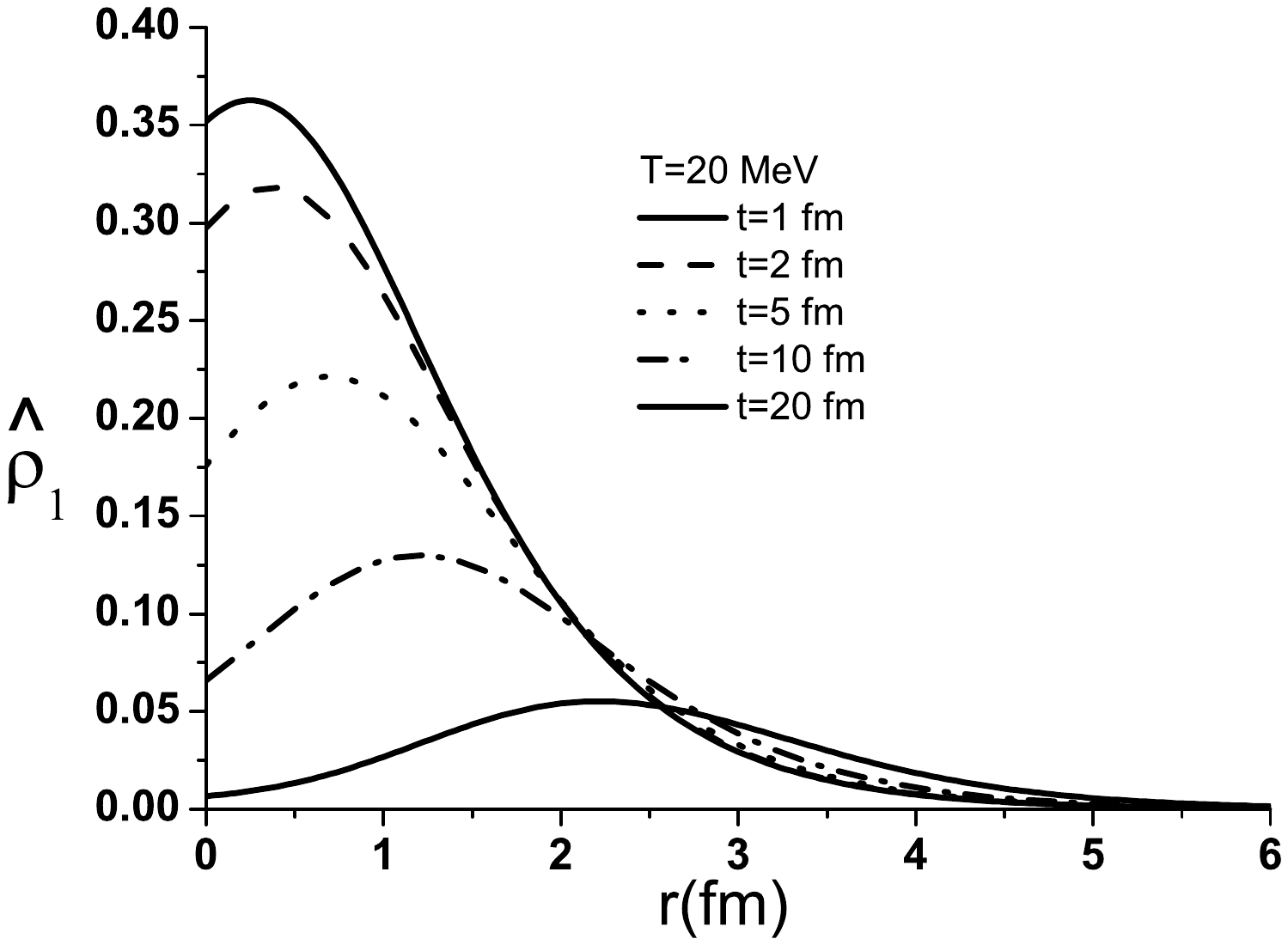}}
\subfigure[ ]{\label{fig:4b}
\includegraphics[scale=0.35]{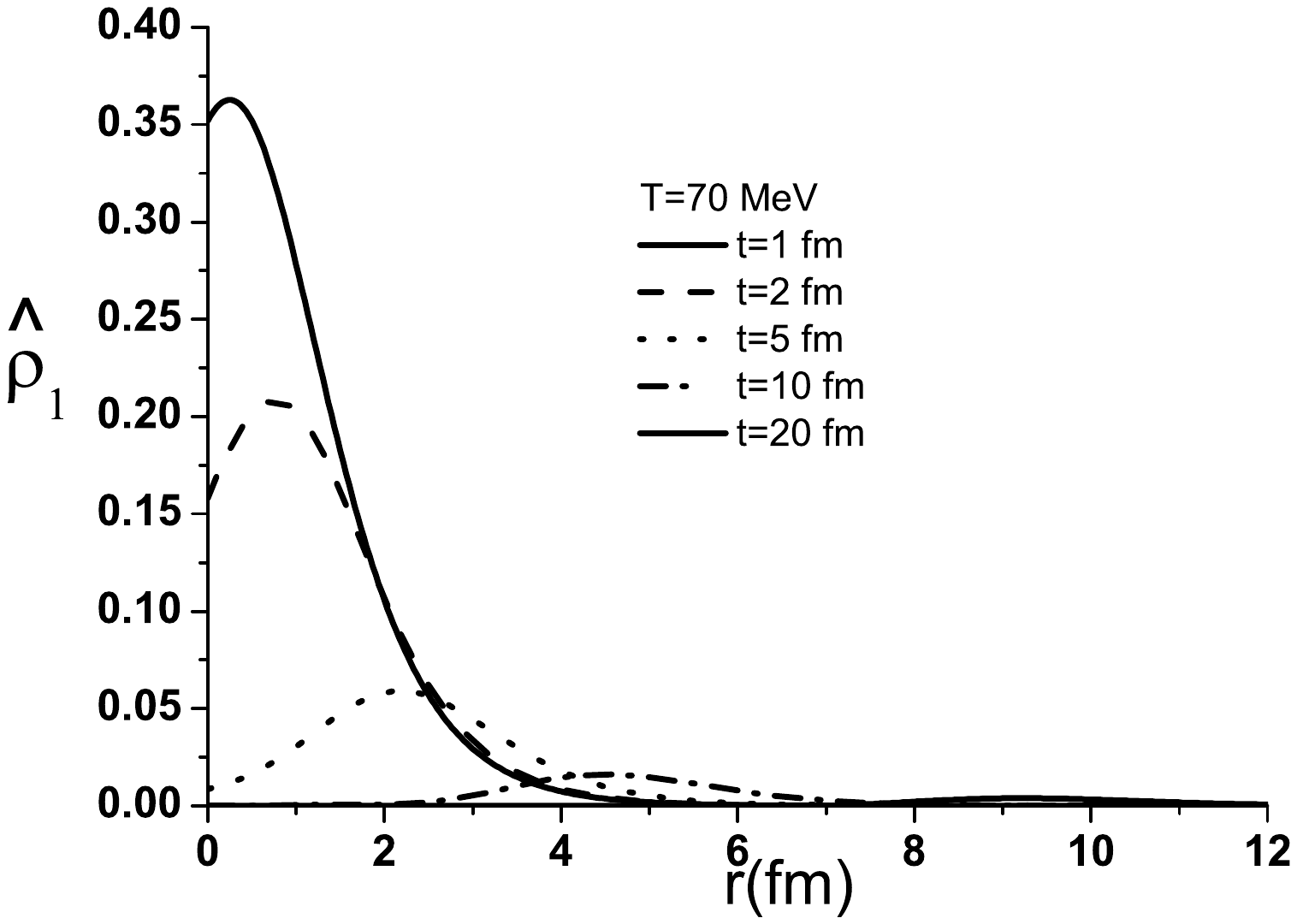}}
\caption[]{Analogous to Fig. \ref{fig3} for  spherical coordinates}
\label{fig4}
\end{figure}

In Fig. \ref{fig3} we show for the linear case and for  the
 ``optimal''  initial condition (\ref{cartsolitz}) the evolution of the
pulse with time for different temperatures. We can see that, increasing the temperature
the pulses move faster and go farther. The same feature can be observed in the spherical
case, as shown in Fig. \ref{fig4} .

\begin{figure}[ht!]
\centering
\subfigure[ ]{\label{fig:5a}
\includegraphics[scale=0.35]{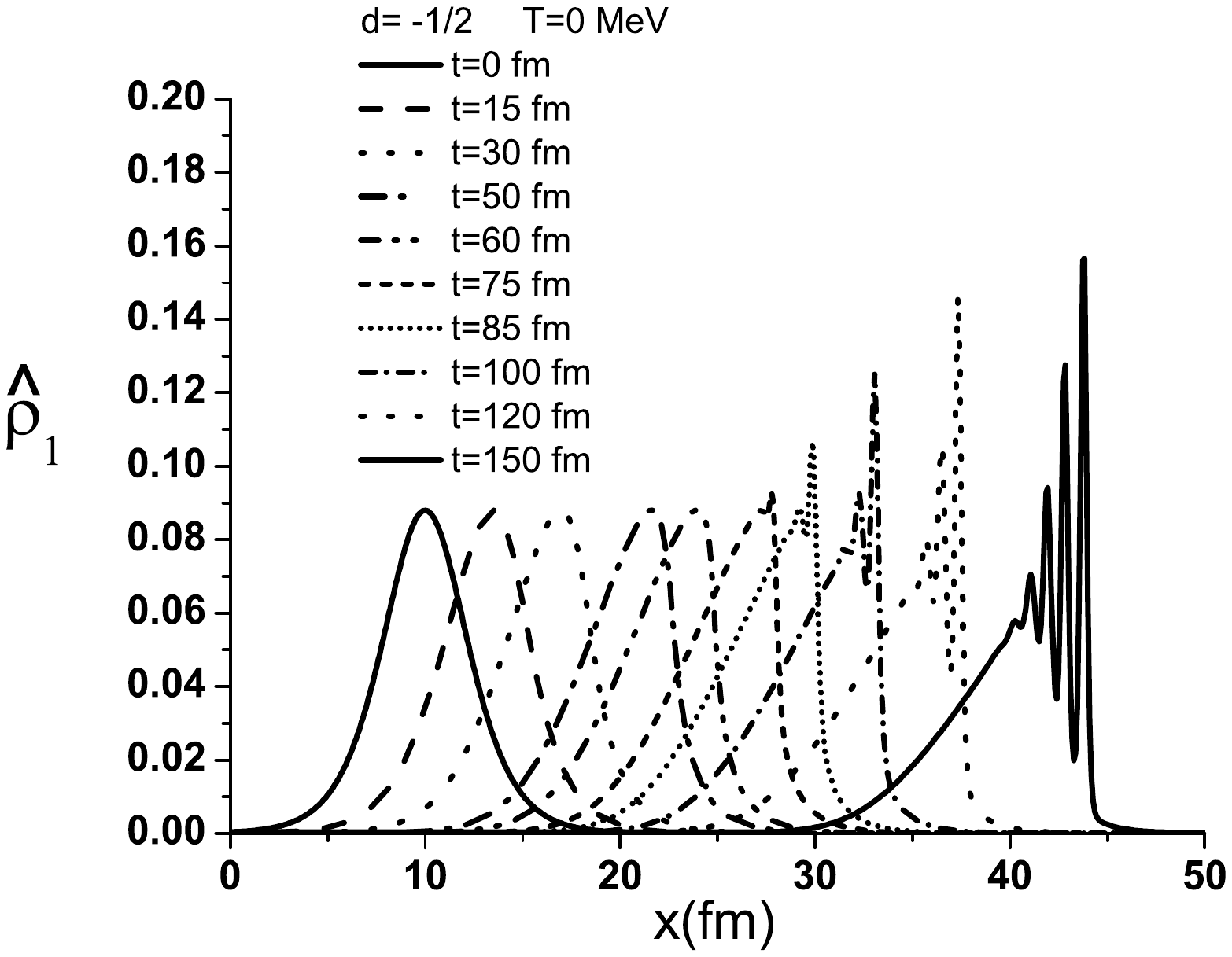}}
\subfigure[ ]{\label{fig:5b}
\includegraphics[scale=0.35]{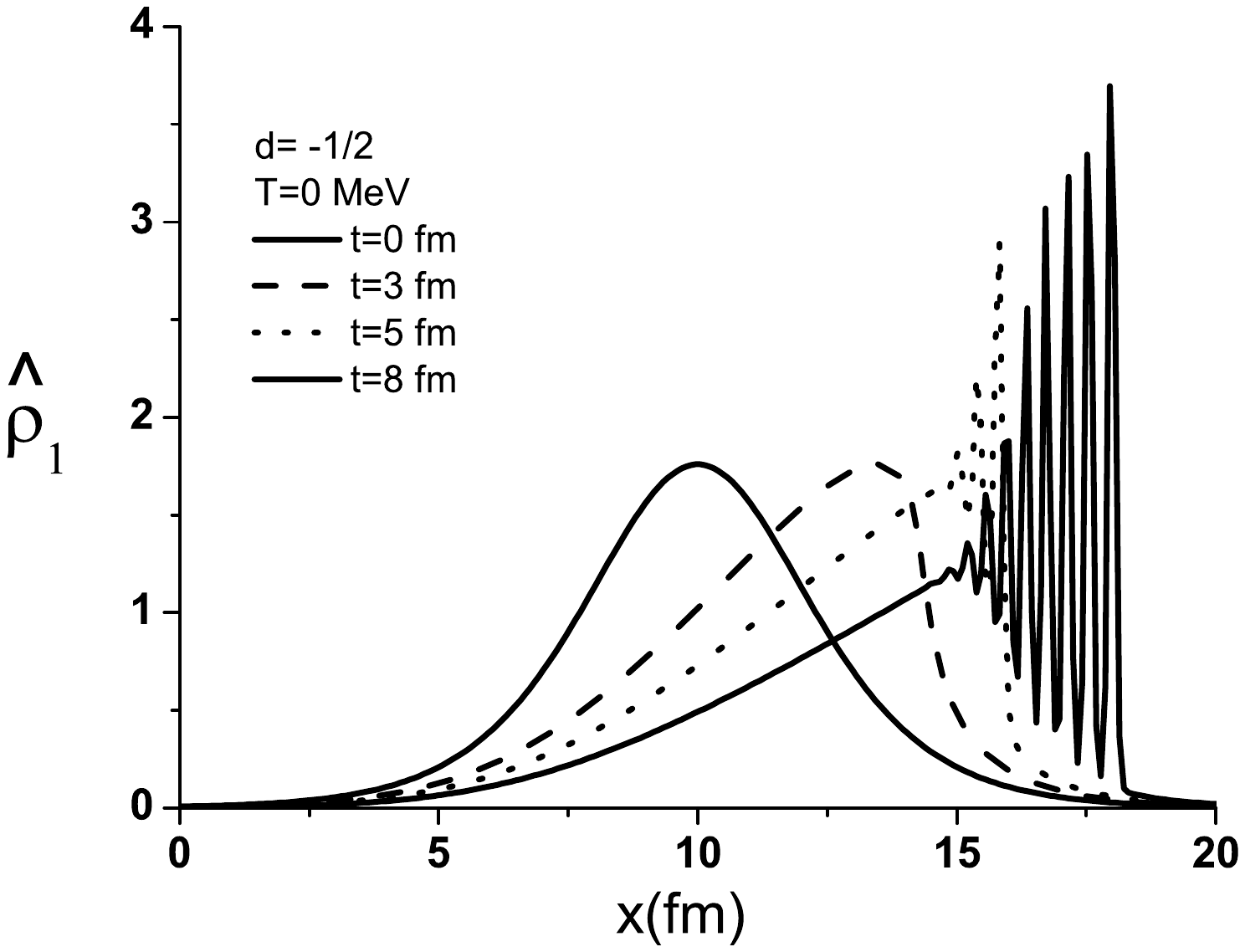}}
\caption[]{a) Shock wave formation in one dimensional Cartesian coordinates.  b) Initial profile multiplied by a factor $20$.}
\label{fig5}
\end{figure}

Setting $d=-{\frac{1}{2}}$ in the  wave equations (\ref{esfkdvtzero})
and (\ref{esfkdvt}) we eliminate
the third order derivative terms. The corresponding wave equations are breaking wave equations.  Out of
smooth initial perturbations, given by (\ref{cartsolitz}), these equations create shock waves. We can see this process in
one dimensional Cartesian coordinates in Fig. \ref{fig5} .  We observe a steepening of the profile until the formation of the
shock, followed by the dispersion of the wave. We see that the higher is the initial
amplitude, the sooner the wave breaking and dispersion occurs.

\begin{figure}[ht!]
\centering
\subfigure[ ]{\label{fig:6a}
\includegraphics[scale=0.35]{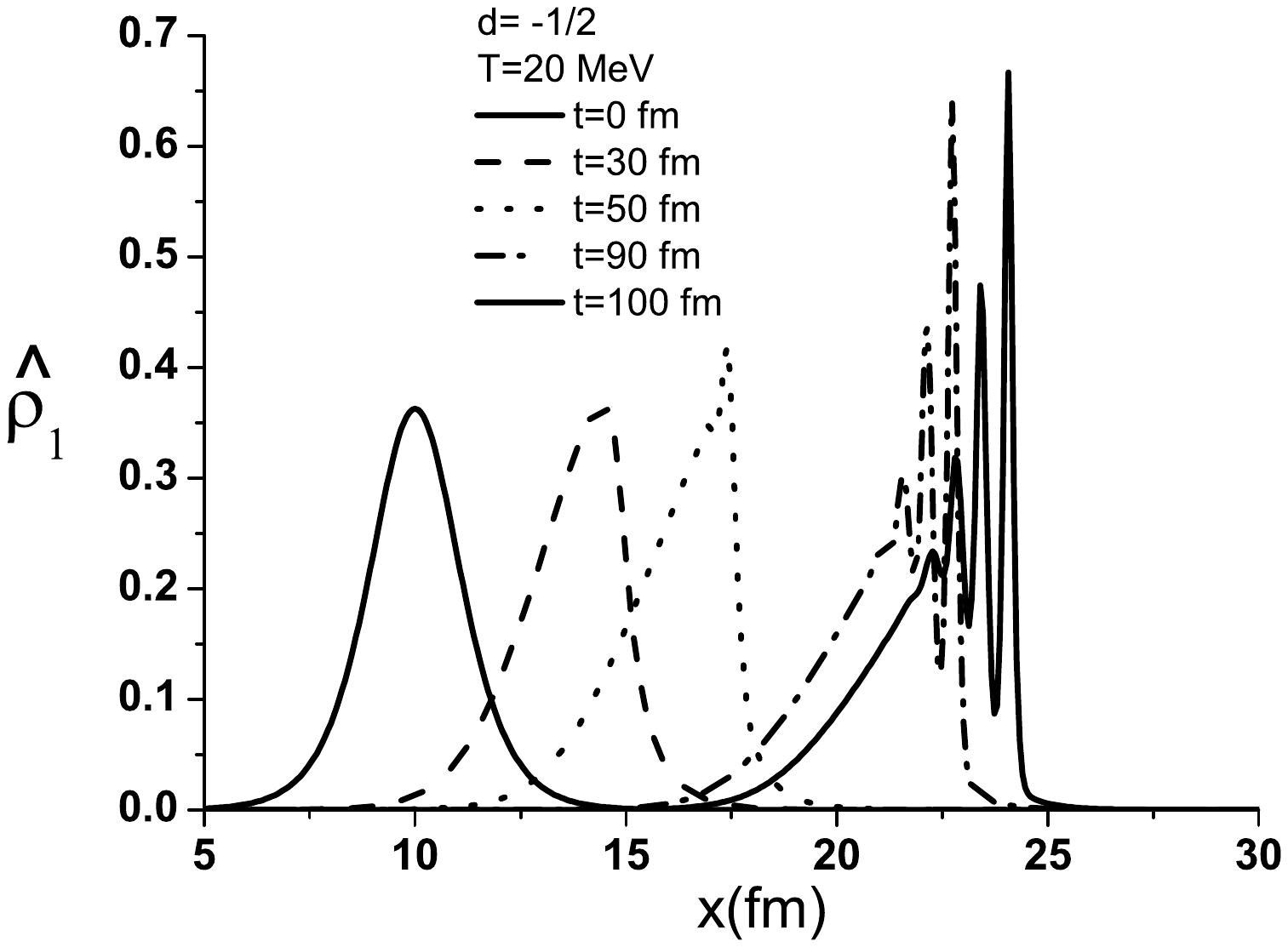}}
\subfigure[ ]{\label{fig:6b}
\includegraphics[scale=0.35]{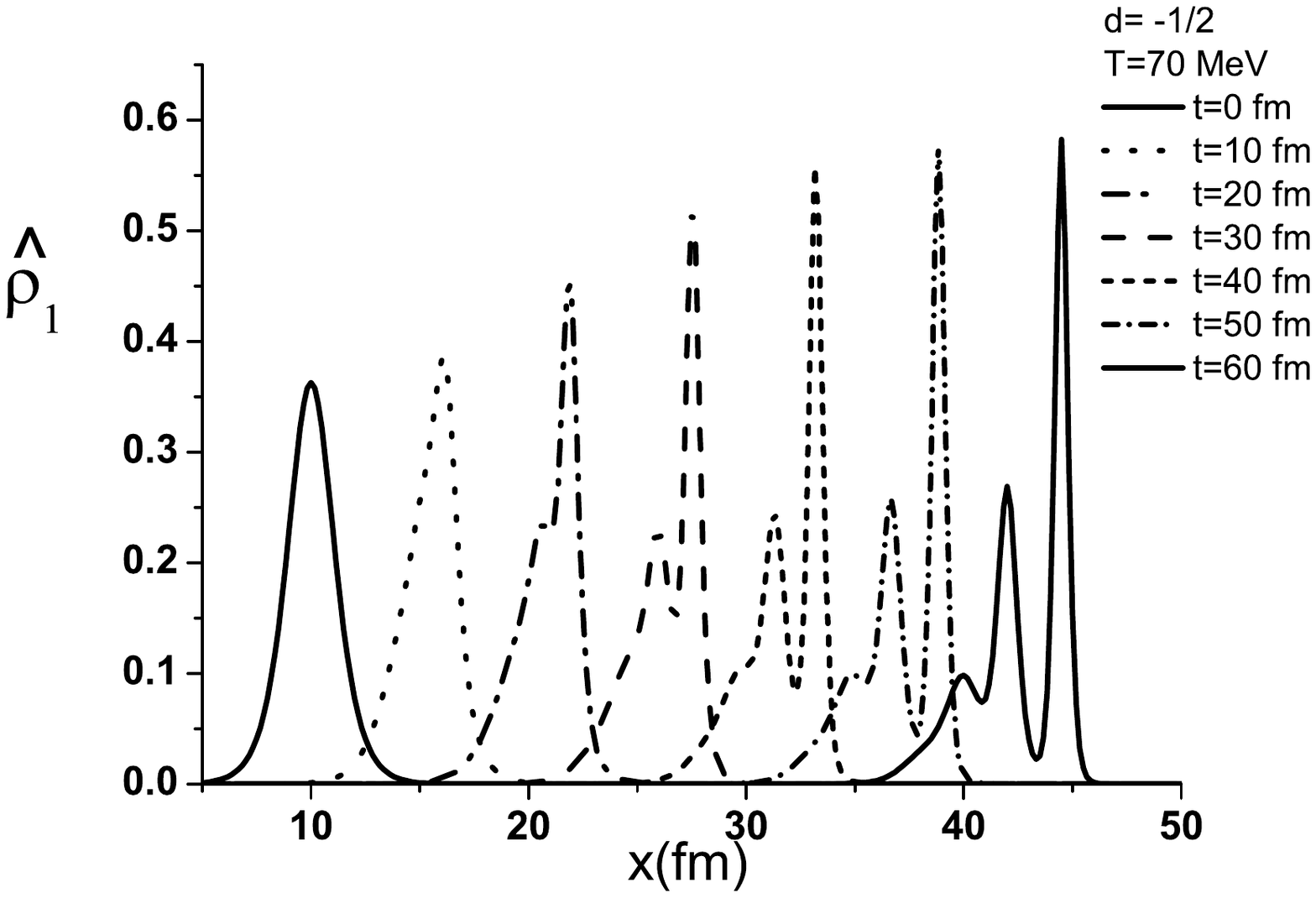}}
\caption[]{Shock wave formation in one dimensional Cartesian
coordinates for different temperatures: a) $T=20$ \, and \, b) $T=70$ $MeV$ .}
\label{fig6}
\end{figure}

In Fig. \ref{fig6} we fix one initial profile and study its time evolution for
two different tempe-ratures. \ref{fig:6a} and \ref{fig:6b} show the development of a shock wave at $T=20$ $MeV$ and $70$ $MeV$ respectively.
As it can be seen, with increasing
temperatures the pulse moves faster and the shock formation and the subsequent dispersive
breaking occurs later.  For the radial case a similar behavior is observed.

\subsubsection{Breaking wave equation in QGP}

The KdV equation can be written as:
\begin{equation}
\frac{\partial u}{\partial t}+
c_s {\frac{\partial u}{\partial t}}
+\beta \frac{\partial}{\partial x}\bigg({\frac{u^{2}}{2}}\bigg) + \mu \frac{\partial ^3u}{\partial x^3}=0
\label{KdV1}
\end{equation}
which has the following analytical soliton solution:
\begin{equation}
f(x,t)={\frac{3(u-c_{s})}{\beta}} \ sech^{2}\Bigg[{\sqrt{{\frac{(u-c_{s})}{4\mu}}}}(x-ut)\Bigg]
\label{exactumKdVdra}
\end{equation}
In the numerical study of  (\ref{bwqcdxitauXt}) and (\ref{bwqcdTfincomrefens}) we use the following soliton-like profile:
\begin{equation}
\hat\rho_{1}(x,t_{0})= A \ sech^{2}\bigg[\frac{x}{B}\bigg]
\label{exactumKdVTZEROlv}
\end{equation}
In this equation $A$ and $B$ represent the amplitude and width of the initial baryon density pulse, respectively.

\begin{figure}[ht!]
\centering
\subfigure[ ]{\label{fig:7a}
\includegraphics[scale=0.35]{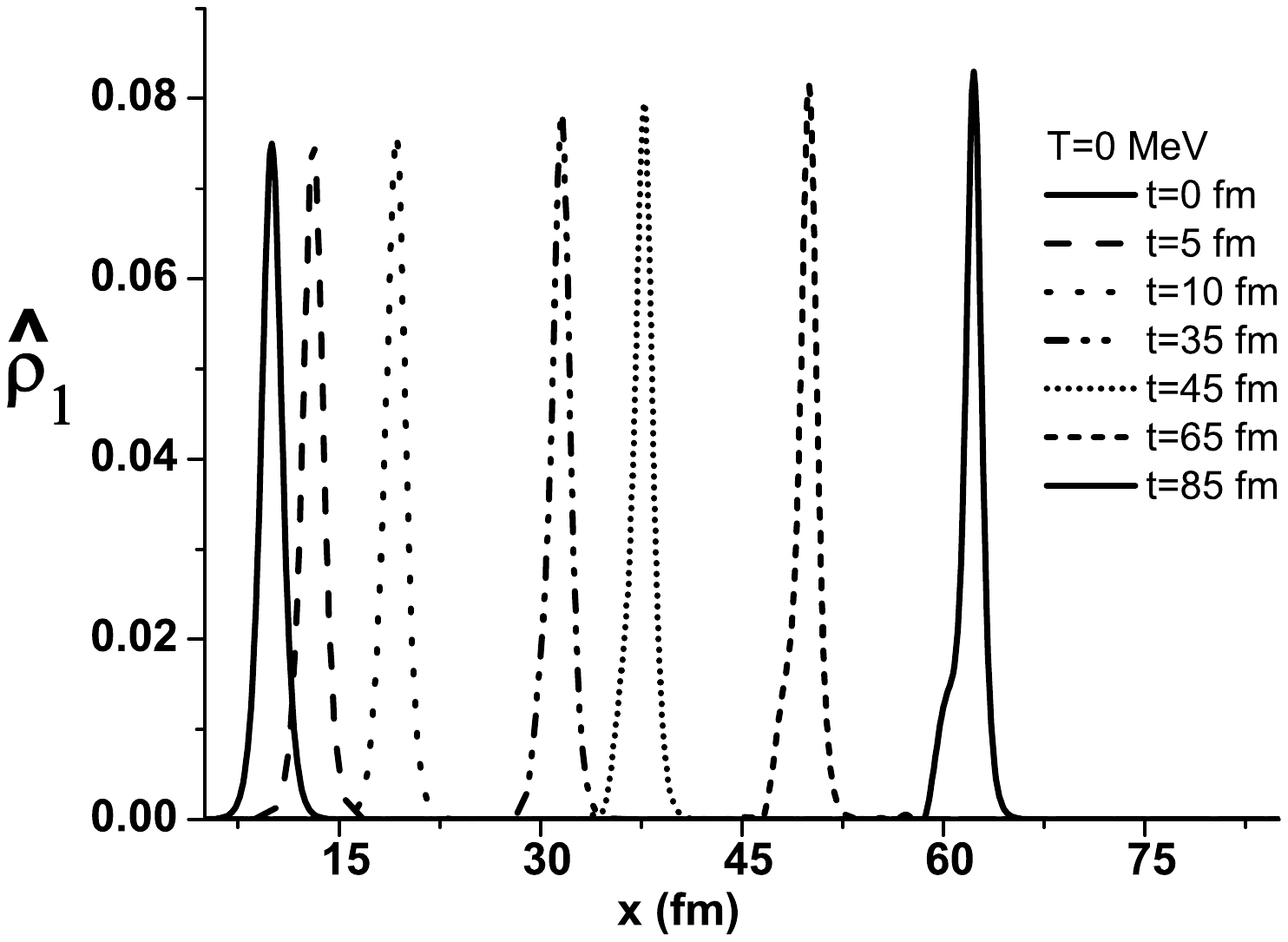}}
\subfigure[ ]{\label{fig:7b}
\includegraphics[scale=0.35]{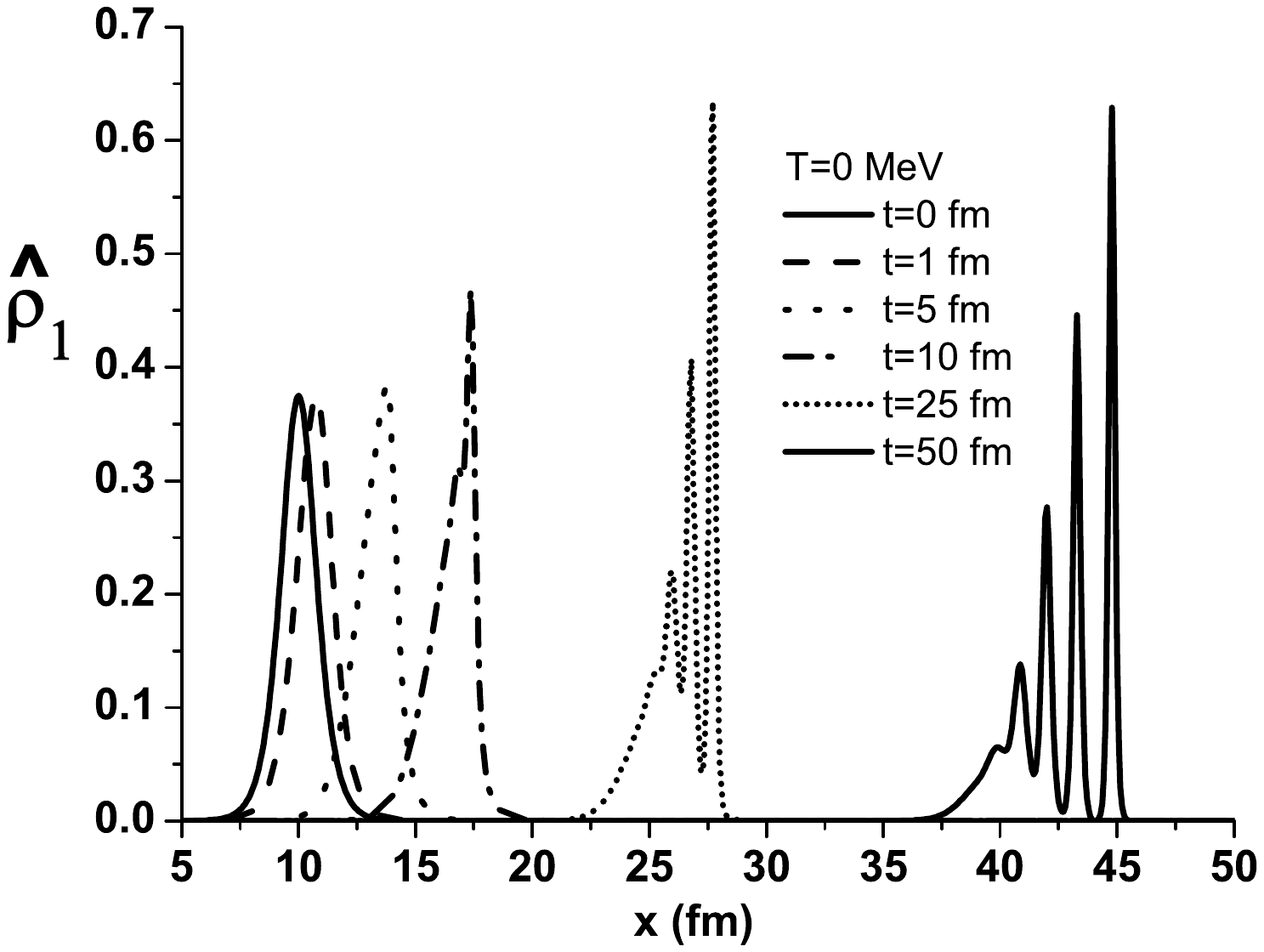}}
\caption[]{Time evolution of a baryon density pulse for cold QGP.}
\label{fig7}
\end{figure}

We present in Fig. \ref{fig7} the numerical solution of (\ref{bwqcdxitauXt}) for different times. In Fig.  \ref{fig:7a} we show for $A=0.075$
and $B=1$ fm and in  \ref{fig:7b} for $A=0.35$  and $B=1$ fm.  It is possible to  observe the evolution of the initial
gaussian-like pulse with the formation of a ``wall'' on the right side. In  \ref{fig:7b} the ``wall'' formation and dispersion occurs much earlier
than in  \ref{fig:7a} due higher initial amplitude.

\begin{figure}[ht!]
\centering
\subfigure[ ]{\label{fig:8a}
\includegraphics[scale=0.35]{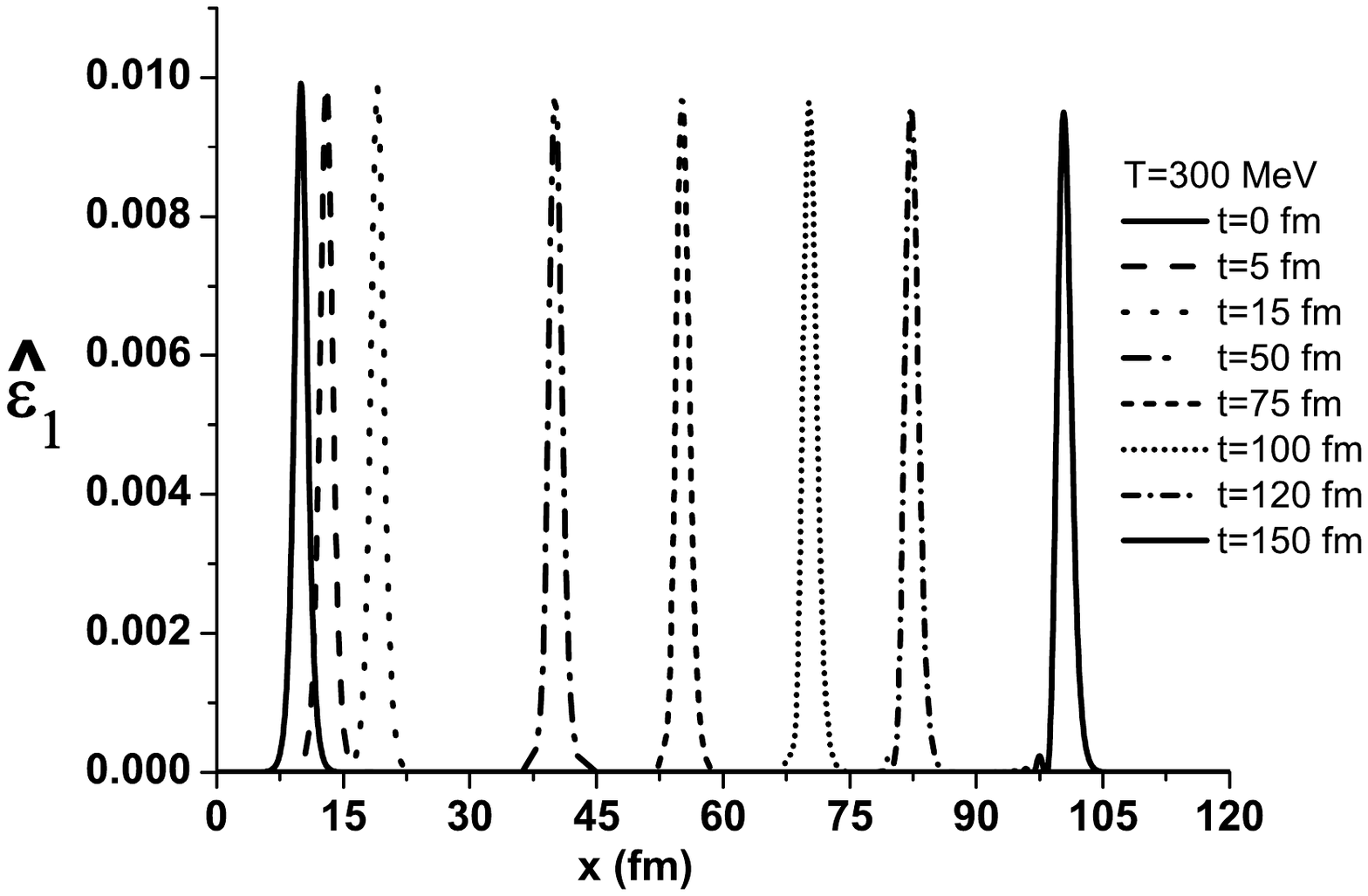}}
\subfigure[ ]{\label{fig:8b}
\includegraphics[scale=0.35]{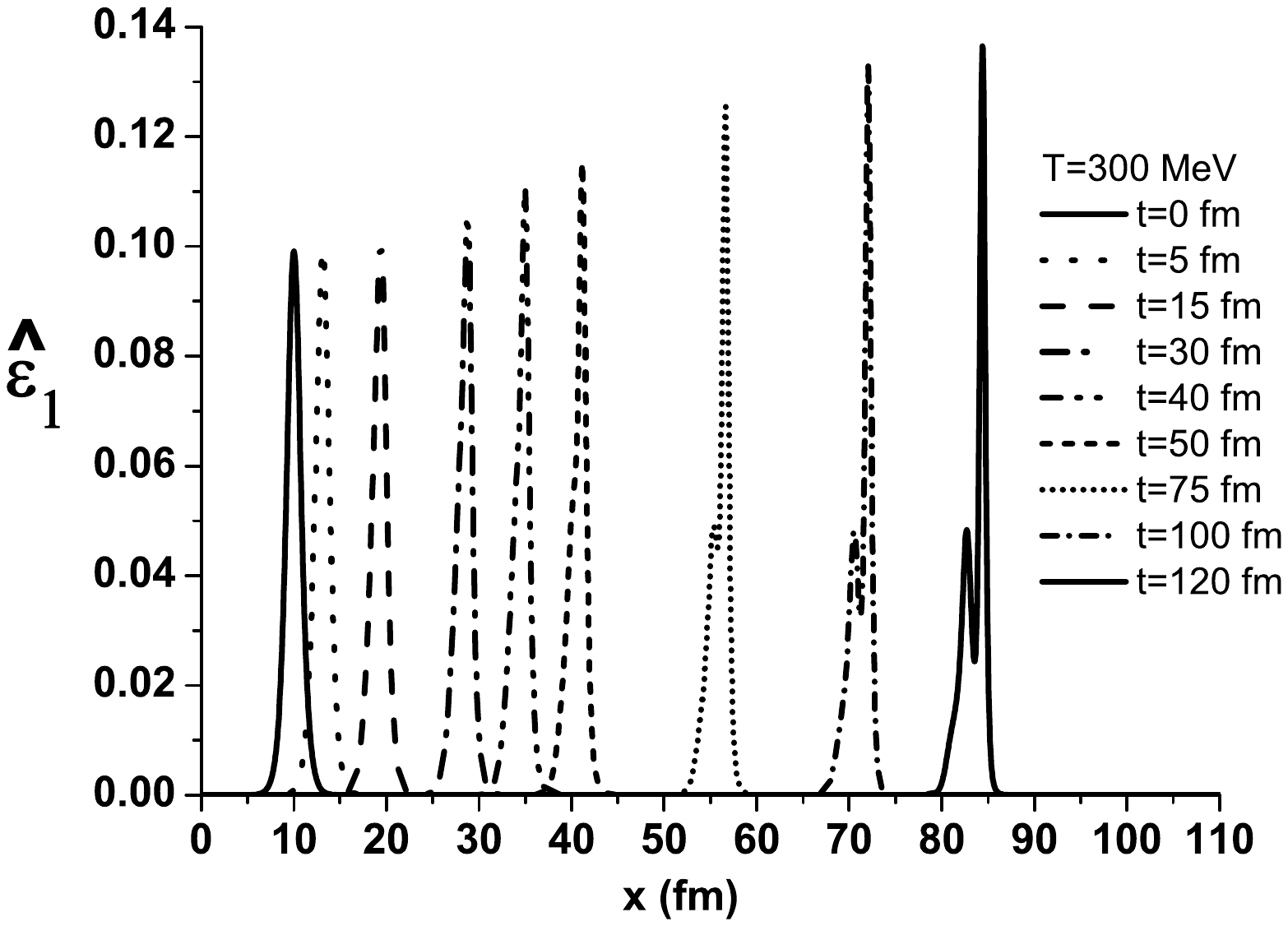}}
\caption[]{Time evolution of an energy density pulse at $T=300$ $MeV$.}
\label{fig8}
\end{figure}

In Fig.  \ref{fig:8a} we show the solution of  (\ref{bwqcdTfincomrefens}) with the initial condition
given by (\ref{exactumKdVTZEROlv}) with $A=0.01$,  $B=1$ fm and $T=300$ $MeV$. Fig. \ref{fig:8b} shows the same but with $A=0.1$ and $B=1$ fm.\
 As in the zero
temperature case, increasing the initial amplitude the breaking process and dispersion
develops earlier.  From  Fig.  \ref{fig:8a} we can conclude that it is possible to find an approximate solitonic behavior even when the differential equation
is not the KdV one.

\bigskip

\subsubsection{KP equation in cold QGP}

We now study the conditions in which the solution (\ref{ckpsol}) must be real and therefore  the constant $h_1$ must be positive. Moreover, following
Refs. \cite{kp2010,moslempp17} we assume that $a^2+b^2=1$ and consider $\hat{\rho_1}$ a normalized perturbation \cite{nos2013},
within the region (in the $u - a$ plane)  Eq.  (\ref{ckpsol}) is well defined and we can have solitons.  The parameters
are chosen to be:  $\rho_{0}= 1$ $fm^{-3}$ , $g = 1.15$ and  $m_G = 460$ $MeV$, which imply $c_s \simeq 0.64$ \cite{nos2013}. The stability
analysis can be made more rigorous with the introduction of the Sagdeev potential \cite{kp2010,moslempp17} by using
$\eta=\xi-ut=ar+bz-d \frac{ c_s \varphi^2 t } {2} -ut$ \ to rewrite equation (\ref{r3dckp}) as an energy balance equation.
For our present purposes the requirements  in \cite{nos2013} are sufficient.

\begin{figure}[ht!]
\centering
\subfigure[ ]{\label{fig:9a}
\includegraphics[scale=0.35]{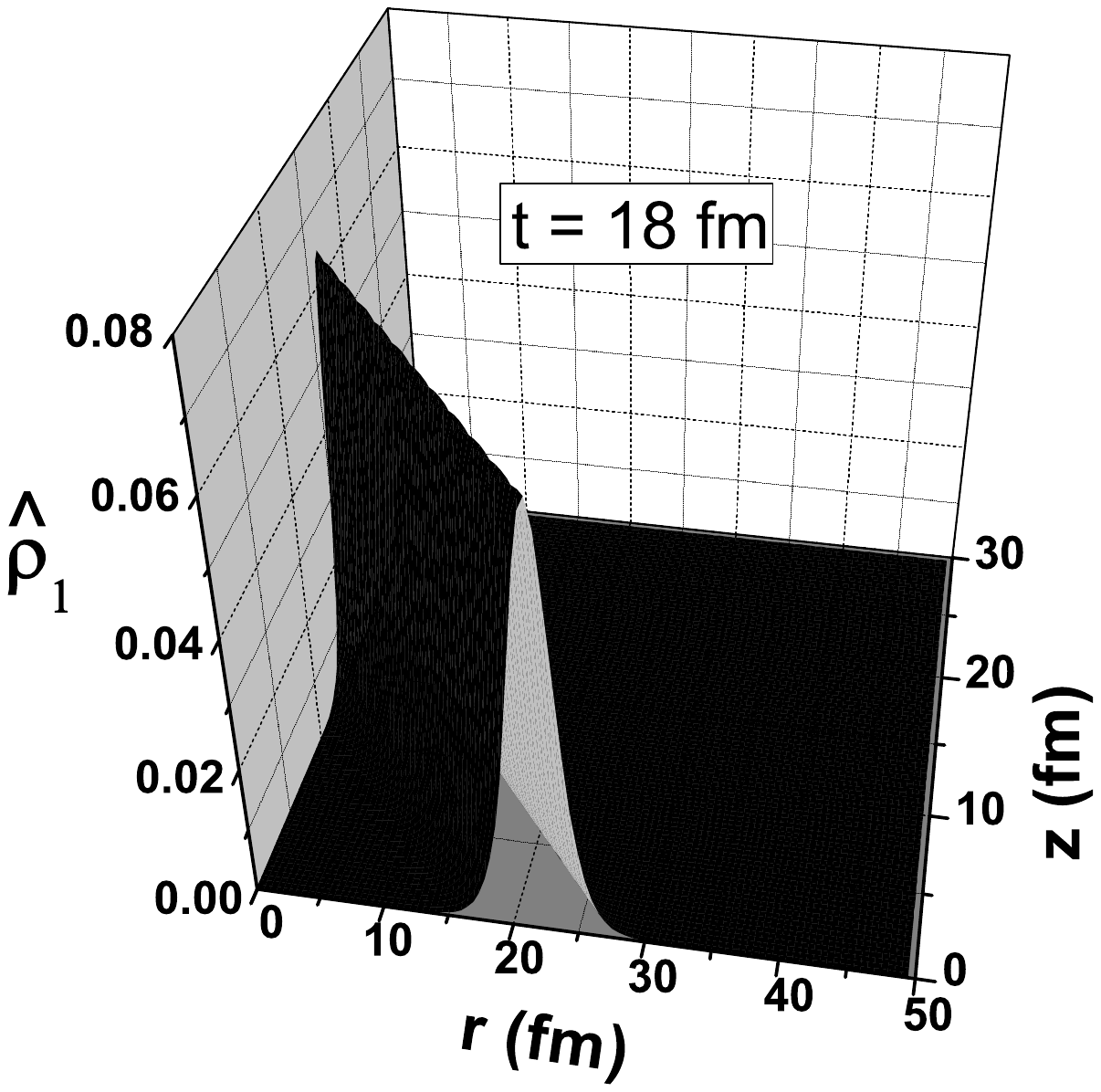}}
\subfigure[ ]{\label{fig:9b}
\includegraphics[scale=0.35]{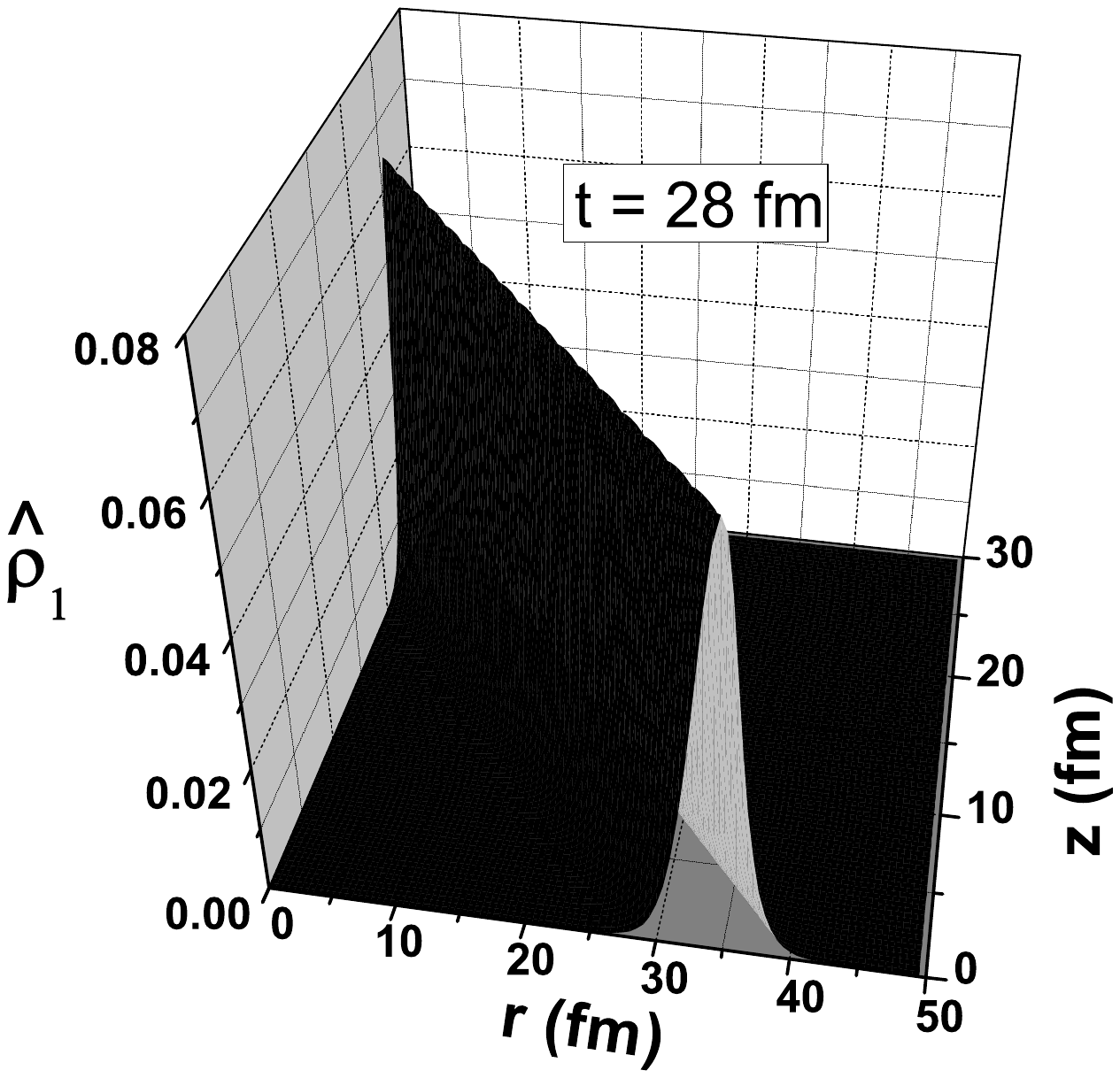}}
\caption[]{Cylindrical soliton time evolution in $r-z$ plane.}
\label{fig9}
\end{figure}

\begin{figure}[ht!]
\centering
\subfigure[ ]{\label{fig:10a}
\includegraphics[scale=0.35]{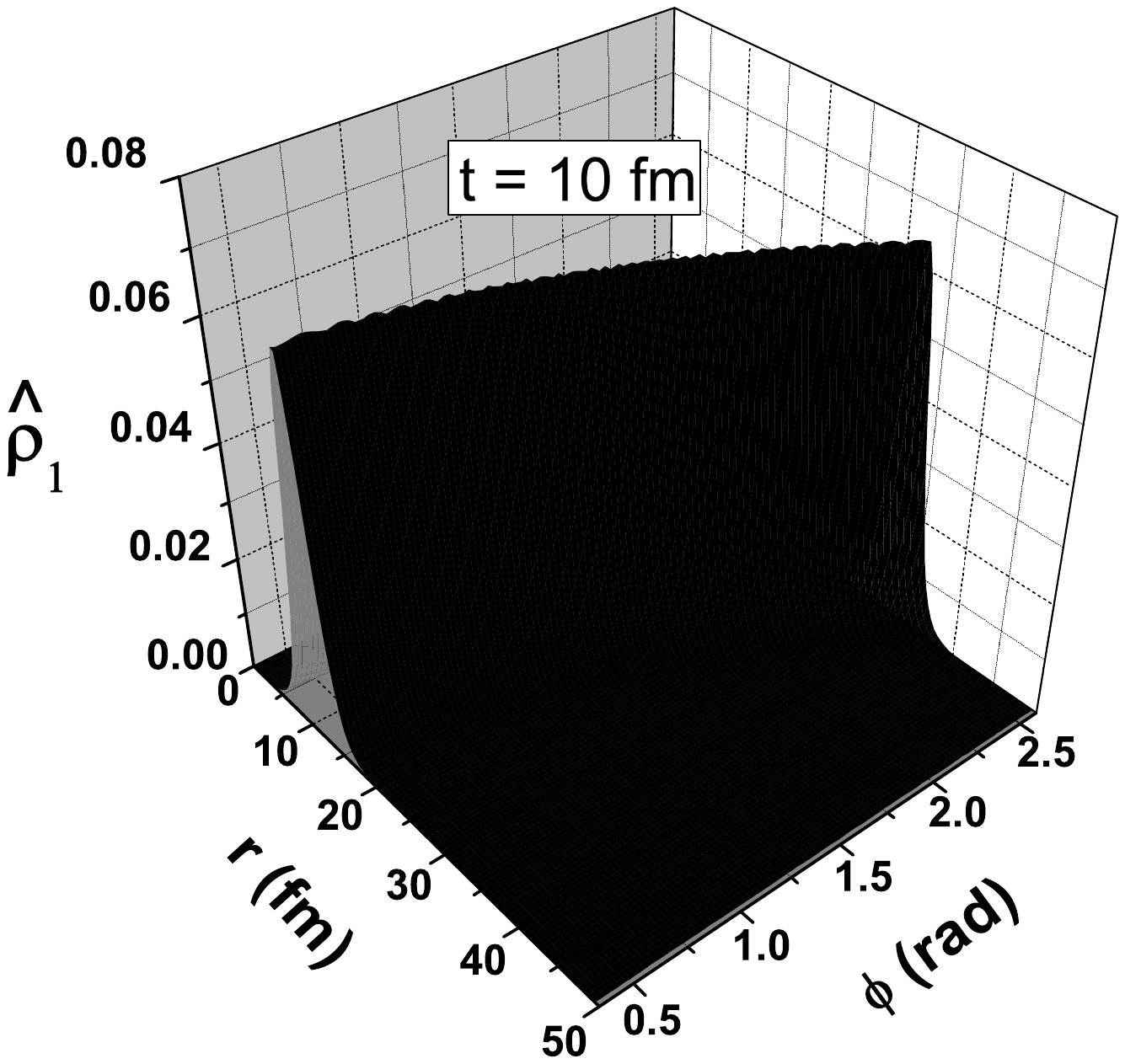}}
\subfigure[ ]{\label{fig:10b}
\includegraphics[scale=0.35]{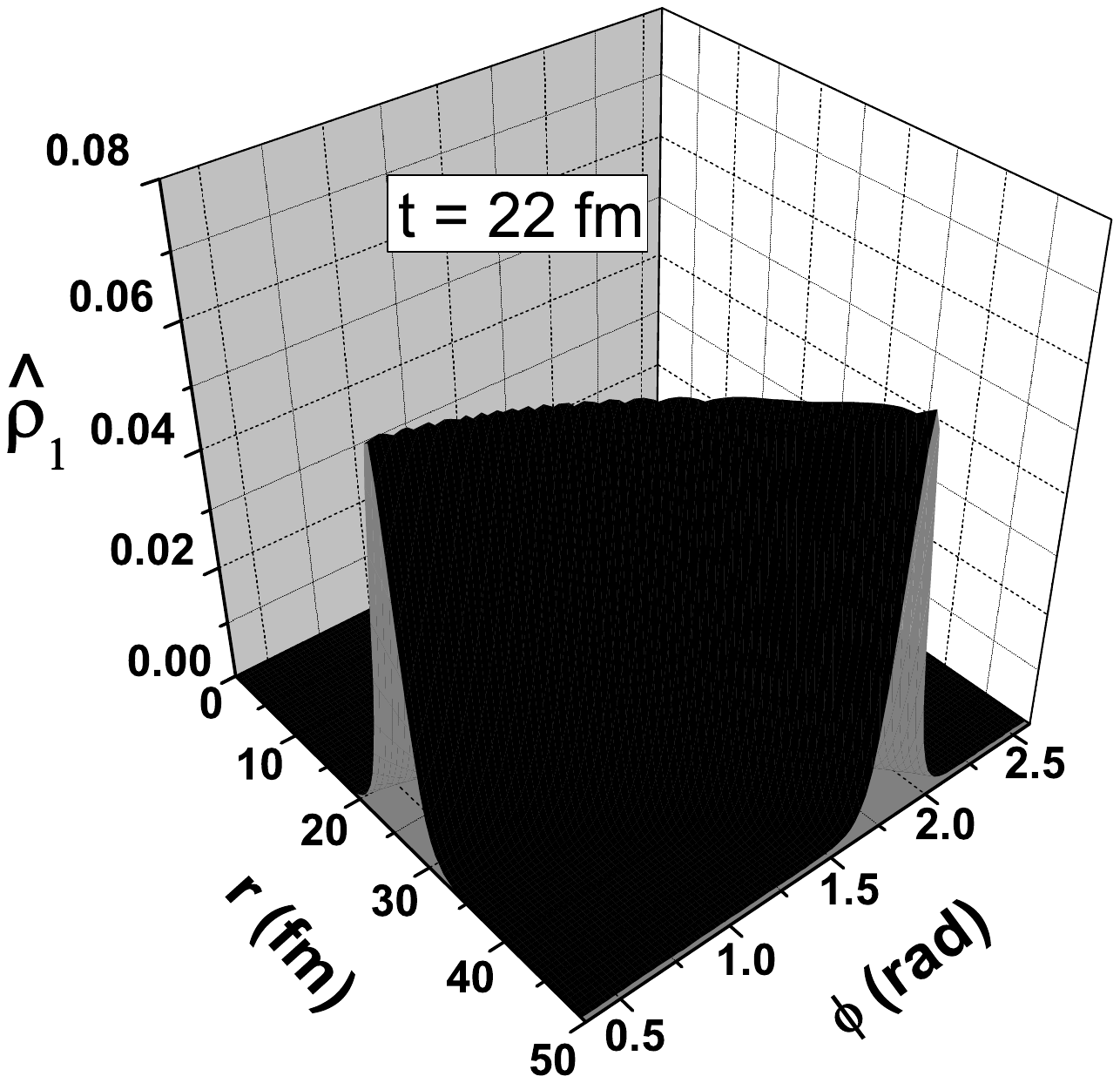}}
\caption[]{Cylindrical soliton time evolution in $r-\varphi$ plane.}
\label{fig10}
\end{figure}

The plot of the soliton evolution is presented in  Fig. \ref{fig9} and in Fig. \ref{fig10}.
We show a plot of (\ref{ckpsol}) with fixed $\varphi=0^{o}$,  $a=0.6$, $b=0.8$, $u=0.73$ and $z$ varying in the range
$0 \, \mbox{fm}  \leq z \,   \leq  30 \,  \mbox{fm} $ which  satisfies  the soliton conditions.
In Fig. \ref{fig:9a} the pulse is observed at  $t=18$ fm whereas in Fig. \ref{fig:9b} at $t=28$ fm.
From the Fig. \ref{fig:9a} we can see that the cylindrical pulse expands outwards in the radial direction. The regions with larger $z$
expand with a delay with respect to the central ($z=0$) region.

Keeping $z=1$ fm fixed,  we show the time evolution of  (\ref{ckpsol}) from $t=10$ fm (Fig. \ref{fig:10a}) to
$t=22$ fm (Fig. \ref{fig:10b}).  The azimuthal angle varies in the range  $20^{o}$ $\leq \varphi \leq $ $150^{o}$.
From the parenthesis in   (\ref{ckpsol}) we can see that the expansion velocity grows with the angle. This asymmetry can be
clearly seen in the figure, where the large angle ``backward'' region moves faster the small angle ``forward'' region. The breaking
of $z$ invariance and  azimuthal symmetry is entangled with the soliton stability \cite{nos2013} and with the physical properties of the system
(contained in the parameters $h_1$, $h_2$ and $c_s$).

\begin{figure}[ht!]
\centering
\subfigure[ ]{\label{fig:11a}
\includegraphics[scale=0.35]{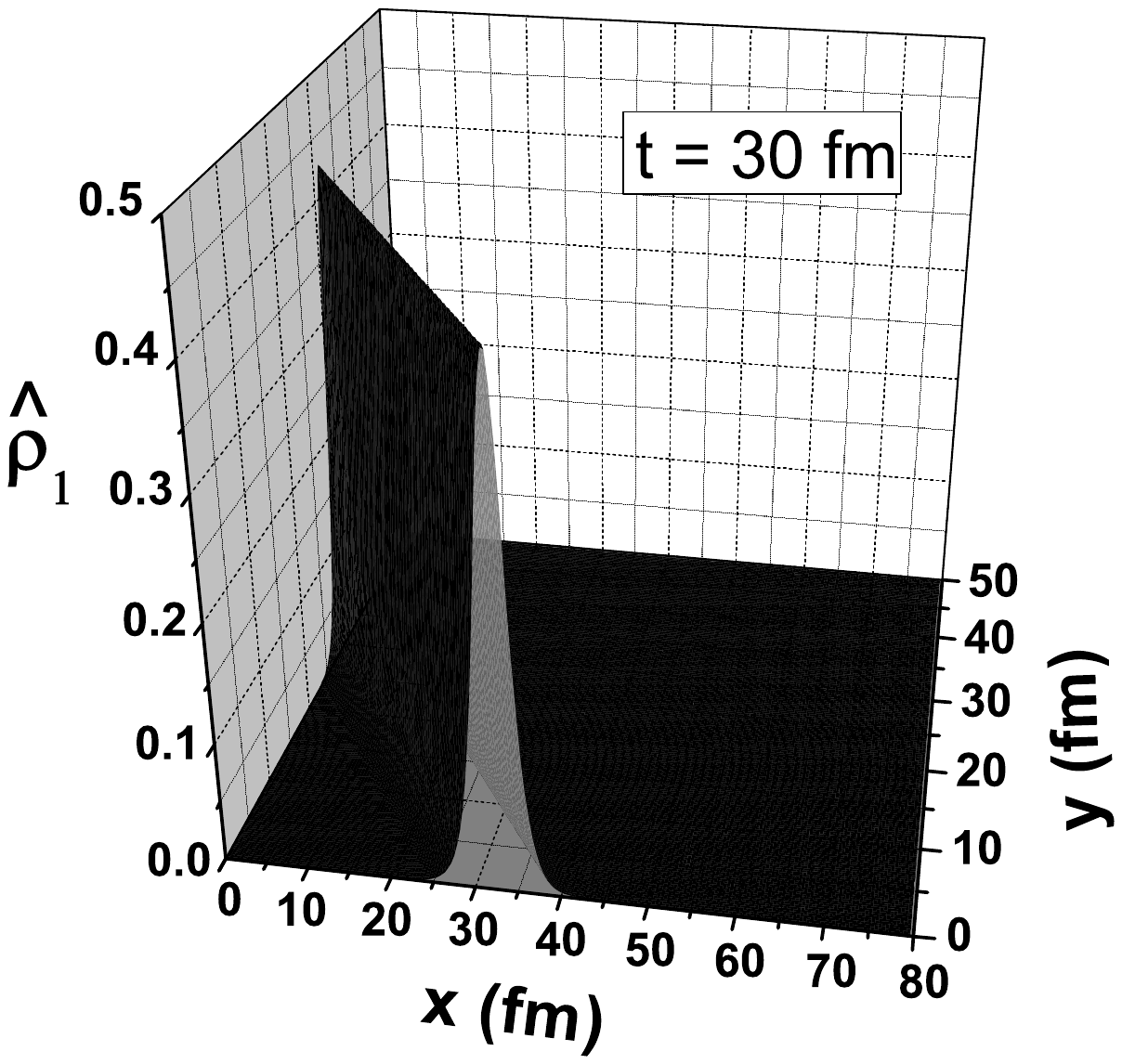}}
\subfigure[ ]{\label{fig:11b}
\includegraphics[scale=0.35]{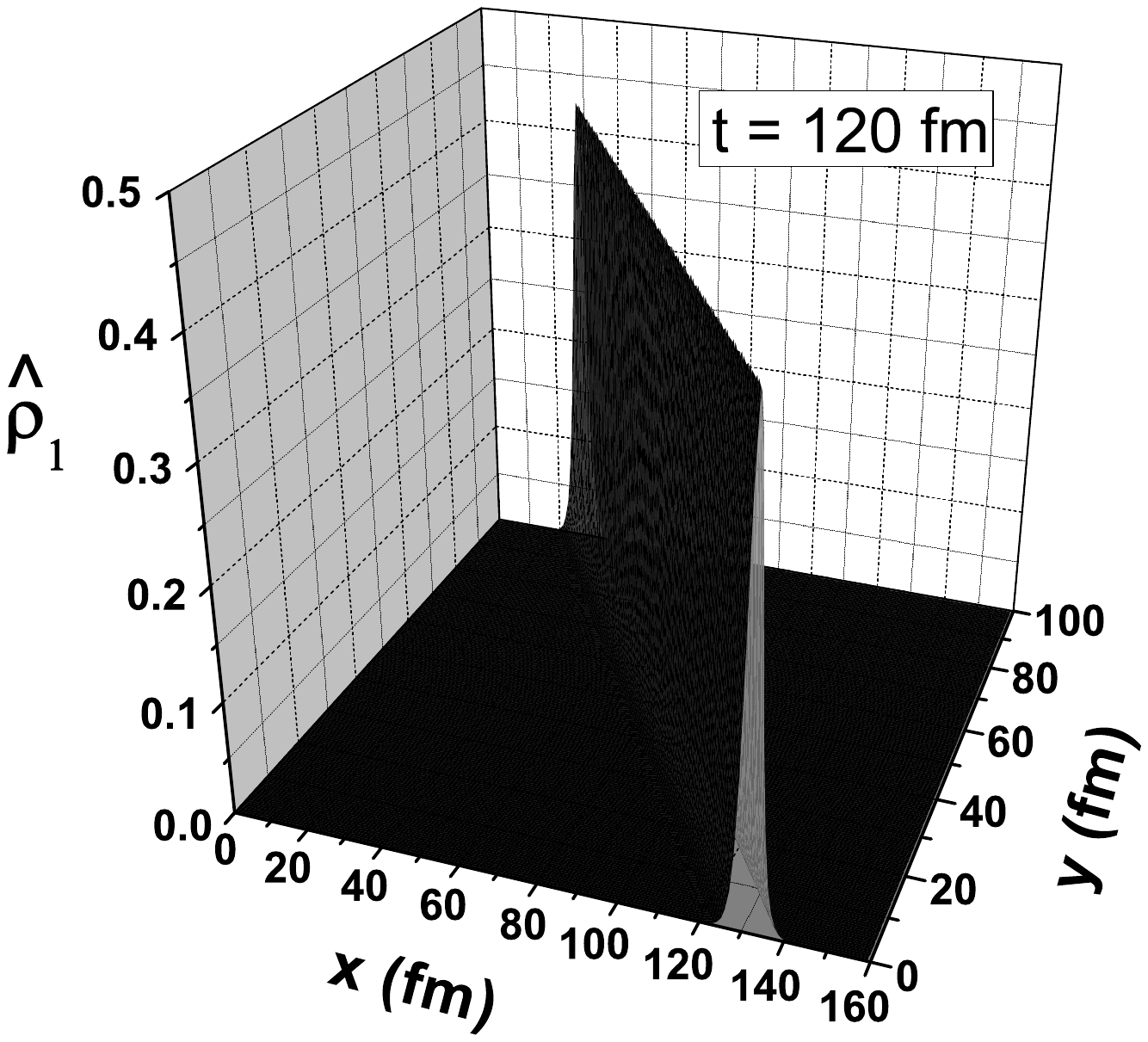}}
\caption[]{Cartesian soliton time evolution in $x-y$ plane.}
\label{fig11}
\end{figure}

We perform the study of the existence condition for the solution (\ref{kpsol}), which must be real and therefore  the constant $U-w$ must be positive.
The parameters are the same:
$\rho_{0}= 1$ $fm^{-3}$ , $g = 1.15$ and  $m_G = 460$ MeV, which imply $c_s \simeq 0.64$ \cite{nos2013}.
We also set ${\mathcal{C}}=0.5$ and  extend the condition in
Refs. \cite{kp2010,moslempp17} to ${\mathcal{A}}^2+{\mathcal{B}}^2+{\mathcal{C}}^{2}=1$ .  As mentioned before, $U > w$ and
again, $\hat{\rho_1}$ is a normalized perturbation \cite{nos2013}
and within the region (in the $U - {\mathcal{A}}$ plane),  (\ref{kpsol}) is well defined and we can have solitons \cite{nos2013}.  The stability analysis
can be performed more rigorously with the introduction of the Sagdeev potential \cite{kp2010,moslempp17} by using ${\mathcal{A}}x+{\mathcal{B}}y+
{\mathcal{C}}\ z-Ut$, to rewrite equation (\ref{kpxyztsual}) as an energy balance equation.
A simple example of soliton evolution is presented in  Fig. \ref{fig11}.
The plot of (\ref{kpsol}) with fixed $z=1$ fm,  ${\mathcal{A}}=0.6$, ${\mathcal{B}} \cong 0.62$, $U=0.66$ and $y$ varying in the range
$0 \, \mbox{fm}  \leq y \,   \leq  50 \,  \mbox{fm} $. The pulse is observed at two times:  $t=30$ fm (Fig. \ref{fig:11a}) and $t=120$ fm (Fig. \ref{fig:11b}).  From the figure we can see that the cartesian pulse expands outwards in the $x$ direction
keeping its shape and form as expected.

\begin{figure}[ht!]
\centering
\subfigure[ ]{\label{fig:12a}
\includegraphics[scale=0.35]{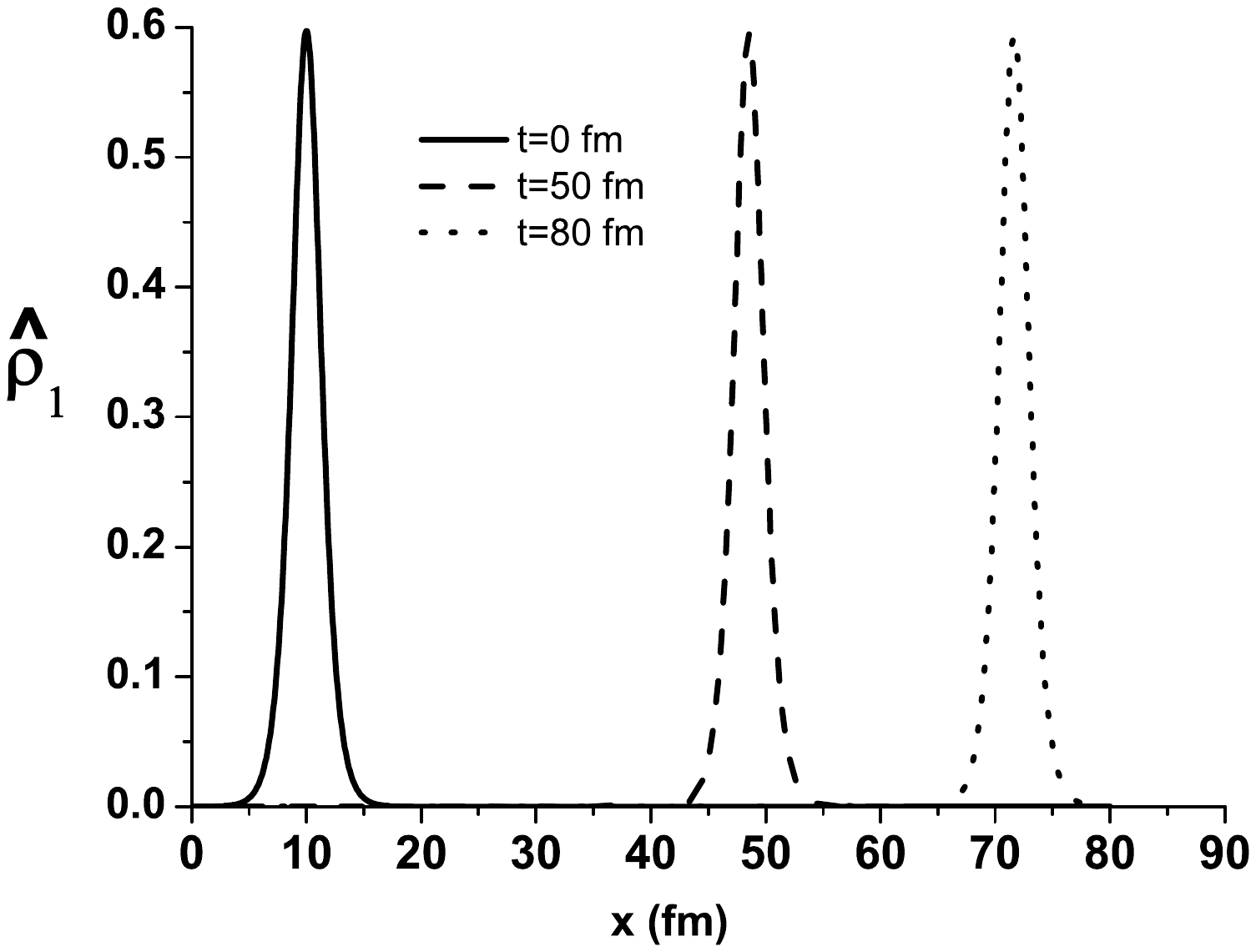}}
\subfigure[ ]{\label{fig:12b}
\includegraphics[scale=0.35]{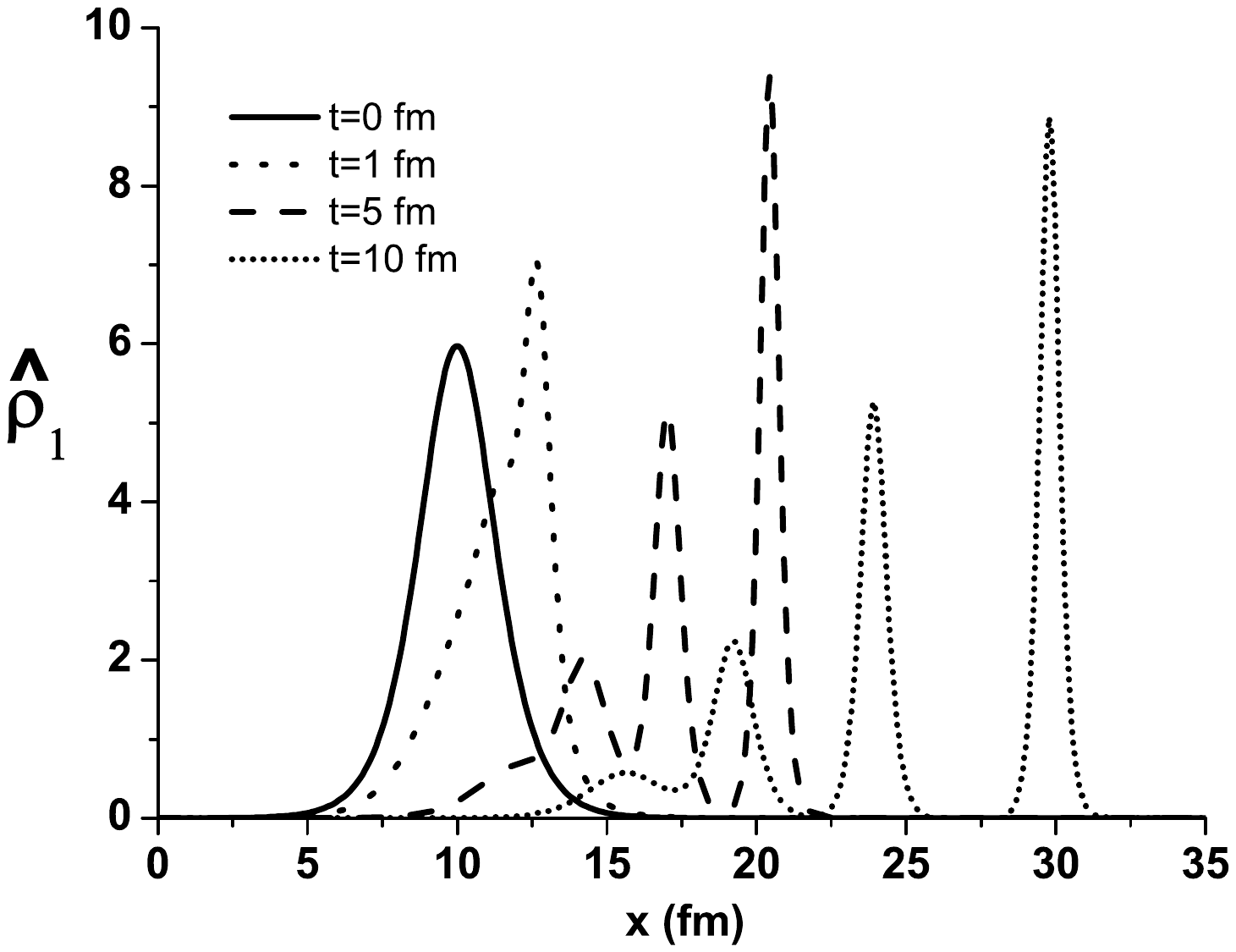}}
\caption[]{a) Soliton propagation in one dimensional Cartesian coordinate and b) Initial profile multiplied by a factor $10$ providing peaks plus ``radiation''.}
\label{fig12}
\end{figure}

Analogously to the nuclear soliton, the particular case of KP (\ref{kpxyztsual}): the KdV (\ref{kdvxt}), has the exact soliton solution (\ref{solitonkdvqcd}).
In Fig. \ref{fig12} we choose $u=0.8$, $\rho_{0}=2 \, fm^{-3}$ and
${c_{s}}^{2}=0.5$.

In Fig. \ref{fig:12a}, the numerical solution of (\ref{kdvxt}) for (\ref{solitonkdvqcd}) as initial condition
is studied for different times.  As expected, the evolution of the initial
gaussian-like pulse as a well defined soliton, keeping its shape and form.
In Fig. \ref{fig:12b}, we show again the numerical solution of (\ref{kdvxt}) for (\ref{solitonkdvqcd}) multiplied by a factor $10$.
Now the initial soliton posses amplitude $6.0$ and starts to develop secondary peaks, which are called ``radiation'' in the literature.  Further time evolution would increase the
strength of these peaks until the complete loss of localization.

\begin{figure}[ht!]
\centering
\subfigure[ ]{\label{fig:13a}
\includegraphics[scale=0.35]{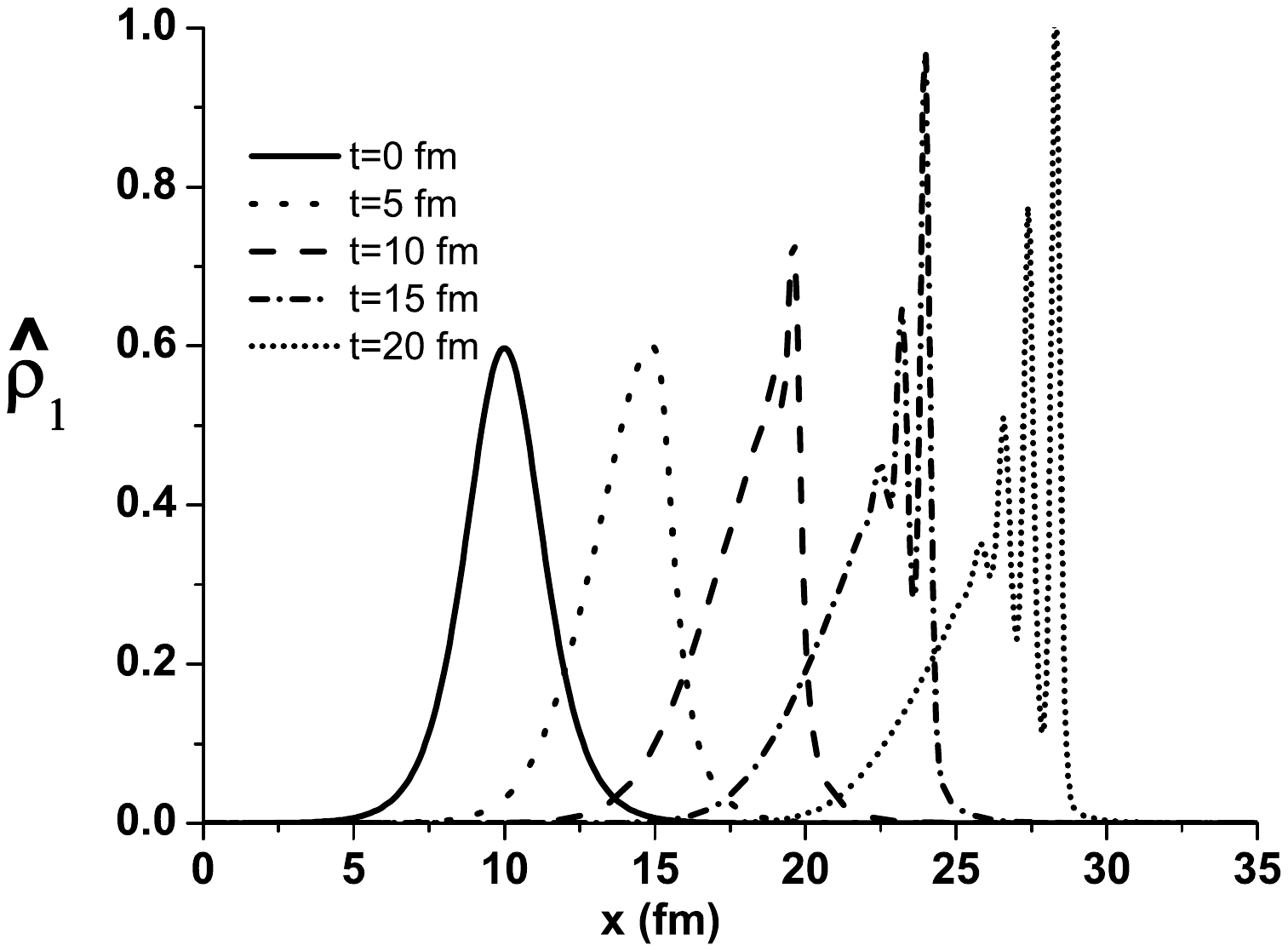}}
\subfigure[ ]{\label{fig:13b}
\includegraphics[scale=0.35]{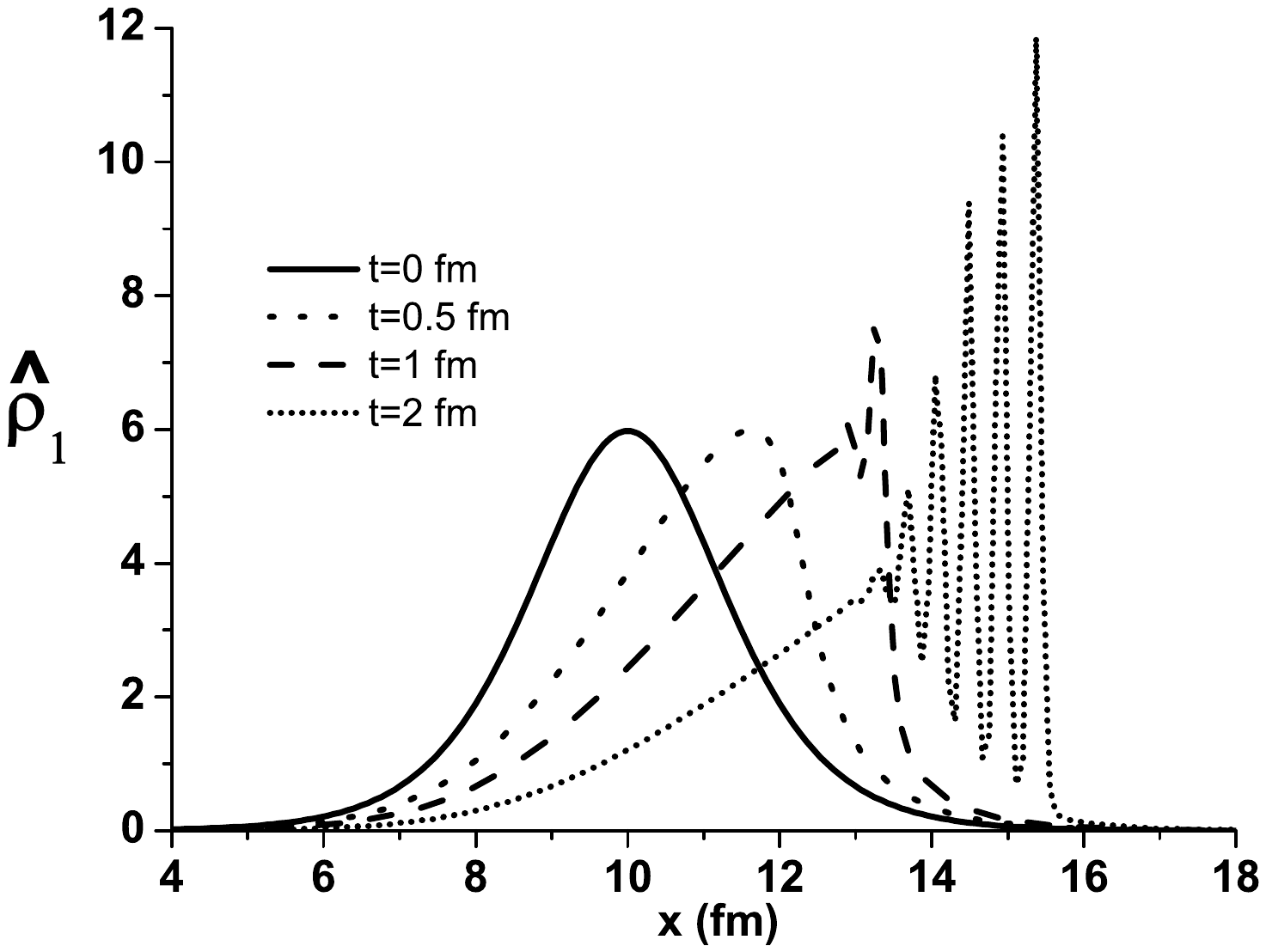}}
\caption[]{a) Breaking wave and b) breaking wave anticipation.}
\label{fig13}
\end{figure}

For the other particular case (\ref{bwmitxitauXt}) of the KdV (\ref{kdvxt}), which we show in
Fig. \ref{fig13}, where again we use (\ref{solitonkdvqcd}) with the same parameters listed above.
Fig. \ref{fig:13a}, we have the breaking following by its dispersion of the initial pulse.
And in Fig. \ref{fig:13b}, we use (\ref{solitonkdvqcd}) multiplied by a factor $10$, which we
observe the anticipation of the breaking following by its dispersion
of the initial pulse much earlier in comparison to Fig. \ref{fig:13a}.

\subsubsection{Burgers equation in hot QGP}

\begin{figure}[ht!]
\centering
\subfigure[ ]{\label{bulk}
\includegraphics[scale=0.35]{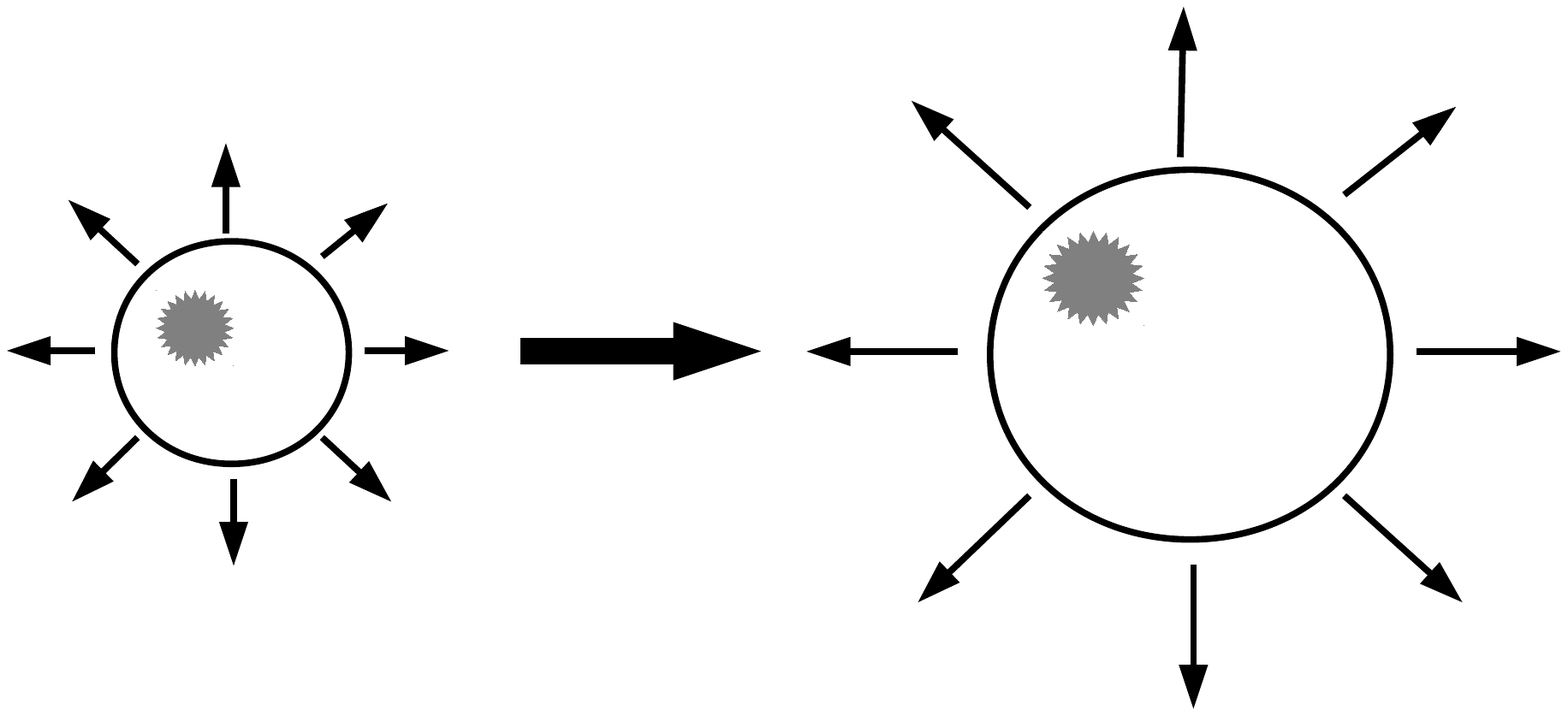}}
\subfigure[ ]{\label{tube}
\includegraphics[scale=0.35]{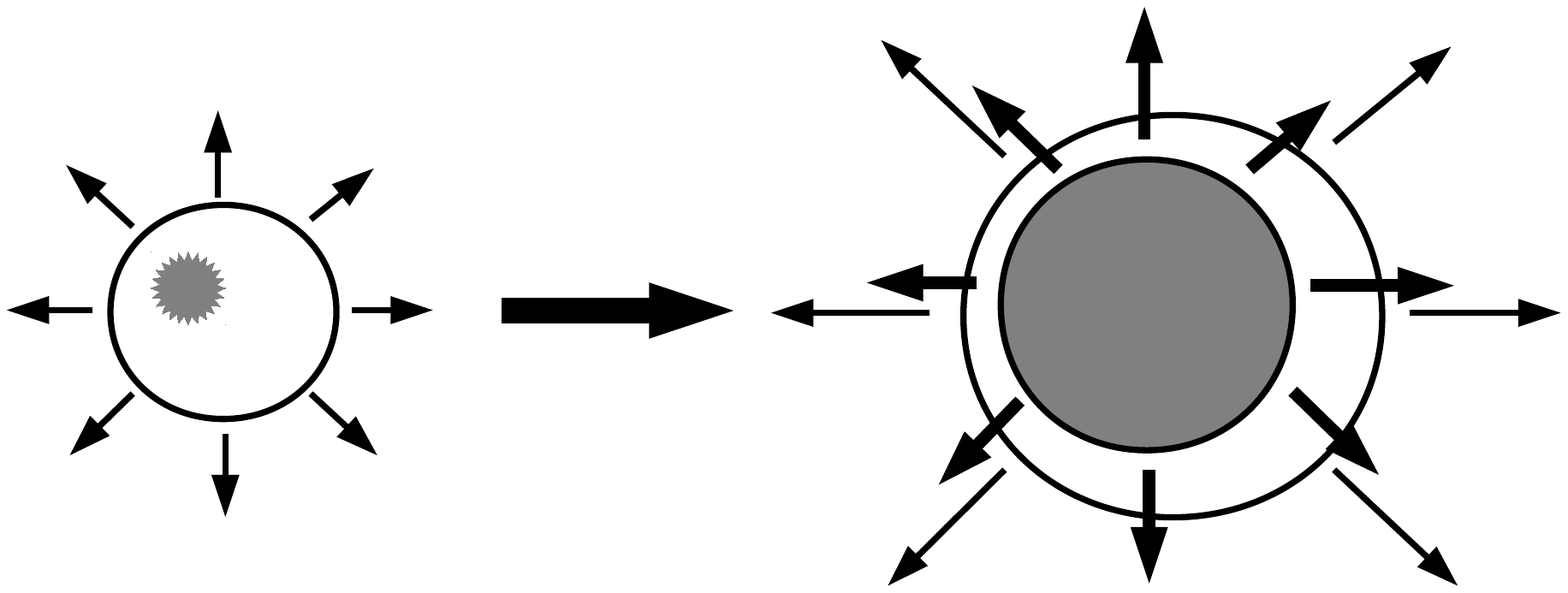}}
\caption[]{The large circle represents the front view of a cylindrical portion of QGP fluid. The small and dark circle represents a flux tube, i.e., a cylindrical perturbation with a higher energy density than the background. a) The tube expands slowly and the background faster. b) The tube
expands much faster than the background. }
\label{figviscous}
\end{figure}

We apply the perturbation concept to study the radial expansion of cylindrical flux tubes in a hot QGP. These tubes are  treated as perturbations in
the energy  density of the system which is formed in heavy ion collisions at RHIC and LHC as  explained in \cite{nos2012}.
During the expansion there is a ``competition'' between the background and the tube.  In Fig. \ref{bulk} the QGP background expands faster and the
tube is "pushed" outwards generating an anisotropic energy (and final particle) distribution.
In Fig. \ref{tube} the opposite occurs: the tube expands faster than the background, generating a different type of anisotropy in the final state.
In principle two and three-particle correlation measurements could distinguish between the two cases. Viscosity may change this picture. As shown in
\cite{nos2012}, a strong viscosity could rapidly damp the tube and reduce any anisotropy.
In Fig. \ref{figviscous} we have two extreme situations and something in-between may occur.
With our formalism we can study quantitatively the evolution of the tube.

We perform the numerical analysis for equations (\ref {burgers_final})
and (\ref{weq}) with the initial condition given  by a gaussian pulse in $\hat\varepsilon_{1}$:
\begin{equation}
\hat\varepsilon_{1}(r) = A \, e^{-r^2 / r^2_0}
\label{condinit}
\end{equation}
where the amplitude $A$ and the approximate width $r_0$ are parameters which depend on the dynamics of flux tube formation.
We shall refer to $r_0$ as the initial ``radius'' of the tube.
The tubes are perturbations, so we expect $A < 1$. According to \cite{nos2012} and references therein, the transverse size of the
tubes is of the order of $1$ fm and in our calculations $r_0 = 0.8 \, fm$.
We consider hot QGP at temperatures $T_{0}= 150 \, MeV$ and $T_{0}=500 \, MeV$ treated as an ideal fluid ($\eta/s = \zeta/s = 0$)
described by (\ref{weq}) and as a viscous fluid ($\eta/s =0.16$ and $\zeta/s = 0$) described by (\ref {burgers_final}) \cite{nos2012}.

\begin{figure}[ht!]
\centering
\subfigure[ ]{\label{fig:14a}
\includegraphics[scale=0.35]{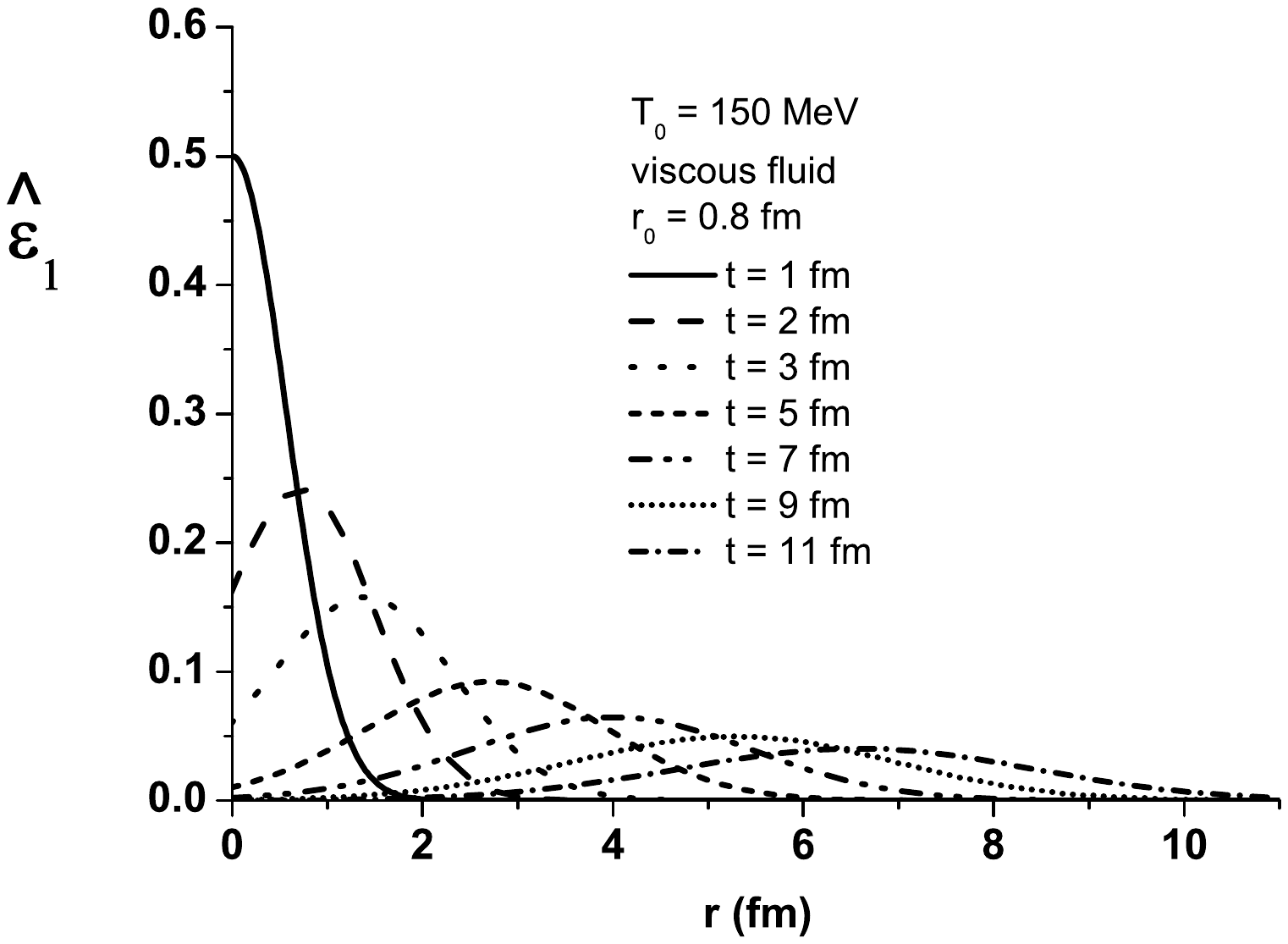}}
\subfigure[ ]{\label{fig:14b}
\includegraphics[scale=0.35]{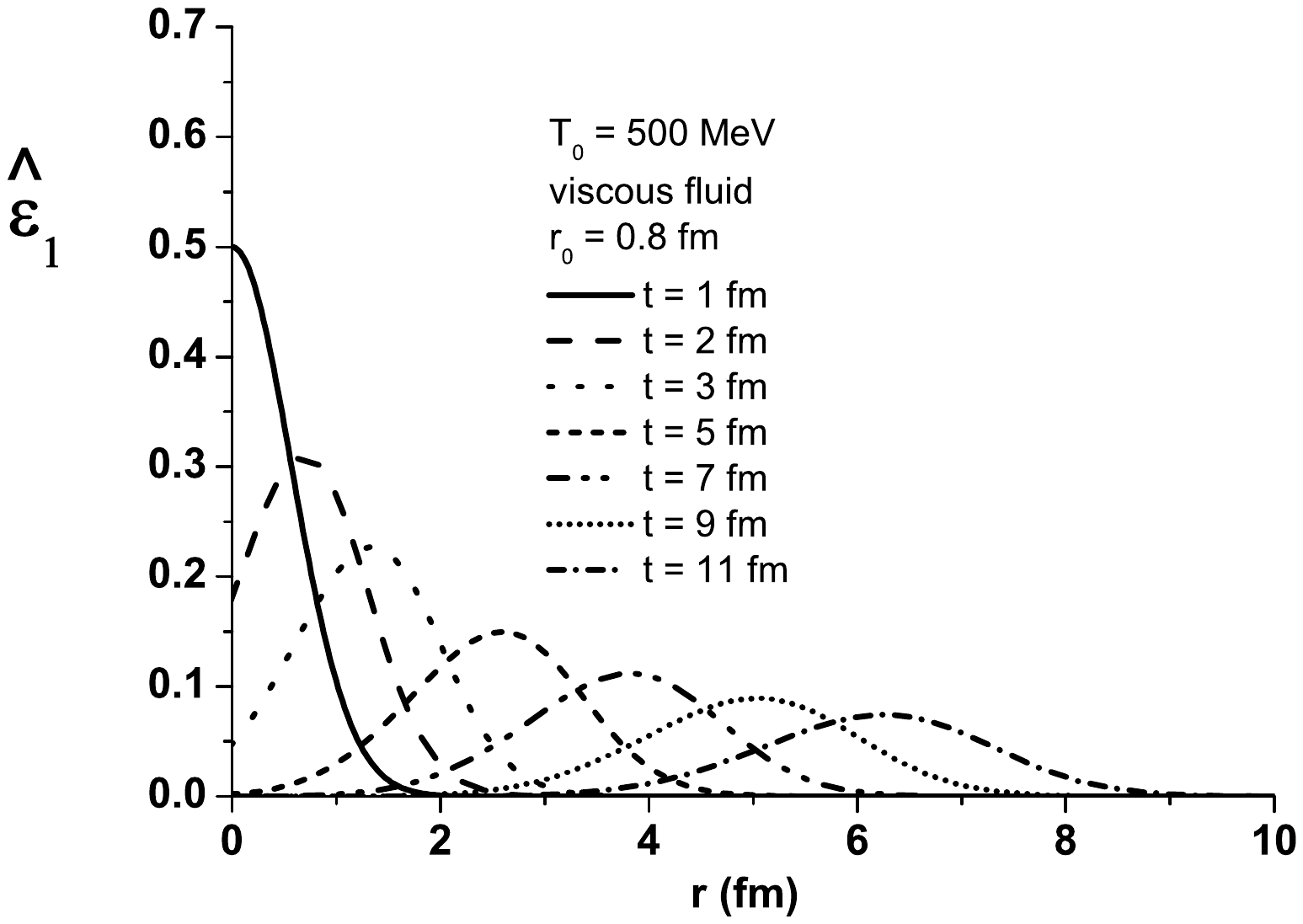}}
\caption[]{Solutions  for several times: a) $T_0=150$ $MeV$ and
b) $T_0=500$ $MeV$.}
\label{fig14}
\end{figure}

In Fig. \ref{fig14} we show numerical solutions of (\ref{burgers_final}) for a viscous fluid
using (\ref{condinit}) and we can observe the increasing temperature favors the tubular
structure survival.

\begin{figure}[ht!]
\centering
\subfigure[ ]{\label{fig:15a}
\includegraphics[scale=0.35]{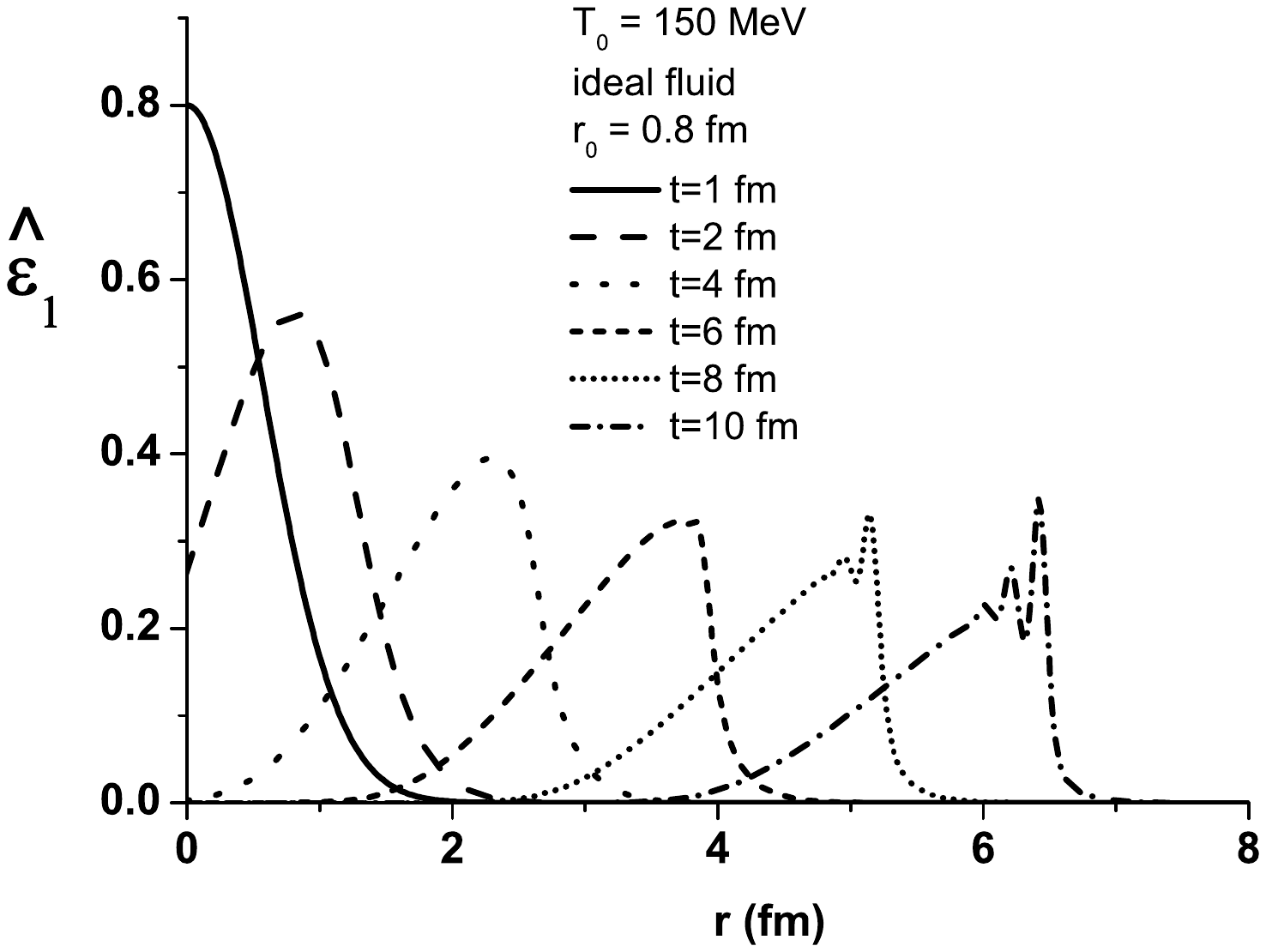}}
\subfigure[ ]{\label{fig:15b}
\includegraphics[scale=0.35]{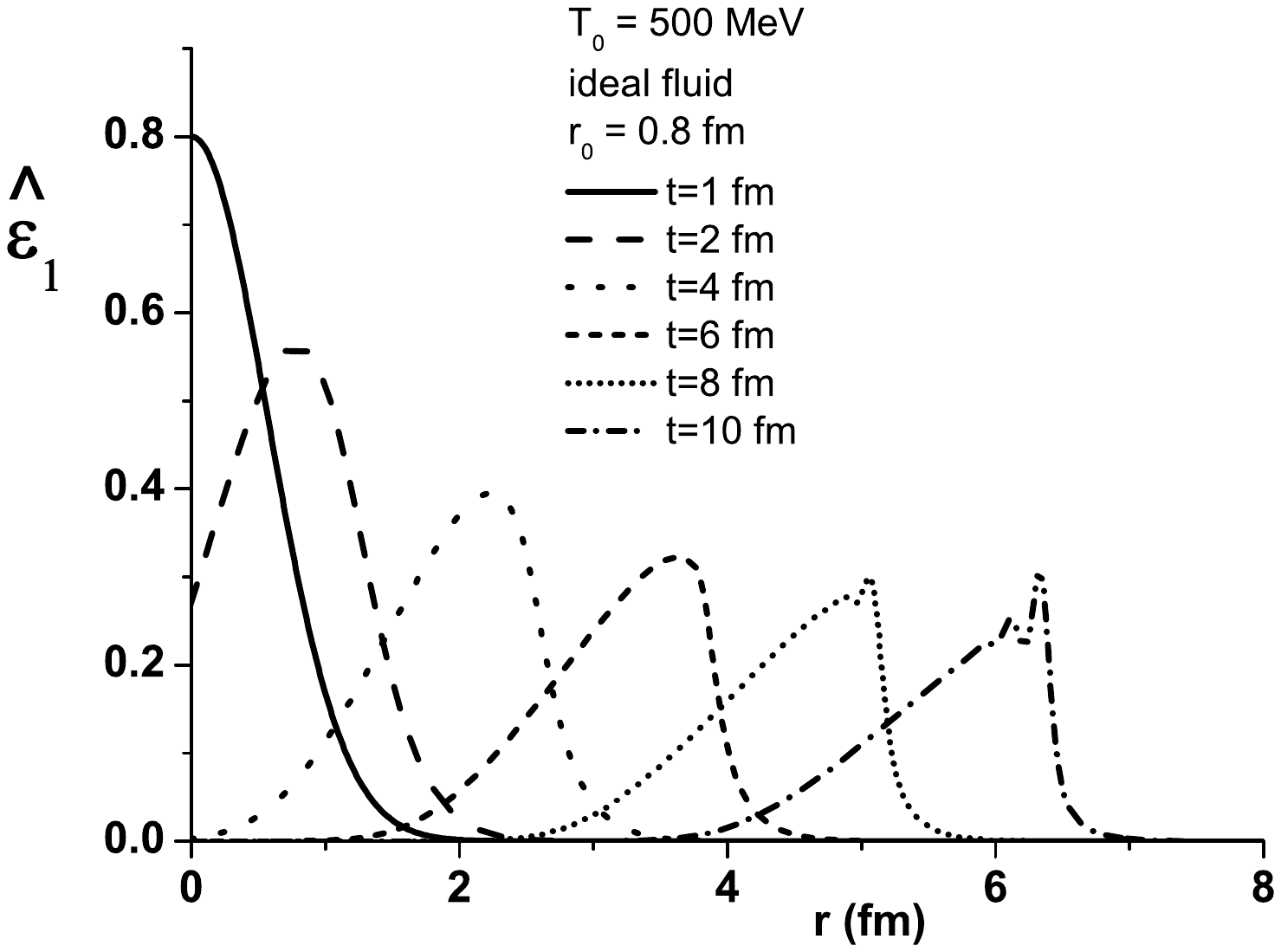}}
\caption[]{Analogous to Fig. \ref{fig14} but for the ideal fluid. a) $T_0=150$ $MeV$ and  b) $T_0=500$ $MeV$.}
\label{fig15}
\end{figure}

In Fig. \ref{fig15} we perform the same study for the ideal fluid (\ref{weq})
with (\ref{condinit}) and we show that breaking with dispersion occurs.

By comparison between Fig. \ref{fig14} and Fig. \ref{fig15} we conclude that viscosity dissipates the breaking followed by dispersion of the pulse. The tube
expands radially with a supersonic velocity and  in less than $4$ $fm/c$  it becomes a ``ring'', with a hole  in the middle.
Moreover, by this time the amplitude is already reduced by a factor two and the tube (or ring) looses the strength to ``push away'' the surrounding matter \cite{nos2012}.

\subsubsection{KP-Burgers in hot QGP}

\begin{figure}[ht!]
\centering
\subfigure[ ]{\label{fig:16a}
\includegraphics[scale=0.35]{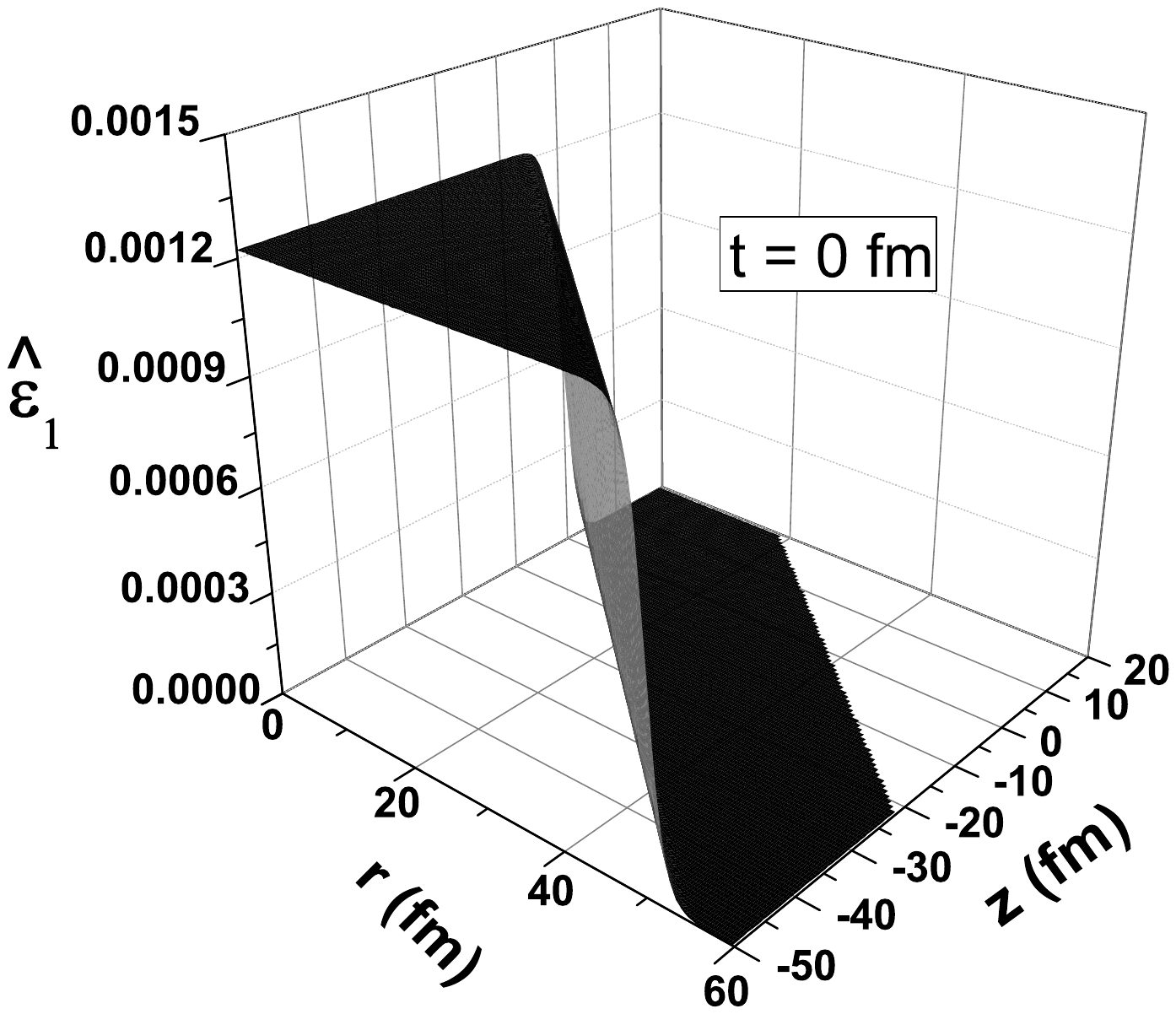}}
\subfigure[ ]{\label{fig:16b}
\includegraphics[scale=0.35]{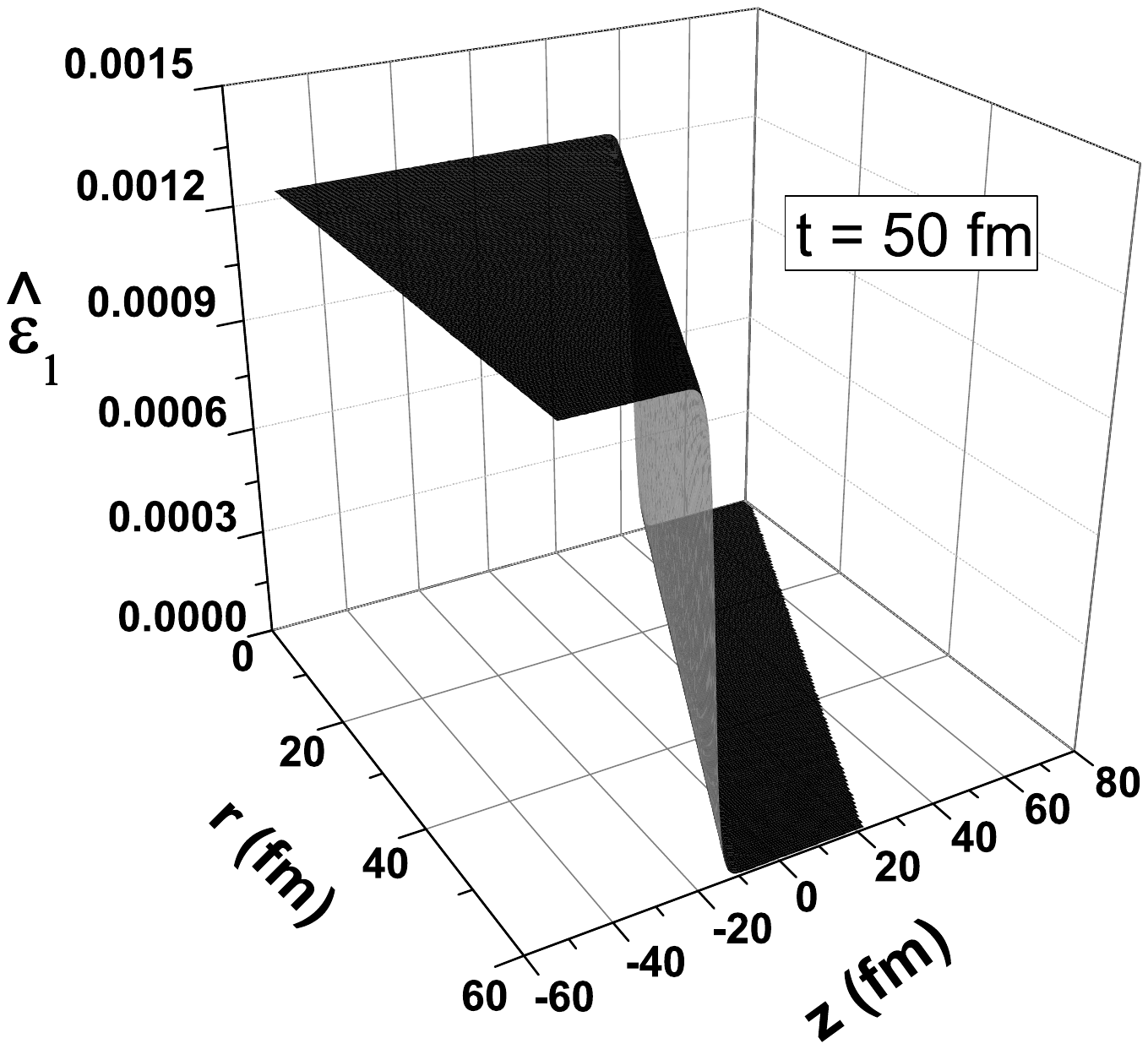}}
\caption[]{Shock wave evolution: a) $t=0$ $fm$ and b) $t=50$ $fm$.}
\label{fig16}
\end{figure}

As an example of time evolution for the analytical solution of the cKP-B equation
(\ref{cburgers3d}) we plot the time evolution of (\ref{burgimpoapair1final})
with $\varphi=0 \, rad$, $D=1$, $A=B=0.5$, $T_{0}=300 \, MeV$ and the viscous fluid with $\eta/s =0.16$ and $\zeta/s = 0$.  The Fig. \ref{fig16} shows the time evolution of the analytical traveling wave.

As it can be seen, (\ref{burgimpoapair1final}) obtained depends directly of the dissipative (viscous) coefficient term of (\ref{cburgers3d}), i.e., it is a
dissipation dominated solution.  The  Fig. \ref{fig16} shows a shock wave propagation in a
time interval of $50$ $fm$.

\section{Conclusion}

The discovery of the quark gluon plasma in the high energy heavy ion colliders brought relativistic hydrodynamics to the main stage of hadron physics.
Encouraged by the vigorous experimental program at CERN theoreticians of hydrodynamics embarked in an ambitious project: the calculation of observables
quantities with relativistic viscous hydrodynamics. The measurement of  two and three particle correlations  may be useful to study the propagation of waves
in the QGP.  Among the sources of waves we have fast partons crossing the medium and also flux tubes formed in the initial stage of heavy ion collisions.
In this work we have emphasized that these waves are most likely nonlinear and should be studied with the appropriate formalism, which, in our opinion, is
the Reductive Perturbation Method.  After making a survey of relativistic hydrodynamics we have presented the RPM in a simple and pedagogical way. The
equation of state of the two most relevant strongly interacting fluids, i.e., nuclear matter and the quark gluon plasma, was discussed. In both cases we have
made an effort to give a pedagogical introduction for non experts. In both cases  we have shown how to obtain a KdV soliton. The main responsible for the
appearance of these solitons are higher order derivative terms in the vector fields appearing in the Lagrangian of the system. In QHD it is enough to relax the
strict mean field approximation, in which all gradients vanish, and allow for spatial inhomogeneities in the vector field. A slightly more careful treatment of
the vector field equation of motion yields the desired term, derived from the Laplacian ${{\nabla}^{2}}{V_0}$.  In QCD, the situation is more complicated
because the gluon is massless. This makes impossible a simple estimate of the corresponding quantity  ${{\nabla}^{2}}{A^{a}_{0}}$. However a careful treatment of the
non-vanishing vacuum condensates leads to a dynamically generated gluon mass, $m_G$, which introduces a mass and a size scale and renders possible the estimate of
the desired Laplacian and the existence of solitons in the QGP.

Combining the equations of hydrodynamics with the equation of state and applying the RPM we have derived several differential equations for the perturbations
in energy and baryon density. These equations connect properties of the waves, such as width and speed, with the microscopic dynamical quantities of the fluids,
such as particle masses and couplings. Several of them have analytical solutions, which were presented and discussed. Some others must be solved numerically.
As expected for nonlinear equations, the results depend very strongly on the initial conditions.
An interesting finding is that, even when we do not have a KdV equation, in many cases the breaking wave equation has very long living localized solutions, which
resemble to solitons.  In some other cases, the initial pulses loose their localization and/or start to present rapid oscillations. All these features may
manifest themselves directly or indirectly in the experimental data. The analysis made here is still qualitative and a closer contact with phenomenology is still
to be made. For now, the obtained results   suggest that viscosity strongly affects the propagation of perturbations in the quark gluon plasma.
In order to confirm this statement the next step is to apply the RPM to the M\"uller-Israel-Stewart theory.

We hope to have convinced the reader that the study of nonlinear waves in hadron physics is an interesting and fast moving field. This study will help to
interpret and understand the data from the LHC.

\section{Appendix: Method of finite differences}

The most general form of a one-dimensional nonlinear wave equation with a second order dissipative term and a third order dispersive terms is given by:
\begin{equation}
\frac{\partial u}{\partial t}+\beta \frac{\partial}{\partial x}\bigg({\frac{u^{2}}{2}}\bigg)+\alpha \frac{\partial ^2u}{\partial x^2}+\mu \frac{\partial ^3u}{\partial x^3}=0
\label{KdV}
\end{equation}
For $\alpha =0$, $\beta$ and $\mu \neq 0$, it is the usual third order nonlinear Kortweg-de Vries equation (KdV). For $\mu =0$, $\beta$ and $\alpha \neq 0$ , it represents the Burgers equations. Finally, when only $\beta \neq 0$, we have the breaking wave equation.
In order to solve this equation numerically, we divide the integration region $ (0 \leq x \leq (n+1)h $ and $ 0 \leq t \leq m \Delta t ) $
with a space step $ h $ and a time step $ \Delta t $. Then, the wave function $ u(x,t) $, solution of equation (\ref{KdV}), assume the discrete values $ u_{i,0},u_{i,1},...,u_{i,j-2},u_{i,j-1},u_{i,j},u_{i,j+1},u_{i,j+2},...,u_{i,n},u_{i,n+1} $ in a given time $ t_i $. The expansion of this solution in Taylor's series leads to:
\begin{equation}
u_{i,j+1}=u_{i,j}+\left(\frac{\partial u}{\partial x}\right)_{i,j}h+{1\over 2!}\left(\frac{\partial ^2 u}{\partial x^2}\right)_{i,j}h^2+{1\over 3!}\left(\frac{\partial ^3 u}{\partial x^3}\right)_{i,j}h^3+O(h^4)
\label{Tayloravantex}
\end{equation}
or, alternatively, to:
\begin{equation}
u_{i,j-1}=u_{i,j}-\left (\frac{\partial u}{\partial x}\right )_{i,j}h+{1\over 2!}\left (\frac{\partial ^2 u}{\partial x^2}\right )_{i,j}h^2-{1\over 3!}\left(\frac{\partial ^3 u}{\partial x^3}\right)_{i,j}h^3+O(h^4)
\label{Tayloraréx}
\end{equation}

We combine these expressions conveniently to obtain finite difference expressions for first, second and third order centered partial space derivatives \cite{chapracanalle}:
\begin{equation}
\left (\frac{\partial u}{\partial x}\right )_{i,j}\approx \frac{u_{i,j+1}-u_{i,j-1}}{2h}=O_xU_i,\,\,j=2,\cdots ,n-1
\label{derivadavantex}
\end{equation}
\begin{equation}
\left (\frac{\partial ^2 u}{\partial x ^2}\right )_{i,j}\approx \frac{u_{i,j+1}-2u_{i,j}+u_{i,j-1}}{h ^2}=O_{xx}U_i\,\,\,\,j=2,\cdots ,n-1
\label{derivadasegunda}
\end{equation}
and
\begin{equation}
\left (\frac{\partial ^3 u}{\partial x ^3}\right )_{i,j}\approx \frac{-u_{i,j+2}+2u_{i,j+1}-2u_{i,j-1}+u_{i,j-2}}{2h ^3}=O_{xxx}U_i\,\,\,\,\,j=2,\cdots ,n-1
\label{derivadaterceira}
\end{equation}
in which we define
\begin{equation}
U_i=\left(
      \begin{array}{c}
        u_{i,2} \\
        \vdots \\
        u_{i,j+1} \\
        u_{i,j} \\
        \vdots \\
        u_{i,n-1} \\
      \end{array}
    \right)
\label{U}
\end{equation}
$ (u_{i,0}=u_{i,1}=u_{i,n}=u_{i,n+1}=0 $ are the boundary values on $x$ axe) and the operators:
\begin{equation}
O_x=\left(
      \begin{array}{ccccc}
        0 & 1/2h & 0 & \cdots & 0 \\
        -1/2h & 0 & 1/2h & \ldots & 0 \\
        \vdots & \ddots & \ddots & \ddots & \vdots \\
        0 & \cdots & -1/2h & 0 & 1/2h \\
        0 & \cdots & 0 & -1/2h & 0 \\
      \end{array}
    \right)
\label{Ox}
\end{equation}

\begin{equation}
O_{xx}=\left(
         \begin{array}{ccccc}
           -2/h^2 & 1/h^2 & 0 & \cdots & 0\\
            1/h^2 & -2/h^2 & 1/h^2 & \cdots & 0\\
           \vdots & \ddots & \ddots & \ddots & \vdots \\
           0 & \cdots & 1/h^2 & -2/h^2 & 1/h^2 \\
           0 & \cdots & 0 & 1/h^2 & -2/h^2 \\
         \end{array}
       \right)
\label{Oxx}
\end{equation}
and
\begin{equation}
O_{xxx}=\left(
          \begin{array}{ccccccc}
            0 & -1/h^3 & 1/2h^3 & 0 & 0 & \cdots & 0 \\
            1/h^3 & 0 & -1/h^3 & 1/2h^3 & 0 & \cdots & 0 \\
            -1/2h^3 & 1/h^3 & 0 & -1/h^3 & 1/2h^3 & \cdots & 0 \\
            \vdots & \ddots & \ddots & \ddots & \ddots & \ddots &  \vdots \\
            0 & \cdots & -1/2h^3 & 1/h^3 & 0 & -1/h^3 & 1/2h^3 \\
            0 & \cdots & 0 & -1/2h^3 & 1/h^3 & 0 & -1/h^3 \\
            0 & \cdots  & 0 & 0 & -1/2h^3 & 1/h^3 & 0 \\

          \end{array}
        \right)
  \label{Oxxx}
        \end{equation}

Analogously, we can expand the solution of equation (\ref{KdV}) around a time $ t_i $:
\begin{equation}
U_{i+1}=U_i+\left (\frac{\partial u}{\partial t}\right )_{i,j}\Delta t+{1\over 2}\left (\frac{\partial ^2 u}{\partial t^2}\right )_{i,j}(\Delta t)^2+O(h^3)
\label{Tayloravantet}
\end{equation}
to obtain
\begin{equation}
\left (\frac{\partial u}{\partial t}\right )_{i,j}\approx \frac{U_{i+1}-U_i}{\Delta t}
\label{dervadavantet}
\end{equation}

When we apply (\ref{dervadavantet}) and the operators (\ref{derivadavantex}), (\ref{derivadasegunda}) and (\ref{derivadaterceira}) in equation (\ref{KdV}), it can be represented equally in times $ t_{i}$ or $ t_{i+1} $, i.e., the following:
\begin{equation}
\frac{U_{i+1}-U_i}{\Delta t}=-\beta O_x[(1/2)U^2_i]-\alpha O_{xx}U_i-\mu O_{xxx}U_i
\label{ti}
\end{equation}
or
\begin{equation}
\frac{U_{i+1}-U_i}{\Delta t}=-\beta O_x[(1/2)U^2_{i+1}]-\alpha O_{xx}U_{i+1}-\mu O_{xxx}U_{i+1}
\label{ti+1}
\end{equation}

As long as the error one makes in both situations is the same, the Crank-Nicolson scheme \cite{cranknicolson} prescribes to take the mean value of these two possibilities, therefore
\begin{equation}
U_{i+1}=U_i-\frac{1}{2}{\Delta t}\beta O_x[(1/2)U^2_i+(1/2)U^2_{i+1}]-\frac{1}{2}{\Delta t}\alpha O_{xx}(U_{i+1}+U_i)-\frac{1}{2}{\Delta t}\mu O_{xxx}(U_{i+1}+U_i)
\label{média}
\end{equation}

Given a initial condition $ U_0 $, $ U_{i+1} $ is the solution of (\ref{média}) at any posterior time. However, this is a set of nonlinear algebraic equations. This problem becomes much more simple if we linearize it. We expand the nonlinear term in a Taylor's series, around the $i-th$ time $ t_i$ \cite{Djidjeli}:
\begin{equation}
\frac{1}{2}U^2_{i+1}=\frac{1}{2}U^2_i+\Delta t\left(\frac{\partial(1/2)u^2}{\partial u}\right)_i\left(\frac{\partial u}{\partial t}\right)_i+O(\Delta t^2)
\label{expansãot}
\end{equation}
which leads to
\begin{equation}
\frac{1}{2}U^2_{i+1}=\frac{1}{2}U^2_i+U_i(U_{i+1}-U_i)
\label{expansãotfinal}
\end{equation}
where we have used (\ref{dervadavantet}).

Replacing this result in (\ref{média}) we get:
\begin{equation}
U_{i+1}=U_i-\frac{1}{2}{\Delta t}\beta O_x(U_iU_{i+1})-\frac{1}{2}{\Delta t}\alpha O_{xx}(U_{i+1}+U_i)-\frac{1}{2}{\Delta t}\mu O_{xxx}(U_{i+1}+U_i)
\label{resultado}
\end{equation}

Using the matrix format of the operators $O_x$ (\ref{Ox}), $O_{xx}$ (\ref{Oxx}) and $O_{xxx}$ (\ref{Oxxx}), equation (\ref{resultado}) becomes a quin-diagonal algorithm:
\begin{equation}
        \left(
          \begin{array}{ccccccc}
            1-2s & -q_{i,j}+s & -p & 0 & 0 & \cdots & 0 \\
            q_{i,j}+s & 1-2s & -q_{i,j}+s & -p & 0 & \cdots & 0 \\
            p & q_{i,j}+s & 1-2s & -q_{i,j}+s & -p & \cdots & 0 \\
            \vdots & \ddots & \ddots & \ddots & \ddots & \ddots &  \vdots \\
            0 & \cdots & p & q_{i,j}+s & 1-2s & -q_{i,j}+s & -p \\
            0 & \cdots & 0 & p &q_{i,j}+s & 1-2s & -q_{i,j}+s \\
            0 & \cdots  & 0 & 0 & p & q_{i,j}+s & 1-2s \\
          \end{array}
        \right)
        \left(
        \begin{array}{c}
         u_{i+1,2} \\
         \vdots \\
         u_{i+1,j-1} \\
         u_{i+1,j} \\
         u_{i+1,j+1} \\
         \vdots \\
         u_{i+1,n-1} \\
         \end{array}
         \right)=
         \left(
        \begin{array}{c}
         r_{i,2} \\
         \vdots \\
         r_{i,j-1} \\
         r_{i,j} \\
         r_{i,j+1} \\
         \vdots \\
         r_{i,n-1} \\
         \end{array}
         \right)
  \label{matriz}
        \end{equation}
in which
\begin{equation}
p=-\frac{1}{4}\mu \frac{\Delta t}{h^3}
\label{p}
\end{equation}
\vspace{0.3cm}
\begin{equation}
q_{i,j}=-\frac{1}{4}\beta\frac{\Delta t}{h}u_{i,j-1}-2p
\label{q}
\end{equation}
\vspace{0.3cm}
\begin{equation}
s=\frac{1}{2}\alpha \frac{\Delta t}{h^2}
\end{equation}
and
\begin{equation}
r_{i,j}=(1+2s)u_{i,j}+pu_{i,j+2}-(2p+s)u_{i,j+1}+(2p-s)u_{i,j-1}-pu_{i,j-2}
\label{r}
\end{equation}

Therefore, from a given initial condition:
\begin{equation}
U_0=\left(
        \begin{array}{c}
         u_{0,2} \\
         \vdots \\
         u_{0,j-1} \\
         u_{0,j} \\
         u_{0,j+1} \\
         \vdots \\
         u_{0,n-1} \\
         \end{array}
         \right)
 \label{initialcondition}
\end{equation}
the set of linear algebraic equations (\ref{matriz}) can be iteratively solved, to obtain the solution $u(x,t)$ of equation (\ref{KdV}) in any posterior time $ t_{i+1}$ is given by:
\begin{equation}
U_{i+1}=\left(
        \begin{array}{c}
         u_{i+1,2} \\
         \vdots \\
         u_{i+1,j-1} \\
         u_{i+1,j} \\
         u_{i+1,j+1} \\
         \vdots \\
         u_{i+1,n-1} \\
         \end{array}
         \right)
 \label{solution}
\end{equation}

We turn now to the two dimensional extension of the KdV equation (\ref{KdVgeneral}), the so called Kadomtsev-Petviashvilli (KP) equation (\ref{KPgeneral}):
\begin{equation}
{\frac{\partial}{\partial x}}\Bigg\{\frac{\partial u}{\partial t}+\frac{\alpha_{1}}{2} \frac{\partial u^2}{\partial x}+\alpha_{2} \frac{\partial ^3u}{\partial x^3}\Bigg\}+
\alpha_{3}\frac{\partial ^2u}{\partial y^2}=0
\label{KPcart2d}
\end{equation}

Repeating all the preceding procedure for this equation, we obtain:
\begin{equation}
O_x U_{i+1}=O_xU_i-\frac{1}{2}{\Delta t}\alpha_1 O_{xx}(U_iU_{i+1})-\frac{1}{2}{\Delta t}\alpha_2 O_{xxxx}(U_{i+1}+U_i)-\frac{1}{2}{\Delta t}\alpha _3 O_{yy}(U_{i+1}+U_i)
\label{KPresultado}
\end{equation}
in which we define,
\begin{equation}
U_i=\left(
      \begin{array}{c}
        u_{i,2,k} \\
        \vdots \\
        u_{i,j+1,k} \\
        u_{i,j,k} \\
        \vdots \\
        u_{i,n-1,k} \\
      \end{array}
    \right)\,\,\,k=1,2,...,l
\label{U}
\end{equation}
$(u_{i,j,0}=u_{i,j,l+1}=0 $ are the boundary values in the $y$ direction) and the operators \cite{chapracanalle}:
\begin{equation}
O_{xxxx}=\left(
          \begin{array}{ccccccc}
            6/h^4 & -4/h^4 & 1/h^4 & 0 & 0 & \cdots & 0 \\
            -4/h^4 & 6/h^4 & -4/h^4 & 1/h^4 & 0 & \cdots & 0 \\
            1/h^4 & -4/h^4 & 6/h^4 & -4/h^4 & 1/h^4 & \cdots & 0 \\
            \vdots & \ddots & \ddots & \ddots & \ddots & \ddots &  \vdots \\
            0 & \cdots & 1/h^4 & -4/h^4 & 6/h^4 & -4/h^4 & 1/h^4 \\
            0 & \cdots & 0 & 1/h^4 & -4/h^4 & 6/h^4 & -4/h^4 \\
            0 & \cdots  & 0 & 0 & 1/h^4 & -4/h^4 & 6/h^4 \\
          \end{array}
        \right)
  \label{Oxxxx}
        \end{equation}
is the finite differences fourth order centered partial space derivative operator, and
 \begin{equation}
O_{yy}=\left(
         \begin{array}{ccccc}
           -2/h_y^2 & 1/h_y^2 & 0 & \cdots & 0\\
            1/h_y^2 & -2/h_y^2 & 1/h_y^2 & \cdots & 0\\
           \vdots & \ddots & \ddots & \ddots & \vdots \\
           0 & \cdots & 1/h_y^2 & -2/h_y^2 & 1/h_y^2 \\
           0 & \cdots & 0 & 1/h_y^2 & -2/h_y^2 \\
         \end{array}
       \right)
\label{Oyy}
\end{equation}
in which $h_y$ is the space step in the $y$ direction.

Replacing the matrix representation of the operators in equation (\ref{KPresultado}) we get:
\begin{equation}
\left(
          \begin{array}{ccccccc}
            c_{i,j,k} & d_{i,j,k} & a & 0 & 0 & \cdots & 0 \\
            b_{i,j,k} & c_{i,j,k} & d_{i,j,k} & a & 0 & \cdots & 0 \\
            a & b_{i,j,k} & c_{i,j,k} & d_{i,j,k} & a & \cdots & 0 \\
            \vdots & \ddots & \ddots & \ddots & \ddots & \ddots &  \vdots \\
            0 & \cdots & a & b_{i,j,k} & c_{i,j,k} & d_{i,j,k} & a \\
            0 & \cdots & 0 & a & b_{i,j,k} & c_{i,j,k} & d_{i,j,k} \\
            0 & \cdots  & 0 & 0 & a & b_{i,j,k} & c_{i,j,k} \\
\end{array}
          \right)\left(
        \begin{array}{c}
         u_{i+1,2,k} \\
         \vdots \\
         u_{i+1,j-1,k} \\
         u_{i+1,j,k} \\
         u_{i+1,j+1,k} \\
         \vdots \\
         u_{i+1,n-1,k} \\
         \end{array}
         \right)=\left(
        \begin{array}{c}
         e_{i,2,k} \\
         \vdots \\
         e_{i,j-1,k} \\
         e_{i,j,k} \\
         e_{i,j+1,k} \\
         \vdots \\
         e_{i,n-1,k} \\
         \end{array}
         \right)\,\,\,\, k=1,2,...,l
  \label{algoritmoKP}
        \end{equation}
in which
\begin{equation}
a=\alpha _2\frac{\Delta t}{h^3}
\label{a}
\end{equation}
\begin{equation}
b_{i,j,k}=1-4a+\frac{\alpha _1 \Delta t}{h}u_{i,j+1,k}
\label{bij}
\end{equation}
\begin{equation}
c_{i,j,k}=6a-\frac{2\alpha _3 h\Delta t}{h_y^2}-\frac{2\alpha _1\Delta t}{h}u_{i,j,k}
\label{cij}
\end{equation}
\begin{equation}
d_{i,j,k}=-1-4a+\frac{\alpha _1 \Delta t}{h}u_{i,j-1,k}
\label{dij}
\end{equation}
$$
e_{i,j,k}=-au_{i,j+2,k}+(1+4a)u_{i,j+1,k}+(\frac{2\alpha _3 h\Delta t}{h_y^2}-6a)u_{i,j,k}+(-1+4a)u_{i,j-1,k}-au_{i,j-2,k}
$$
$$
-\frac{2\alpha _3 h\Delta t}{h_y^2}(u_{i+1,j,k+1}+u_{i+1,j,k-1}+u_{i,j,k+1}+u_{i,j,k-1})
$$

The algorithm (\ref{algoritmoKP}) represents $l$ sets of linear algebraic equations. As long as each one of these sets are self-consistent, they must be solved iteratively, until the answer converges to the solution $(U^{i+1})^f$ after $f$ iterations. To stop the iterations, we can use the criterion \cite{Qcao}:
\begin{equation}
\frac{\|(U^{i+1})^f-(U^{i+1})^{f-1}\|}{\|(U^{i+1})^{f-1}\|}<\epsilon
\end{equation}

In what concerns the stability of this numerical method, the conservation of some quantities such as:
\begin{equation}
P(t)=\int_{-\infty} ^\infty \int_{-\infty} ^\infty udxdy
\label{pt}
\end{equation}
and
\begin{equation}
E(t)=\frac{1}{2}\int_{-\infty} ^\infty \int_{-\infty} ^\infty u^2dxdy
\label{pt}
\end{equation}
are frequently used as a criterion to verify its reliability. In refs. \cite{teukolsky} and \cite{iitaka}, this criterion
is used to show analytically that these numerical methods based on the Crank-Nicolson scheme are unconditionally stable.

\bigskip

\section{Acknowledgements}
This work has been partially funded by FAPESP, CNPq, CAPES and USP (NAP-QCD program).
The authors are grateful to J. Noronha, Y. Hama, F. Grassi, S. B.  Duarte, T. Kodama and
M. Munhoz for stimulating discussions.

\end{document}